\documentclass[floatfix,twocolumn,prd,showpacs,preprintnumbers,amsmath,nofootinbib,amssymb,superscriptaddress]{revtex4-1}

\usepackage{color} 

\definecolor{green}{rgb}{0,.5,0}
 \newcommand{\red}[1]{{ #1}}
\usepackage{graphicx}% Include figure files
\usepackage[subfigure]{graphfig}
\usepackage{epsfig}
\usepackage{dcolumn}% Align table columns on decimal point
\usepackage{amsmath}
\usepackage{multirow}
\usepackage{booktabs}   %%表格横线宏包
\usepackage{diagbox}
\usepackage{colortbl}
\usepackage{soul}
\usepackage{slashed}
\usepackage{physics}
\usepackage{bm}
\usepackage{tikz} %%画图工具
\usepackage[colorlinks,linkcolor=blue,citecolor=blue,urlcolor=blue]{hyperref}
\usepackage{soul}
\definecolor{red}{rgb}{1,0,0}

\def\bal#1\eal{\begin{align}#1\end{align}}
\newcommand{\MSbar}{{\overline{\text{MS}}}}

\def\MSbar{\overline{\mathrm{MS}}}

\def\RIMOM{\mathrm{RI/MOM}}
\def\RIpMOM{\mathrm{RI'/MOM}}
\def\JA{J_{A}^{\mu}}
\def\Js{J_{A,s}^{\mu}}
\def\Jns{J_{A,ns}^{\mu}}
\def\TR{{\displaystyle \mathrm{T}_{F}}}

\immediate\write18{texcount -inc -incbib -sum borra.tex > /tmp/wordcount.tex}

\begin{document}

\title{Realization of all-to-all fermion propagator for the first principle high accuracy strong interaction prediction}

\author{Zhi-Cheng Hu}
\affiliation{Institute of Modern Physics, Chinese Academy of Sciences, Lanzhou, 730000, China}
\affiliation{Institute of Theoretical Physics, Chinese Academy of Sciences, Beijing 100190, China}
\affiliation{University of Chinese Academy of Sciences, School of Physical Sciences, Beijing 100049, China}

\author{Ji-Hao  Wang}
\affiliation{Institute of Theoretical Physics, Chinese Academy of Sciences, Beijing 100190, China}
\affiliation{University of Chinese Academy of Sciences, School of Physical Sciences, Beijing 100049, China}

\author{Long Chen}
\affiliation{School of Physics, Shandong University, Jinan, Shandong 250100, China}

\author{Xiangyu Jiang}
\affiliation{CAS Key Laboratory of Theoretical Physics, Institute of Theoretical Physics, Chinese Academy of Sciences, Beijing 100190, China}
\affiliation{Department of Physics, Indiana University, Bloomington, Indiana 47405, USA}

\author{Liuming Liu}
\affiliation{Institute of Modern Physics, Chinese Academy of Sciences, Lanzhou, 730000, China}
\affiliation{University of Chinese Academy of Sciences, School of Physical Sciences, Beijing 100049, China}

\author{Shi-Hao Su}
\affiliation{Institute of Modern Physics, Chinese Academy of Sciences, Lanzhou, 730000, China}
\affiliation{University of Chinese Academy of Sciences, School of Physical Sciences, Beijing 100049, China}

\author{Peng Sun}
\email[Corresponding author: ]{pengsun@impcas.ac.cn}
\affiliation{Institute of Modern Physics, Chinese Academy of Sciences, Lanzhou, 730000, China}

\author{Yi-Bo Yang}
\email[Corresponding author: ]{ybyang@itp.ac.cn}
\affiliation{University of Chinese Academy of Sciences, School of Physical Sciences, Beijing 100049, China}
\affiliation{CAS Key Laboratory of Theoretical Physics, Institute of Theoretical Physics, Chinese Academy of Sciences, Beijing 100190, China}
\affiliation{School of Fundamental Physics and Mathematical Sciences, Hangzhou Institute for Advanced Study, UCAS, Hangzhou 310024, China}
\affiliation{International Centre for Theoretical Physics Asia-Pacific, Beijing/Hangzhou, China}

\begin{abstract}
We propose a ``blending" algorithm that projects the all-to-all fermion propagator onto spatial low-frequency modes (LFM) combines the projection with a stochastic estimate of spatial high-frequency modes (SHFM) at each time slice.
This approach enables the calculation of  correlation functions at arbitrary points for arbitrary hadron states in strongly interacting quantum field theories (QFT) with fermions, such as quantum chromodynamics (QCD).
Specifically, LFM allows the construction of spatially extended hadron states below a certain energy threshold by diagonalizing multi-fermion interpolation fields. Meanwhile, the local interactions required for N-point correlation functions in QFT can be approximated in an unbiased manner through a reweighted summation of both LFM and SHFM contributions. 
To demonstrate the efficiency of this algorithm, we obtained  
    $g_A^u=0.8408(86)$, $g_A^d= -0.3929(86)$, $g_A^s=-0.0381(57)$, $g_A^{u+d+s}=0.410(20)$ and $g_A^{u-d}=1.2337(84)$ 
for the nucleon at $m_{\pi}=135$ MeV and $a=0.077$ fm using 41  configurations. 
We also provide a consistency check of the pion electric form factor and charge radius derived from 3-point and 4-point correlation functions is also provided.
\end{abstract}

\maketitle

{\bf Introduction}: 
Lattice Quantum Chromodynamics (LQCD), as a first-principle approach of QCD, provides stringent constraints on hadron structure and serves as a crucial component for testing the Standard Model from first-principle calculations. 
Taking the proton's axial vector coupling \( g^{u-d}_A \) as an example, it is directly related to the neutron's weak decay \( n \rightarrow p e \bar{\nu} \).
The latest FLAG review reports  \( 1.265(20) \)~\cite{FlavourLatticeAveragingGroupFLAG:2024oxs, Liang:2018pis,Park_2021ypf,QCDSFUKQCDCSSM:2023qlx,Bali:2023sdi,Djukanovic:2024krw} for the LQCD calculation using the \( N_f = 2 + 1 \) flavor dynamical fermion configurations and \( 1.263(10) \)~\cite{chang2018per,Walker-Loud:2019cif,Jang:2023zts,Alexandrou:2023qbg} for the \( N_f = 2 + 1 + 1 \) case.

However, to achieve higher precision and compare with the global average \( 1.2756(13)\) \cite{10.1093/ptep/ptaa104} from the Particle Data Group (PDG), the QED correction, suggested to be 2\%~\cite{PhysRevLett.129.121801} is essential. Although the isospin symmetry breaking (ISB) effect is \( \mathcal{O}((m_d - m_u)^2 / \Lambda^2_{\text{QCD}}) \) \cite{chang2018per} for \( g_A \) and thus negligible regarding the current experimental uncertainty, it can have significant corrections to various hadron mass differences~\cite{BMW:2014pzb, Giusti:2017dmp}, and the hadron vacuum polarization (HVP) of the muon \( g-2 \)~\cite{Borsanyi:2020mff}.

Within the framework of perturbative expansion in \( \alpha \), the QED correction to an N-point correlation function can be represented as an (N+2)-point function, incorporating a pair of vertices \( \sum_q e_q \bar{q} \gamma_{\mu} A^{\mu} q \) at leading order. Similarly, the ISB corrections corresponds to an (N+1)-point function with an additional vertex \( (m_u - m_d)(\bar{u}u - \bar{d}d)/2 \). However, the exact computation of all-to-all fermion propagators incurs substantial computational and storage costs, scaling as \( \mathcal{O}(V^2) \) for a lattice volume \( V \equiv N_L^3 \times N_T \). Consequently, projecting the fermion propagator onto specific sources becomes necessary. Among the available techniques—such as the sequential source method \cite{Bernard:1985tm}, stochastic approaches with low-mode substitution (LMS)~\cite{Yang:2015zja,xQCD:2010pnl}, and field sparsening \cite{Li:2020hbj}—the field sparsening method, which projects the propagator onto random points across all time slices, is particularly advantageous for N-point (N \( \ge \) 4) correlation functions~\cite{Ma:2023kfr}.

In addition, there exists an important class of N-point functions that can be decomposed into several quark loops connected to the external hadron states via gluon exchanges. These diagrams are crucial for determining the sea quark contributions to hadron properties such as mass, momentum, and spin, and also QED and ISB correction at $\alpha^2_s$ order. However, their evaluation requires massive computation on additional propagators, typically using noise grid sources (e.g., \cite{XQCD:2013odc}) combined with enhancement techniques such as Split-even~\cite{Giusti:2019kff} and/or LMS.

Despite these advancements, extracting hadron properties from N-point functions is complicated by excited-state contamination. The distillation method~\cite{physrevd.80.054506,Morningstar:2011ka}, which projects the propagator onto its low-momentum modes, has proven effective in mitigating excited-state effects while enhancing statistical precision. This technique has been widely adopted in recent LQCD studies of the hadron spectrum (e.g.,~\cite{Edwards:2011jj,Dudek:2012xn,Yan:2024gwp}). However, its application to N-point functions is very limited, as the high-momentum modes are excluded in the projected propagators. 
Although the distillation method can effectively isolate excited states like \( \pi N \) or \( \pi\pi N \) in the nucleon’s spectrum~\cite{Hackl:2024whw}, it does not eliminate their contamination in nucleon matrix elements (e.g.,~\cite{Wang:2023omf,Alexandrou:2024tps,Barca:2024hrl})—limiting its usefulness for improving accuracy in such calculations.

In this work, we present a novel stochastic method for estimating the all-to-all fermion propagator, enabling high-precision calculations of N-point correlation functions—including those with disconnected quark diagrams—while suppressing excited-state contamination through the distillation technique. As demonstrations, we report sub-percent determinations of the axial coupling constant $g_A$,  and the pion vector form factor from four-point functions.

{\bf Framework of blending method:} To compute quark propagators and hadron correlation functions efficiently, we work in a reduced linear subspace of the lattice. For a fixed time slice, let $\mathcal{L}$ be the vector space of dimension $N_c N_L^3$ associated with the lattice sites and color degrees of freedom. The distillation method~\cite{physrevd.80.054506} constructs a low-rank approximation of the quark propagator by projecting onto the subspace $\mathcal{L}_1 \subset \mathcal{L}$ spanned by the $N_e$ lowest-lying eigenvectors $\{v_{\lambda_i}\}$ of the gauge-covariant spatial Laplacian. While this captures the long-distance physics relevant for constructing hadron interpolating operators, it discards short-distance modes essential for local current insertions.

We extend this framework by introducing a ``blending space" that incorporates both the controlled low-mode subspace $\mathcal{L}_1$ and a stochastic representation of its orthogonal complement $\mathcal{L}_2 = \mathcal{L} / \mathcal{L}_1$. We construct an orthonormal basis for the blending space as:
\[
\{\phi_i\}_{i=1}^{N_e + N_{\mathrm{st}}} \equiv \{v_{\lambda_1}, \dots, v_{\lambda_{N_e}}, \eta_1, \dots, \eta_{N_{\mathrm{st}}}\},
\]
where the $\{\eta_j\}$ are $N_{\mathrm{st}}$ random vectors uniformly sampled from $\mathcal{L}_2$. The identity operator on $\mathcal{L}$ can then be approximated as
\begin{align}
 \hat{I} = \sum_{i=1}^{\dim \mathcal{L}} |V_i\rangle\langle V_i| \approx \sum_{k=1}^{N_e + N_{\mathrm{st}}} \Omega^{(1)}_k |\phi_k\rangle\langle\phi_k|,   
\end{align}
with the weights
\[
\Omega^{(1)}_k = \begin{cases} 1 & \text{for } k \le N_e, \\ \frac{\dim \mathcal{L}_2}{N_{\mathrm{st}}} & \text{for } k > N_e. \end{cases}
\]
The exact treatment of the first $N_e$ modes preserves the distillation subspace essential for constructing low-lying hadron states. The stochastic modes in $\mathcal{L}_2$ are weighted to yield an unbiased estimate of the identity on the complementary subspace, with the weight $\dim \mathcal{L}_2 / N_{\mathrm{st}}$ ensuring the correct expectation value. This approximation becomes exact in the limit $N_{\mathrm{st}} \to \dim \mathcal{L}_2$.

The operator $M_{\mathcal{O}}$ in a quark bilinear current $\mathcal{O} = \int d^3x\, d^3y\, \bar{q}(x) M_{\mathcal{O}}(x,y) q(y)$, can also be projected onto the blending space. Inserting the approximate identity on both sides of the current yields
\begin{align}
M_{\mathcal{O}}(x,y) \approx \sum_{i,j=1}^{N_e+N_{\mathrm{st}}} |\phi_i(x)\rangle \mathcal{O}_{ij} \langle \phi_j(y)|,
\end{align}
where $\mathcal{O}_{ij} \equiv \Omega^{(2)}_{ij} \int d^3z\, d^3w\, \langle\phi_i(z)| M_{\mathcal{O}}(z,w) |\phi_j(w)\rangle$. The weight tensor
\[
\Omega^{(2)}_{ij} = \begin{cases} 
1 & i,j \le N_e, \\
\frac{\dim \mathcal{L}_2}{N_{\mathrm{st}}} & i \le N_e < j \text{ or } j \le N_e < i, \\
\frac{\dim \mathcal{L}_2(\dim \mathcal{L}_2-1)}{N_{\mathrm{st}}(N_{\mathrm{st}}-1)} & i,j > N_e,\ i\neq j, \\
\frac{\dim \mathcal{L}_2}{N_{\mathrm{st}}} & i=j> N_e
\end{cases}
\]
corrects for double counting and ensures the estimator remains unbiased. The distinct treatment of diagonal stochastic terms reflects that the identity on $\mathcal{L}_2$ is recovered as the outer product of random vectors, whose expectation value requires careful normalization. For currents connecting different time slices ($x_4 \neq y_4$), the weight factor can be simplified to $\Omega^{(2)}_{ij} = \Omega^{(1)}_i \Omega^{(1)}_j$.

Following the distillation paradigm, we compress the all-to-all quark propagator $S$ by projecting it onto the blending space. The resulting ``blended propagator" is
\[
S_{ij}(t_1,t_2) = \int d^3x\, d^3y\, \langle\phi_i(\vec{x},t_1)| S(\vec{x},t_1;\vec{y},t_2) |\phi_j(\vec{y},t_2)\rangle,
\]
which reduces the storage from $\mathcal{O}((4N_c N_L^3 N_T)^2)$ to $\mathcal{O}((4N_T(N_e+N_{\mathrm{st}}))^2)$ for the projected propagator plus $\mathcal{O}(N_T N_c N_L^3 (N_e+N_{\mathrm{st}}))$ for the basis vectors. This compression enables flexible and efficient computation of correlation functions with arbitrary interpolating operators and current insertions.

   \begin{figure}
        \includegraphics[width=0.46 \textwidth]{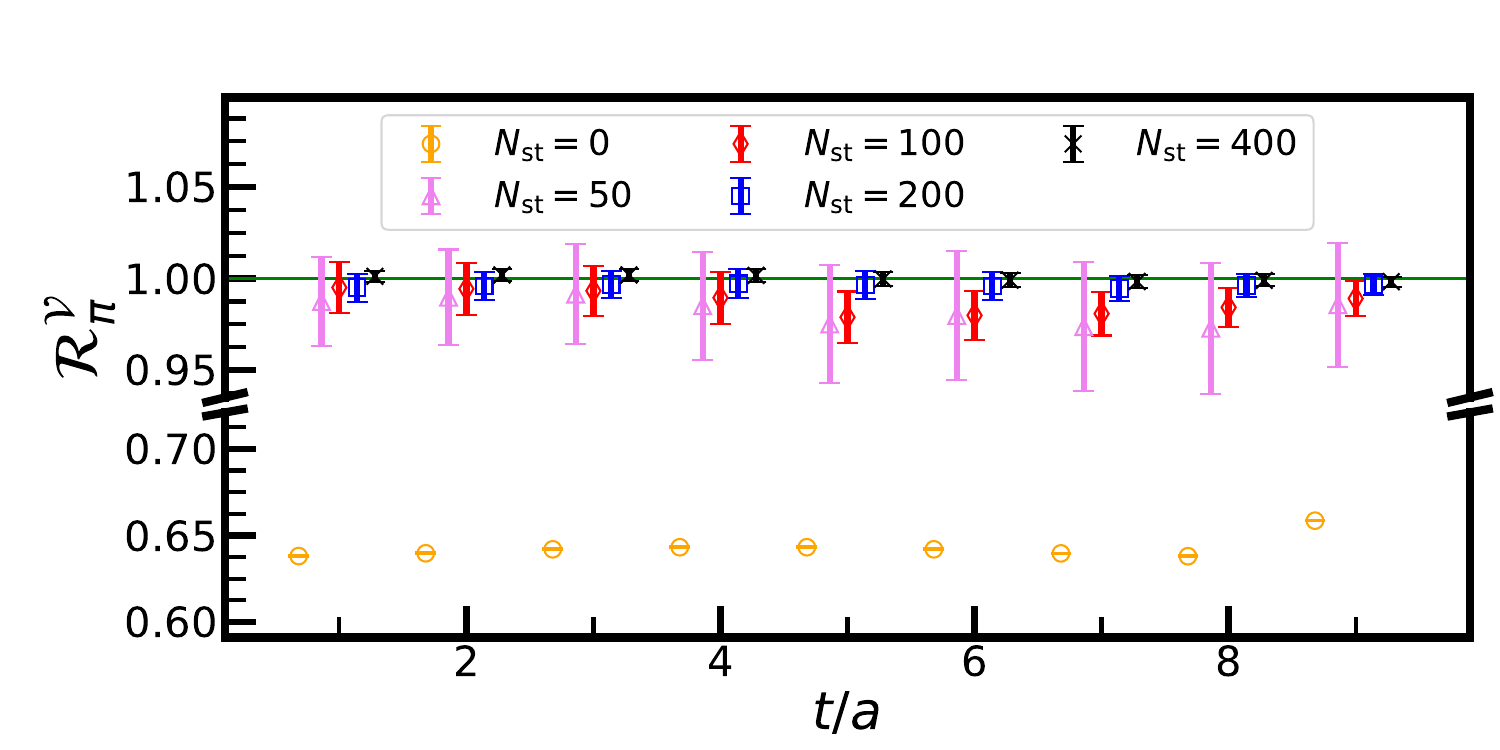}
        \caption{
            The $\pi^+$ matrix element of the conserved vector current operator with \( N_{\rm e} = 100 \) and $n_{\rm cfg} =90$ on C24P29 ensemble, varies the current inserted position $t/a$ with different number of $N_{\rm st}$.
        }
        \label{Fig:conserved_current}
    \end{figure}   

To assess whether the calculation projected onto the blending space provides a reliable estimate of the true value, one can examine the following ratio:
\begin{align}
{\cal R}_H^{\cal O}&(t,t_f)\equiv \frac{\int \dd^3 x \dd^3 y\dd^3 z\langle H(\vec{x},t_f){\cal O}(\vec{y},t)H^{\dagger}(\vec{z},0)\rangle}{\int \dd^3 x\dd^3 z\langle H(\vec{x},t_f)H^{\dagger}(\vec{z},0)\rangle}\nonumber\\
&=\langle {\cal O}\rangle_H+{\cal O}(e^{-\delta mt},e^{-\delta m(t_f-t)},e^{-\delta mt_f}) \label{eq:3pt_over_2pt}
\end{align}
where $H$ is a hadron interpolation field, and $\langle {\cal O}\rangle_H$ is the  ${\cal O}$ matrix element in the ground state of $H$. For the case of the discretized conserved vector current for the Wilson-like fermion,
\begin{align}
{\mathcal V}&\equiv \int \dd^3 x \dd^3 y \bar{q}(x)V^c(x,y)q(y),\nonumber\\
V^c(x,y) &= \frac{1}{2} \big[
    \delta_{x,y+\hat{n}_4a}(1+\gamma_4) U^\dagger_4(y) \nonumber\\
    &\quad\quad - \delta_{x+\hat{n}_4a,y} (1-\gamma_4) U_4(x) 
\big],
\end{align}
with \( U_4(x) \) denoting the gauge link in the temporal direction, ${\cal R}_{H}^{{\mathcal V}}$ corresponds to the vector charge for any $t_f> t> 0$ and choice of $H$. In Fig.~\ref{Fig:conserved_current}, we show this ratio with $t_f/a=10$ of the $\pi^+$ interpolation field using the 2+1 flavor CLQCD ensemble C24P29~\cite{Hu:2023jet}, with a pion mass \( m_{\pi} \simeq 290 \) MeV, $a=0.105$ fm and lattice volume \( N_L^3 \times N_T = 24^3 \times 72 \). We use  \( N_{\rm e} = 100 \) low-lying eigenvectors of the Laplace operator (with 20 steps of the stout smearing with $\rho=0.125$) to construct the external pion state, and 400 additional random vectors ($\sim$ 1\% of $[{\mathcal L}_2]=N_cN_L^3-N_{\rm e}=41372$) in the complement subspace to approximate the conserved current. Using 90 configurations, ${\cal R}_{\pi}^{{\mathcal V}}$ is independent of $t$, and the average over $t/a\in[1,9]$ gives 1.0002(29) which agrees perfectly with the theoretical expectation with a 0.3\% uncertainty. 
In contrast, when using only the vectors in the subspace \( \mathcal{L}_1 \) (i.e., \( N_{\rm st} = 0 \)), the matrix element is not a constant and equals \( 0.6432(05) \) at $t=t_f/2$, exhibiting a significant bias. This demonstrates the necessity of sampling in the subspace \( \mathcal{L}_2 \) to obtain unbiased results.

The mathematical proof and more numerical evidence on the unbiasedness of the blending method can {be} found in the Supplemental Materials~\cite{supplemental}.

\begin{table}                 
\caption{lattice spacing $a$, lattice volume $N_L^3\times N_T$, pion mass $m_{\pi}$, number of configurations $n_{\rm cfg}$,  $N_{\rm e}$ and $N_{\rm st}$ of two ensembles~\cite{Hu:2023jet,CLQCD:2024yyn} used in this work.}  
\begin{tabular}{c | c c  c | c c c } 
\hline
Symbol & $a$ (fm) & $\tilde{L}^3\times \tilde{T}$ & $m_{\pi}$ (MeV) & $n_{\rm cfg}$ & $N_{\rm e}$ & $N_{\rm st}$\\
\hline 
C24P29 & 0.10524(05)(62) &  $24^3\times 72$ &292.3(1.0) & 90 & 100 & 400\\
F64P13 & 0.07753(03)(45) &  $64^3\times 128$ &134.1(1.5) & 41 & 140 & 60\\
\hline
\end{tabular}  
\label{tab:ensem}
\end{table}

{\bf Applications:} 
Building upon our introductory discussions, precise determinations of hadron structure necessitate three key components: (1) accurate computation of connected N-point functions, (2) robust control of excited-state contamination, and (3) reliable evaluation of disconnected quark loop contributions. In the following sections, we will systematically examine each of these aspects through concrete examples. For our calculations of N-point functions, we employ the C24P29 ensemble previously used for the conserved current study, as these cases with non-vanishing ${\cal O}_{ij}$ matrix elements require a relatively large number of stochastic samples ($N_{\rm st}$). The remaining two components will be computed using the F64P13 ensemble at lattice spacing $a=0.077$ fm~\cite{CLQCD:2024yyn}. The complete specifications of both ensembles are provided in Table~\ref{tab:ensem}, and we iterated over all source-time slices in the calculations to suppress the statistical uncertainties.

\begin{figure}
    \centering
    \includegraphics[width=0.45 \textwidth]{pion_FF_v2.pdf}
    \caption{
        The figure shows the form factors extracted from the four-point and three-point functions, respectively.
        The green and black bands represent the results parameterized using the z-expansion.
       For better visualization, data points of $f_{\pi\pi}^{\rm 3pt}$ were slightly shifted to the left.
     }
    \label{Fig:Pion Form Factor}
\end{figure}

(1) First of all, we consider the connected diagram contribution of the $\pi^{\pm}$ electric form factor $f_{\pi\pi}(Q^2)$ using three-point functions and also four-point functions.
In this work, we only consider the four-point function with two $V_4\equiv\bar{d} \gamma_4 d$ currents inserted in the same quark line which is the most challenging case in four-point functions using the traditional sequential method, and ignored the reversed-time-order part which contributes to the excited state contamination only.
In such a case, the $f_{\pi\pi}(Q^2)$ can be obtained from the four-point function as,
\begin{align}
    f_{\pi\pi}&(Q^2)=\lim_{t_f\gg t_2 \gg t_1 \gg 0} 
    \Bigg\{  
        \int \dd^3 x \dd^3 y\dd^3 z e^{{\rm i}\vec{p}\cdot(\vec{z}-\vec{y})}\nonumber\\
        &\langle H_\pi(\vec{x},t_f)V_4(\vec{y},t_2)V_4(\vec{z},t_1)H_\pi^{\dagger}(\vec{0},0)\rangle e^{(E_{\pi}-m_{\pi})(t_2-t_1)}\nonumber\\&/\int \dd^3 x\langle H_\pi(\vec{x},t_f)H_\pi^{\dagger}(\vec{0},0)\rangle
    \Bigg\}^{\frac{1}{2}} 
    \frac{2\sqrt{m_{\pi}E_{\pi}}}{m_{\pi}+E_{\pi}},
\end{align}
where $Q^2=|\vec{p}|^2-(E_{\pi}-m_{\pi})^2$ and $E_{\pi}=\sqrt{m_{\pi}^2+|\vec{p}|^2}$. 

    \begin{figure*}
    \includegraphics[width=0.45\linewidth]{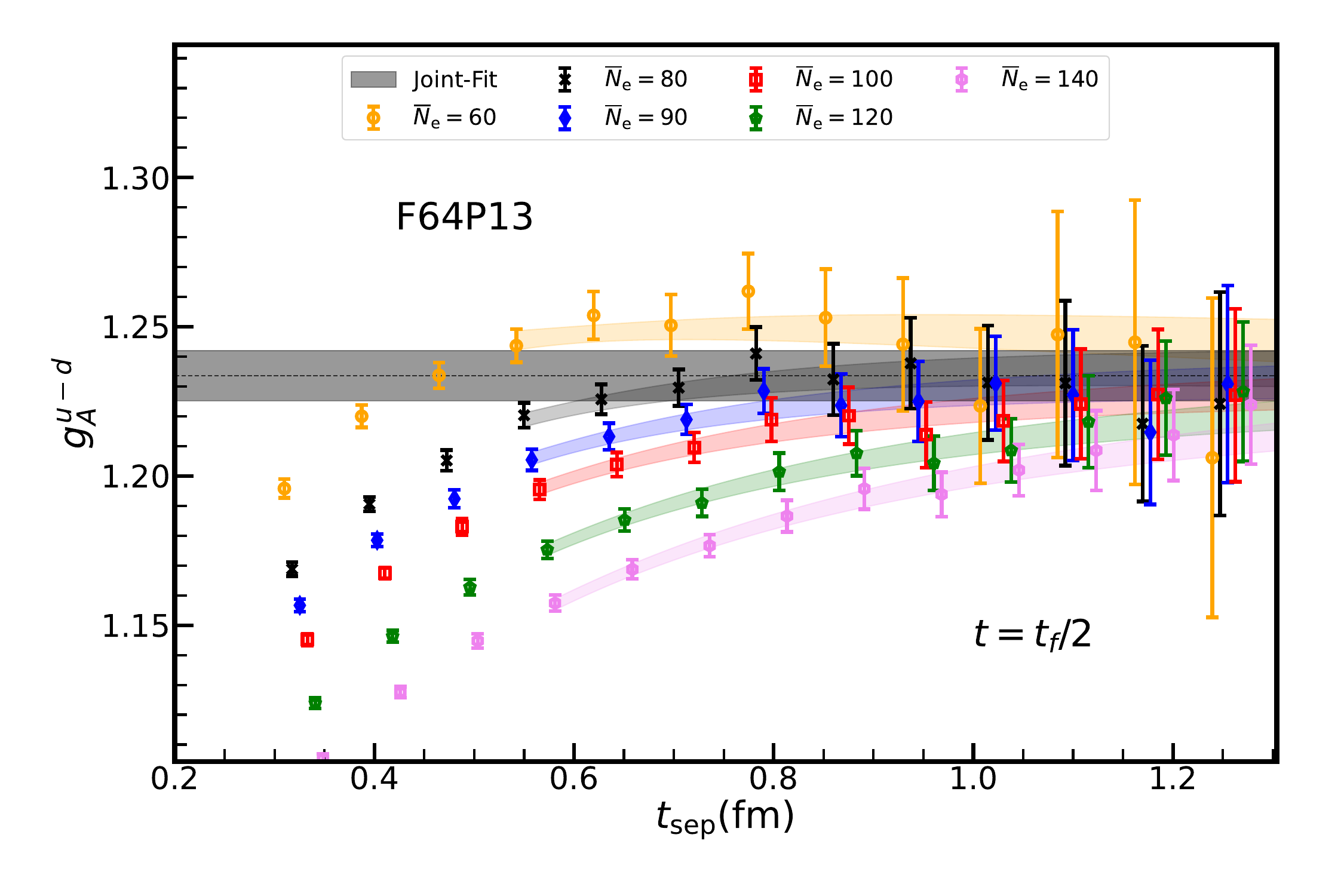}
    \includegraphics[width=0.45\linewidth]{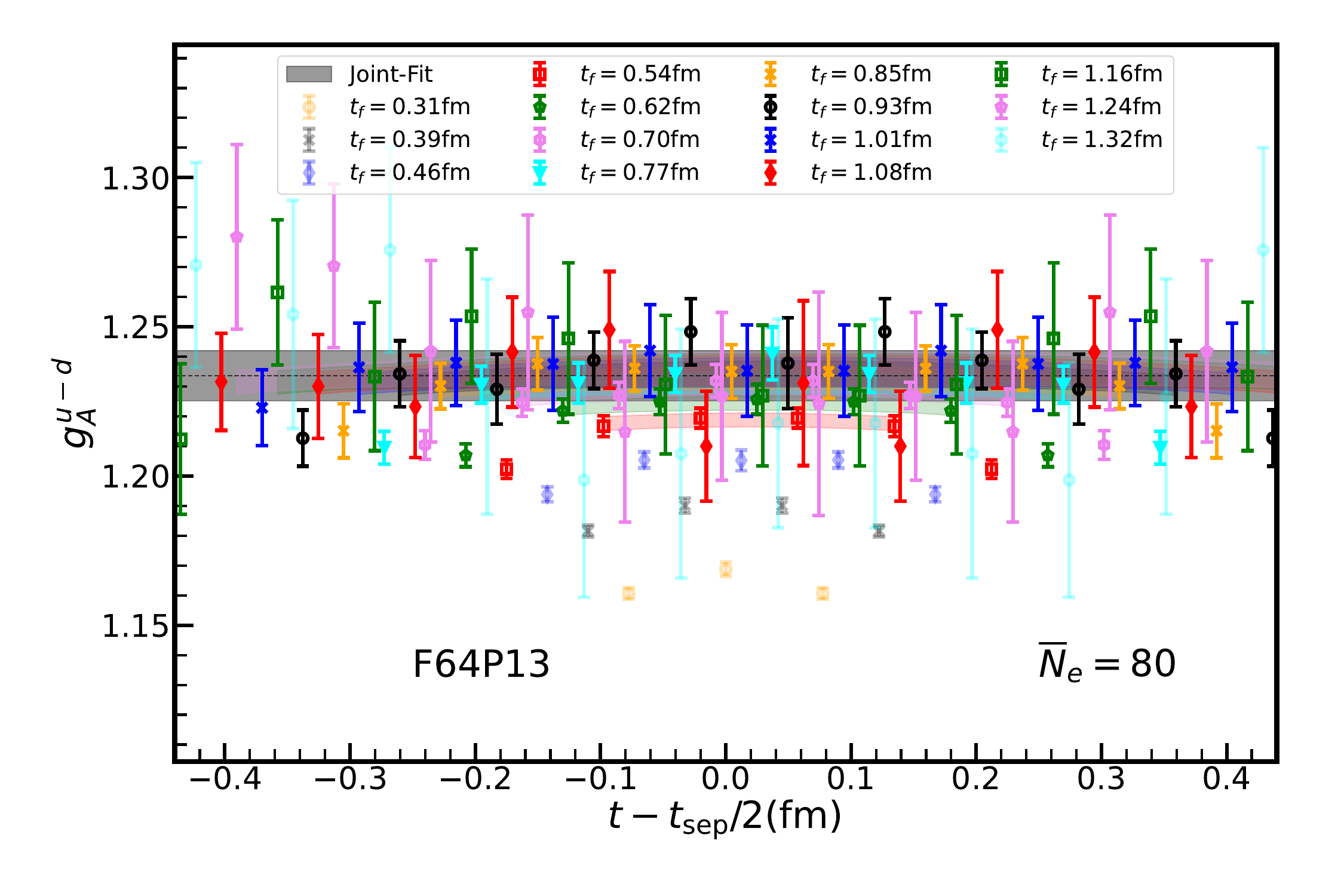}
        \caption{
            ${\cal R}_N^{A_{u-d}}(t_f,t_f/2)$ with different $\bar{N}_e$ used for the nucleon interpolator (left panel) and ${\cal R}_N^{A_{u-d}}(t_f,t)$ for $\bar{N}_e=80$ which has good suppression of excited-state contamination (right panel) on the CLQCD F64P13 ensemble with 41 configurations.
            The gray bands correspond to that from the joint fit of the data with different $\bar{N}_e$. 
        }
        \label{Fig:GA}
    \end{figure*}

Based on the calculation detailed in Supplemental Material~\cite{supplemental}, the form factors calculated from the three-point and four-point functions on the C24P29 ensemble using $t_f=$~2~fm, are shown in Fig.~\ref{Fig:Pion Form Factor}, and both of them are consistent with each other within the uncertainty. Using the z-expansion \cite{zExpansion}, the charge radius obtained from the three-point function is $\langle r^2_{\pi} \rangle=0.332(29)~{\rm fm^2}$, while the four-point function gives $ 0.329(33) ~{\rm fm^2}$. Both results are consistent with each other and in agreement with the most precise result 0.322(08) fm$^2$~\cite{Wang:2020nbf} with similar pion mass using the overlap fermion, which used 3.4 times as many configurations and also LMS to compute the long-distance contribution from the exact all-to-all propagators.

(2) The nucleon iso-vector axial charge, $g^{u-d}_A \equiv \langle A_{u-d} \rangle_N$ with $A_{q}\equiv\bar{q}\gamma_5S\!\!\!/ q$ and $S$ being the polarization vector of nucleon, can serve as an exemplary case for studying excited-state contamination (ESC) in hadron matrix elements. While lattice QCD calculations can determine the corresponding three-point function to two-point function ratio ${\cal R}_N^{A_{u-d}}$ with relatively small statistical uncertainties, the results are particularly susceptible to systematic effects from excited-state contributions including the Roper-like states and also multi-particle states involving a nucleon plus one or more pions~\cite{Bar:2016uoj,Hansen:2016qoz}. It makes $g_A$ both a challenging and informative benchmark for examining excited-state systematics.

{The blending method enables the selection of multiple values of $\bar{N}_{\rm e} \le N_{\rm e}$ for the external nucleon state, allowing the relative weights of excited and ground states to be varied without incurring additional inversions costs in fermion propagator generation. For the physical pion mass ensemble F64P13 at $a = 0.078$ fm which is used here for illustration, we choose $N_{\rm e} = 140$ which balances the two competing effects: excited-state contamination in \(g_A^{u-d}\) grows with \(N_{\rm e}\), while statistical uncertainty decreases. The axial current $A_{u-d}$ is consistently constructed using the full blending space $N_{\rm e} + N_{\rm st}$, with $N_{\rm st} = 60$ found to be sufficient to saturate the statistical uncertainty.
The left panel of} Figure~\ref{Fig:GA} displays ${\cal R}_N^{A_{u-d}}(t_f,t_f/2)$ as a function of $t_f$, which we parameterize as:
\begin{align}
        &{\cal R}_H^{{\cal O}} (t_f,t) =
            \big\{\langle {\cal O}\rangle_H   \notag
                \\ 
                &\ +\sum_{i=1,2}[c_{i0}d_i({\bar{N}_e}) (e^{-\Delta_{i} (t_f-t)}+e^{-\Delta_{i}  t}) +c_{ii}d_i^2({\bar{N}_e}) e^{-\Delta_{i} t_f}] \notag \\
&\ +c_{21}d_1({\bar{N}_e})d_2({\bar{N}_e}) (e^{-\Delta_{1} (t_f-t)-\Delta_{2} t} +e^{-\Delta_{2}(t_f-t)-\Delta_{1} t})\big\}\notag\\
&\quad /
\left[1+d_1^2({\bar{N}_e})e^{-\Delta_1 t_f}+d_2^2({\bar{N}_e})e^{-\Delta_2 t_f}\right].
    \label{Eq:R_H_Ne}
\end{align}
where $d_{1,2}$ and $\Delta_{1,2}$ are also constrained by the two-point function $C_2(t)=Ze^{-m_0t}(1+d_1^2e^{-\Delta_1 t_f}+d_2^2e^{-\Delta_2 t_f})$. 
Data points with different colors represent different $\bar{N}_{\rm e}$ values for the external state. We observe that while larger {\color{red}$\bar{N}_{\rm e}$} reduces statistical uncertainty, it simultaneously increases excited-state contamination, manifesting as stronger $t_f$ dependence in ${\cal R}_N^{A_{u-d}}$. {For the case $\bar{N}_{\rm e}=80$, all values of ${\cal R}_N^{A_{u-d}}(t_f,t)$ in the region $t_f > 0.6$ fm and $t \in [a, t_f - a]$ agree with the fitted band within statistical uncertainty, indicating satisfactory saturation of ESC.}

{The blending method provides a versatile framework for controlling excited-state contamination (ESC) through multiple independent analysis strategies, all yielding consistent results for the axial charge. A joint three-state fit incorporating data across various $\bar{N}_{\rm e}$, $t$, and $t_f$ values gives $g_A^{u-d}=1.2337(84)$, eliminating the ambiguity and systematic bias associated with selecting a specific $\bar{N}_{\rm e}$. 
}

Our numerical tests reveal that 
{the blending method reduces the required number of inversions by at least one order of magnitude compared to traditional sequential or stochastic methods~\cite{Alexandrou:2022dtc,Bali:2023sdi,Jang:2023zts} to achieve comparable statistical precision for $g_A^{u-d}$ at the physical pion mass. At higher pion masses, although the reduction in inversions is less pronounced relative to existing approaches~\cite{Bali:2023sdi,DallaBrida:2018tpn,chang2018per} (reference for the other results at 0.08fm and 300 MeV), the blended propagator enables its reuse without requiring additional inversions for more general purposes, such as the precise estimation of disconnected diagrams discussed below. Further details on the inversions cost comparison are provided in the Supplemental Material~\cite{supplemental}.}

\begin{figure}
    \centering
    \includegraphics[width=0.45 \textwidth]{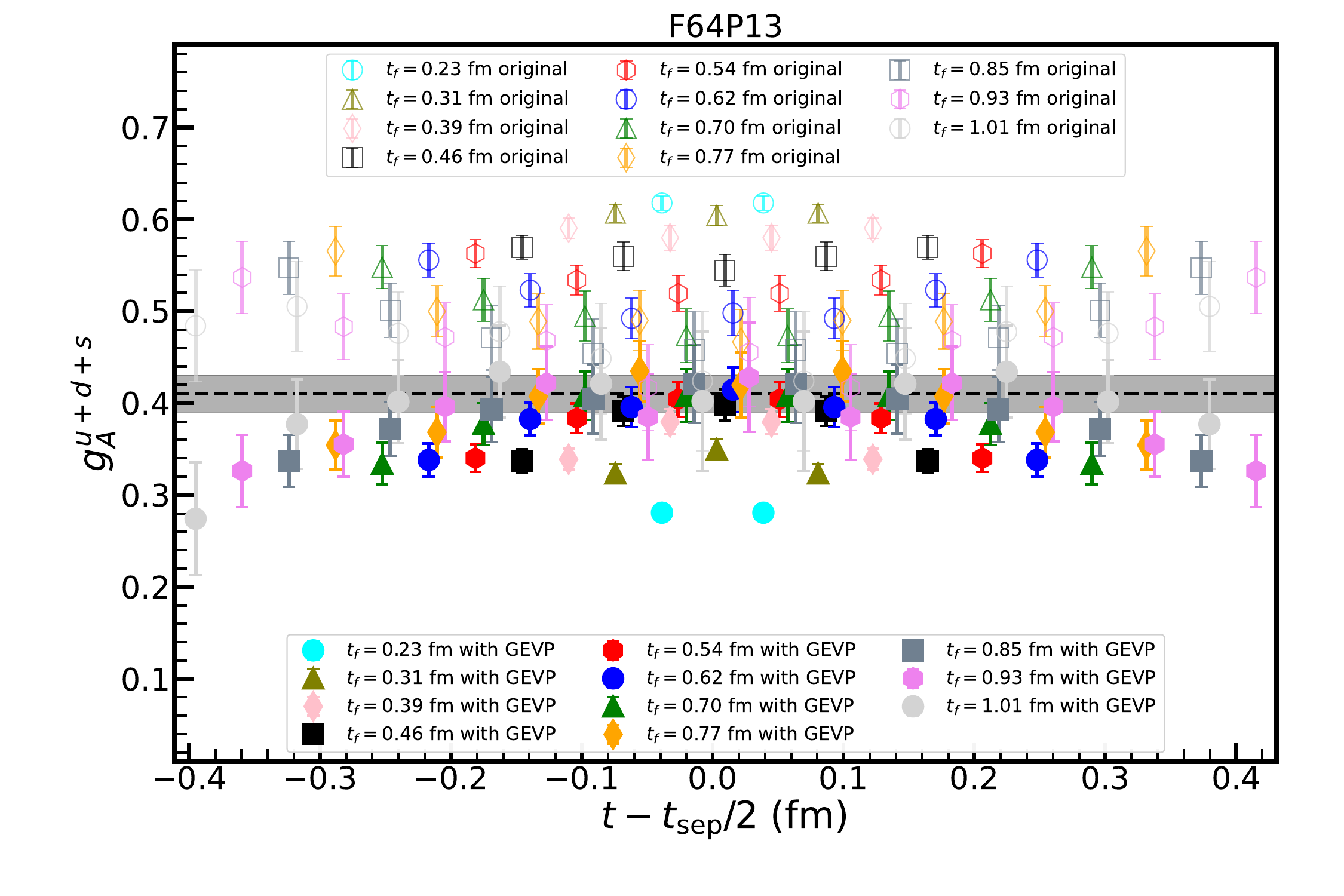}
    \caption{
             Ratio \({\cal R}_N^{A_{u+d+s}}(t_f,t)\) on the F64P13 ensemble for different source‑sink separations \(t_f\) and insertion times \(t\), before (hollow symbols) and after (filled symbols) solving the GEVP with the operator basis \(\{\mathcal{N}, \mathcal{N}{A_{u+d+s}},\mathcal{N}{A_{u+d-2s}}\}\). The gray band indicates the extracted ground‑state matrix element.
     }
    \label{fig:ga_gevp}
\end{figure}

(3) {We also compute the disconnected quark loop contributions to \(g_A^{u+d+s}\) and the associated singlet renormalization on the F64P13 ensemble, using the same propagators as in the \(g_A^{u-d}\) calculation. After applying the 3-loop singlet renormalization factor \(Z^{\rm s}_A/Z^{\rm ns}_A = 1.006(10)\) at \(\overline{\mathrm{MS}}\) 2 GeV~\cite{supplemental}, we obtain the renormalized ratio \({\cal R}_N^{A_{u+d+s}}(t_f,t_f/2)\) shown as the hollow data point in Fig.~\ref{fig:ga_gevp}. Since the excited-state contamination (ESC) in this case is insensitive to \(\bar{N}_{\rm e}\), we take \(\bar{N}_{\rm e}=N_{\rm e}\) to minimize the statistical uncertainty.

A key advantage of the blending method is that it enables the inclusion of additional 5-quark interpolators at no extra inversion cost on the necessary contractions of 300+ quark diagrams~\cite{Wang:2025nsd,supplemental}. 
We exploit this flexibility by extending the basis with ``current involved" interpolators~\cite{Barca:2024hrl,Barca:2025det} ${\cal N}A_{u+d+s}$ and ${\cal N}A_{u+d-2s}$ with $I(J^P)=\frac12(\frac12^+)$ projection, which couples strongly to the \( (\mathcal{N} f_1 )_{\rm s-wave}\) scattering state~\cite{supplemental}.

With this enriched basis, we perform a $3 \times 3$ generalized eigenvalue problem (GEVP) analysis to systematically isolate the ground-state contribution. 
As shown by the filled data points in Fig.~\ref{fig:ga_gevp}, the resulting ratio \({\cal R}_N^{A_{u+d+s}}(t_f,t)\) exhibits dramatically reduced ESC. 
The signal stabilizes already for \(t_f \ge 0.4\)~fm  and \(t \in [2a, t_f-2a]\), with consistent plateaus across the entire fit range. Two-state fits of GEVP-improved interpolators with AIC from different $t_f^{\rm min}$ yield \(g_A^{u+d+s}=0.410(20) \) at physical point ensemble F64P13.

We obtained $g_A^{u+d}=0.448(15)$ and  $g_A^{s}=-0.0381(57)$ from  the same GEVP analysis.  
Combining these with isovector charge \(g_A^{u-d}=1.2337(84)\), we extract the individual flavor contributions at the physical pion mass (\(a=0.078\) fm): \(g_A^u =0.8408(86)  \)  and \(g_A^d = -0.3829(86)\). Compared with previous studies~\cite{Alexandrou:2017oeh,Lin:2018obj,Alexandrou:2019brg}, our results achieve significantly higher precision despite using much smaller inversions~\cite{supplemental}—a direct consequence of the blending method’s ability to perform a full GEVP analysis, which effectively suppresses ESC and enables a clean extraction of ground-state matrix elements.}

{\bf Summary:}
We propose a novel “blending method” for the unbiased {and efficient approximation of all-to-all quark propagators. This approach enables high-precision calculations of hadronic \(N\)-point functions while offering systematic control over excited-state contamination. Building on the advantages of the distillation framework, the blending method allows the construction of correlation functions with multiple interpolating operators sharing the same quantum numbers. This, in turn, facilitates the robust extraction of ground-state matrix elements via a variational analysis (GEVP) with significantly reduced statistical noise.

A key advantage of the method emerges at the physical pion mass, where the computational cost required to achieve a given precision can be one order of magnitudes smaller than the traditional methods. This makes sub-percent precision predictions for various nucleon matrix elements attainable in the near future as demonstrated by $g_T^{u-d}$~\cite{Wang:2025nsd} after the continuum, chiral and infinite volume extrapolations. 
Meanwhile, the blending method provides a way to compute matrix elements for multi-body scattering analysis, opening up the possibility of completely resolving the contamination from low-energy scattering states such as $\mathcal{N}\pi$  and $\mathcal{N}\pi\pi$ in the future.
}
However, the computational cost remains prohibitive for Overlap fermions~\cite{Chiu:1998eu,Liu:2002qu}, where inversion costs are significantly higher. Further optimizations are therefore necessary to enhance the efficiency of the blending method for broader applicability.

Beyond the precision frontier of LQCD, the efficient computation of N-point functions and the suppression of excited-state contamination remain central challenges in various areas, including parton physics via large momentum effective theory~\cite{Ji:2013dva,Ji:2020ect}, lattice cross-section calculations \cite{Ma:2017pxb}, and studies of elastic and resonance structures of hadron through the hadronic tensor \cite{Liu:1993cv,Liu:1999ak,Liang:2019frk,Liang:2023uai}. The blending method can help address these challenges  and advance our understanding of hadron structure and dynamics.

\section*{Acknowledgement}
 We thank the CLQCD collaborations for providing us their gauge configurations with dynamical fermions~\cite{Hu:2023jet}, which are generated on HPC Cluster of ITP-CAS, the Southern Nuclear Science Computing Center(SNSC) and the Siyuan-1 cluster supported by the Center for High Performance Computing at Shanghai Jiao Tong University. 
We thank Siyang Chen, Ying Chen, Xu Feng, Luchang Jin, Jian Liang, Yushan Su and Gen Wang for valuable comments and suggestions. 
The calculations were performed using the PyQUDA package~\cite{Jiang:2024lto} with QUDA~\cite{Clark:2009wm,Babich:2011np,Clark:2016rdz} through HIP programming model~\cite{Bi:2020wpt}. The numerical calculation were carried out on the ORISE Supercomputer, HPC Cluster of ITP-CAS and Advanced Computing East China Sub-center. This work is supported in part by National Key R\&D Program of China No.2024YFE0109800, NSFC grants No. 12525504, 12293060, 12293062, 12435002, 12447101, 12205171, 12235008 and 12321005, the science and education integration young faculty project of University of Chinese Academy of Sciences, the Strategic Priority Research Program of Chinese Academy of Sciences, Grant No.\ XDB34030303 and YSBR-101, and the Department of Science and Technology of Shandong province No.~tsqn202312052 and~2024HWYQ-005.

\bibliography{ref}

\clearpage
% \documentclass[floatfix,twocolumn,prd,showpacs,preprintnumbers,amsmath,nofootinbib,amssymb,superscriptaddress]{revtex4-1}

% \usepackage{color} 
% % \newcommand{\blue}[1]{{\color{blue} #1}}
% \definecolor{green}{rgb}{0,.5,0}
% % \newcommand{\green}[1]{{\color{green} #1}}
% % \definecolor{red}{rgb}{1,0,0}
% %\newcommand{\red}[1]{{ #1}}
% % \newcommand{\red}[1]{{ #1}}
% \usepackage{graphicx}% Include figure files
% \usepackage[subfigure]{graphfig}
% \usepackage{epsfig}
% %\usepackage{epstopdf}
% \usepackage{dcolumn}% Align table columns on decimal point
% \usepackage{amsmath}
% \usepackage{multirow}
% \usepackage{booktabs}   %%表格横线宏包
% \usepackage{diagbox}
% \usepackage{colortbl}
% \usepackage{soul}
% \usepackage{slashed}
% \usepackage{physics}
% \usepackage{bm}
% \usepackage{tikz} %%画图工具
% %hubl
% \usepackage[colorlinks,linkcolor=blue,citecolor=blue,urlcolor=blue]{hyperref}

% \newcommand{\dslash}[1]{{#1\!\!\!/}}
% % \def\bea{\begin{eqnarray}}
% \def\epsl{\epsilon}
% % \def\eea{\end{eqnarray}}
% \newcommand{\nblr}{\overleftrightarrow{D}}
% \def\bal#1\eal{\begin{align}#1\end{align}}
% \newcommand{\MSbar}{{\overline{\text{MS}}}}
% \newcommand{\MOM}{\text{RI/MOM}}

% \immediate\write18{texcount -inc -incbib -sum borra.tex > /tmp/wordcount.tex}

% \begin{document}

\begin{widetext}

\section*{Supplemental materials}

\begin{table}[!h]                
\caption{lattice spacing $a$, lattice volume $N_L^3\times N_T$, pion mass $m_{\pi}$, number of configurations $n_{\rm cfg}$,  $N_{\mathrm e}$ and $N_{\mathrm st}$ of two ensembles~\cite{Hu:2023jet,CLQCD:2024yyn} used in the $g_A$ calculations.}  
\begin{tabular}{c | c c  c | c c c } 
\hline
Symbol & $a$ (fm) & $\tilde{L}^3\times \tilde{T}$ & $m_{\pi}$ (MeV) & $n_{\rm cfg}$ & $N_{\mathrm e}$ & $N_{\mathrm st}$\\
\hline 
C24P29 & 0.10524(05)(62) &  $24^3\times 72$ &292.3(1.0) & 90 & 100 & 400\\
F48P30 & 0.07753(03)(45) &  $48^3\times 96$ &300.4(1.2) & 40 & 100 & 200\\
F64P13 & 0.07753(03)(45) &  $64^3\times 128$ &134.1(1.5) & 41 & 140 & 60\\
\hline
\end{tabular}  
\label{tab:ensem_sm}
\end{table}

\red{The results presented in these supplementary materials are based on numerical calculations for the ensembles, details of which are given in the Table.~\ref{tab:ensem_sm}.}

Rest part of the supplemental materials are organized into four sections. Section~\ref{sec:Proof} provides a mathematical justification of the blending method along with numerical evidence from multipoint functions. Section~\ref{sec:pionff} details the calculation of the pion form factor. Section~\ref{sec:gAFit} presents the high-precision determination of $g_A^{u,d,s}$ at the physical pion mass, and also their 3-loop perturbative matching to the $\overline{\mathrm{MS}}$ scheme 2 GeV. Section~\ref{sec:cost_compare} compares the computational inversions cost of the blending method with that of traditional approaches in the literature.

\subsection{Justification of the blending method}\label{sec:Proof}

    \subsubsection{Mathematical Proof}

For a fixed time slice of a 4D hypercubic lattice of dimensionless size \( N_L^3 \times N_T \), we define the space \( \mathcal{L} \) as the vector space associated with the lattice sites and color degrees of freedom, with dimension \( [\mathcal{L}] = N_c N_L^3 \). Within \( \mathcal{L} \), a ``blending space" is constructed by combining the eigenvectors \( \{ \phi_i       \equiv v_{\lambda_i} \}_{i=1}^{N_{\mathrm e}} \) of the discrete Laplace operator spanning \( \mathcal{L}_1 \), and  
a set of \( N_{\mathrm st} \) orthogonal random vectors \(\{ \phi_{N_e+j} \equiv\eta_j \}_{j=1}^{N_{\mathrm st}} \) sampled uniformly from \( \mathcal{L}_2 \equiv  \mathcal{L} / \mathcal{L}_1 \). One can generate the $N_{\mathrm st}$ random vectors, deflate the eigenvectors $v_{\lambda_i}$ from them, and then apply Gram-Schmidt orthogonalization to obtain the orthonormalized vectors $\eta_j$.

We first select a complete orthogonal basis set
$\{ \phi_{1},\phi_{2},\ldots,\phi_{N_e},e_1,\ldots,e_{\lfloor\mathcal{L}_{2}\rfloor} \}$,
then $\phi_{N_e+j}$ can be expanded as linear combinations of $e_i$,
$\phi_{N_e+j} = \sum_{i=1}^{[\mathcal{L}_2]} c_{i}^{(j)}e_{i}$ and defined as a vector in this basis set, 
\begin{align}
 \bm{c}^{(j)} = \left(0,0,...,0,c_{1}^{(j)},c_{2}^{(j)}, \ldots,c_{\lfloor\mathcal{L}_{2}\rfloor}^{(j)}\right).   
\end{align}
Based on this definition, the probability density function (PDF) $\mathbb{P}(\phi_{N_{\mathrm e}+N_{\mathrm st}},\phi_{N_{\mathrm e}+N_{\mathrm st}-1},\ldots,\phi_{N_{\mathrm e}+1})$ of the noise vector in the constructed orthogonal space can be formally expressed as,
\begin{align}
    \mathbb{P}(\phi_{N_{\mathrm e}+N_{\mathrm st}},\phi_{N_{\mathrm e}+N_{\mathrm st}-1},...,\phi_{N_{\mathrm e}+1}) 
    = \frac{1}{\prod_{n=1}^{N_{\mathrm st}}S_{2([\mathcal{L}_{2}]-n)}}\prod_{ 0 \leq i<j \leq N_{\mathrm st}}\delta(\bm{c}^{(i)*} \cdot \bm{c}^{(j)}) \prod_{n=1}^{N_{\mathrm st}} \delta(|| \bm{c}^{(n)} ||-1),
    \label{eq:PDF}
\end{align}
where $S_{2({\cal L}_2-n)}$ denotes the surface area of a unit sphere in $2({\cal L}_2-n)$-dimensional space, the $|| . ||$ denotes the norm of the vector. Since \eqref{eq:PDF} only contain the inner product and norm of noise vectors, it should be invariant under unitary transformations on $\mathcal{L}_2$, and we will see this invariance can strongly constrain the structure of the mathematical expectation of noise vectors. 

To demonstrate the unbiasedness of our estimator, we shall verify the following second-order relation first:
\begin{align}       \mathbb{E}\left [\sum_{i=1}^{N_e+N_{\mathrm st}}\sum_{j=1}^{N_e+N_{\mathrm st}}\Omega^{(2)}_{ij}|\phi_{i} \rangle  \langle \phi_{i} | \otimes |\phi_{j} \rangle\langle \phi_{j} | \right ] = \hat{I} \otimes \hat{I},
\label{eq:completness}
\end{align}
where $\mathbb{E}[.]$ denotes the mathematical expectation. Expressing the noise vector in terms of a complete orthogonal basis requires computing two key expectations: the second moment $\mathbb{E}[c_{i}^{(\alpha)} c_{j}^{(\alpha)*}]$ and the fourth moment $\mathbb{E}[c_{i}^{(\alpha)} c_{j}^{(\alpha)*} c_{k}^{(\beta)} c_{l}^{(\beta)*}]$.

Due to unitary invariance, the expectation $\mathbb{E}[c_{i}^{(\alpha)} c_{j}^{(\alpha)*}]$ should be proportional to $\delta_{ij}$. Combining the normalization condition $||\bm{c}^{\alpha}||=1$, we have
\begin{align}
    \mathbb{E}[c_{i}^{(\alpha)} c_{j}^{(\alpha)*}] = \frac{1}{[\mathcal{L}_2]}\delta_{ij},
\end{align}
Therefore, the eigen-to-noise contribution ($i\le N_{\mathrm e}$, $j>N_{\mathrm e}$) of Eq.~(\ref{eq:bld_final}) is,
\begin{equation}
\begin{aligned}
     \mathbb{E}\left [ \frac{[\mathcal{L}_2]}{N_{\mathrm st}}\sum _{i=1}^{N_e}\sum_{j=1}^{N_{\mathrm st}}|\phi_{i}\rangle \langle \phi_{i}| \otimes |\phi_{N_e+j} \rangle \langle \phi_{N_e+j}| \right ] 
    & = \frac{[\mathcal{L}_2]}{N_{\mathrm st}}\sum _{i=1}^{N_e}\sum_{j=1}^{N_{\mathrm st}} \sum_{n,m=1}^{[\mathcal{L}_2]} \mathbb{E} \left [ c_{n}^{(N_e+j)} c_{m}^{(N_e+j)*} \right ] |\phi_{i}\rangle \langle \phi_{i}|\otimes|e_{n} \rangle \langle e_{m}| \\
    & = \sum _{i=1}^{N_e}\sum_{n=1}^{[\mathcal{L}_2]} |\phi_{i}\rangle \langle \phi_{i}| \otimes |e_{n} \rangle \langle e_{n}|.
\end{aligned}  
\end{equation}

Due to the constraints from the unitary transformation invariance
\begin{align}
    \mathbb{E}[c_{i}^{(\alpha)} c_{j}^{(\alpha)*} c_{k}^{(\beta)} c_{l}^{(\beta)*}] 
    = 
    U_{i i^{'}}U^{\dagger}_{j^{'}j}U_{k k^{'}}U^{\dagger}_{ l^{'}l}
    \mathbb{E}[c_{i^{'}}^{(\alpha)} c_{j^{'}}^{(\alpha)*} c_{k^{'}}^{(\beta)} c_{l^{'}}^{(\beta)*}]
    ,
    \label{eq:constrain}
\end{align}
and the relation of the unitary transformation,
\begin{align}
    & U_{i i^{'}}U^{\dagger}_{j^{'}j}U_{k k^{'}}U^{\dagger}_{ l^{'}l}
    \delta_{i^{'}j^{'}}\delta_{k^{'}l^{'}}
     = U_{ii^{'}}U^{\dagger}_{i^{'}j}U_{kk^{'}}U^{\dagger}_{k^{'}l}= \delta_{ij}\delta_{kl}, \\
    & U_{i i^{'}}U^{\dagger}_{j^{'}j}U_{k k^{'}}U^{\dagger}_{ l^{'}l}
    \delta_{i^{'}l^{'}}\delta_{j^{'}k^{'}} = U_{ii^{'}}U^{\dagger}_{i^{'}l}U_{kk^{'}}U^{\dagger}_{k^{'}j}= \delta_{il}\delta_{kj}.
\end{align},
the expression of $\mathbb{E}[c_{i}^{(\alpha)} c_{j}^{(\beta)} c_{k}^{(\alpha)*}c_{l}^{(\beta)*}]$ can only be
\begin{align}
    \mathbb{E}[c_{i}^{(\alpha)} c_{j}^{(\alpha)*} c_{k}^{(\beta)} c_{l}^{(\beta)*}]  =
    A\delta_{ij}\delta_{kl} + B\delta_{il}\delta_{kj}.
    \label{eq:expression}
\end{align}

Using the normalization $|| \bm{c}^{(\alpha)} ||= || \bm{c}^{(\beta)} || =1$, and taking the trace of $ij$ and $kl$ in Eq.~\eqref{eq:expression}, we have 
\begin{align}
    \mathbb{E}[||\bm{c}^{(\alpha)}||^2||\bm{c}^{(\beta)}||^2] 
    =
    [\mathcal{L}_2]^{2}A+[\mathcal{L}_2]B=1.
\end{align}
Using the orthogonality $\bm{c}^{(\alpha)*} \cdot \bm{c}^{(\beta)} = \delta_{\alpha\beta}$ and taking the trace of $il$ and $kj$ in Eq.~\eqref{eq:expression}, we also have
\begin{align}
    \mathbb{E}[|\bm{c}^{(\alpha)*} \cdot \bm{c}^{(\beta)}|^2] 
    =[\mathcal{L}_2]A+[\mathcal{L}_2]^{2}B= 
    \begin{cases}
    1, & \alpha = \beta,\\
    0, & \alpha \neq \beta,
    \end{cases}
\end{align}
where $| . |$ is used to denote the norm of a complex number which is different from the norm of a vector $|| . ||$ .

Solving for $A$ and $B$, we obtain 
\begin{align}
    \mathbb{E}[c_{i}^{(\alpha)} c_{j}^{(\alpha)*} c_{k}^{(\beta)} c_{l}^{(\beta)*}] =
    \begin{cases}
    \frac{1}{[\mathcal{L}_2]([\mathcal{L}_2]+1)} (\delta_{ij}\delta_{kl} + \delta_{il}\delta_{kj}), & \alpha = \beta. \\
    \frac{1}{[\mathcal{L}_2]^{2}-1} \delta_{ij}\delta_{kl} -  \frac{1}{[\mathcal{L}_2]([\mathcal{L}_2]^{2}-1)}\delta_{il}\delta_{kj}, & \alpha \neq \beta,
    \end{cases}
\end{align}
and the noise-to-noise contribution ($i,j> N_{\mathrm e}$) of Eq.~(\ref{eq:bld_final}) becomes,
\begin{align}
    & \mathbb{E}\left [\frac{[\mathcal{L}_2]([\mathcal{L}_2]-1)}{N_{\mathrm st}(N_{\mathrm st}-1)}\sum_{i \neq j}^{N_{\mathrm st}}|\phi_{N_e+i} \rangle \langle \phi_{N_e+i} | \otimes |\phi_{N_e+j} \rangle \langle \phi_{N_e+j} | + \frac{[\mathcal{L}_2]}{N_{\mathrm st}}\sum_{i}^{N_{\mathrm st}} |\phi_{N_e+i} \rangle \langle \phi_{Ne+i} | \otimes |\phi_{N_e+j} \rangle \langle \phi_{Ne+j} | \right ] \nonumber \\ 
    = 
    & (\frac{[\mathcal{L}_2]([\mathcal{L}_2]-1)}{[\mathcal{L}_2]^{2}-1}+\frac{[\mathcal{L}_2]}{[\mathcal{L}_2]([\mathcal{L}_2]+1)})\sum_{n,m=1}^{[\mathcal{L}_2]} |e_n \rangle \langle e_n| \otimes |e_m \rangle \langle e_m| + (-\frac{[\mathcal{L}_2]([\mathcal{L}_2]-1)}{[\mathcal{L}_2]([\mathcal{L}_2]^{2}-1)}+\frac{[\mathcal{L}_2]}{[\mathcal{L}_2]([\mathcal{L}_2]+1)})\sum_{n,m=1}^{[\mathcal{L}_2]} |e_n \rangle \langle e_m| \otimes |e_m \rangle \langle e_n| \nonumber\\
    = 
    & \sum_{n,m=1}^{[\mathcal{L}_2]} |e_n \rangle \langle e_n| \otimes |e_m \rangle \langle e_m|.
\end{align}  

Eventually the final result with all the terms is:
\begin{align}
    & \mathbb{E}\left [ \sum_{i=1}^{N_e+N_{\mathrm st}}\sum_{j=1}^{N_e+N_{\mathrm st}}\Omega^{(2)}_{ij}|\phi_{i} \rangle |\phi_{j} \rangle \langle \phi_{i} | \langle \phi_{j} | \right ] \nonumber \\
    = 
    & \sum_{i=1}^{N_e}\sum_{j=1}^{N_e} |\phi_i \rangle \langle \phi_i| \otimes |\phi_j \rangle \langle \phi_j| + \sum_{i=1}^{N_e}\sum_{m=1}^{[\mathcal{L}_2]}|\phi_i \rangle \langle \phi_i| \otimes |e_m \rangle \langle e_m| + \sum_{n=1}^{[\mathcal{L}_2]}\sum_{j=1}^{N_e}|e_n \rangle \langle e_n| \otimes |\phi_j \rangle \langle \phi_j| + \sum_{n,m=1}^{[\mathcal{L}_2]} |e_n \rangle \langle e_n| \otimes |e_m \rangle \langle e_m| \nonumber \\
    =
    & (\sum _{i=1}^{N_e}|\phi_{i}\rangle \langle \phi_{i}| + \sum_{n=1}^{\mathcal{L}_2}|e_{n} \rangle \langle e_{n}|) \otimes (\sum_{j=1}^{N_e}|\phi_{j}\rangle \langle \phi_{j}| + \sum_{m=1}^{[\mathcal{L}_2]}|e_{m} \rangle \langle e_{m}|) = \hat{I} \otimes \hat{I},
\end{align} 
as we claimed in Eq.~\eqref{eq:completness}. Given ~\eqref{eq:completness}, we can derive \red{(with $\mathcal{O}_{ij}$ include the weights $\Omega^{(2)}_{ij}$ in its defintion)}:
\begin{align}
\mathbb{E}\left [\sum_{i=1}^{N_e+N_{\mathrm st}}\sum_{j=1}^{N_e+N_{\mathrm st}}|\phi_{i} (x)\rangle \mathcal{O}_{ij} \langle \phi_{j} (y)| \right ] 
& = \sum_{i=1}^{N_e+N_{\mathrm st}}\sum_{j=1}^{N_e+N_{\mathrm st}} \int \text{d}^{3}z \text{d}^{3}w \ \Omega_{ij} \mathbb{E}\left [|\phi_{i} (x)\rangle \langle \phi_{i}(z)|M_{\mathcal{O}}(z,w) | \phi_{j}(w)\rangle \langle \phi_{j} (y)| \right ] \nonumber \\ 
& = \int \text{d}^{3}z \text{d}^{3}w \hat{I}_{xz} M_{\mathcal{O}}(z,w) \hat{I}_{wy} = M_{\mathcal{O}}(x,y),\label{eq:bld_final}
\end{align}
This ensures that the weighted sum of $M_{\mathcal{O}}$
in the blending space is an unbiased estimate. The proof for more non-trivial cases can be obtained using the similar procedure.

For the composite operators with the elements located at multiple time slices, since the probabilities are independent  on the different time slices, the fourth moments can be computed as:
\begin{align}
    \mathbb{E} \left[  c_i^{(\alpha)}(t_1)c_j^{(\alpha)*}(t_1) c_k^{(\beta)}(t_2)c_l^{(\beta)*}(t_2) \right] 
    & = \mathbb{E} \left[  c_i^{(\alpha)}(t_1)c_j^{(\alpha)*}(t_1)\right] \mathbb{E} \left[ c_k^{(\beta)}(t_2)c_l^{(\beta)*}(t_2) \right] = \frac{1}{[\mathcal{L}_2]^2}\delta_{ij}\delta_{kl}.
\end{align}
Then, we can check that using both weights $\Omega_{i}^{(1)}\Omega_{j}^{(1)}$ and $\Omega_{ij}^{(2)}$ yields the same result. For $\Omega_{i}^{(1)}\Omega_{j}^{(1)}$, there are
\begin{align}
    \frac{[\mathcal{L}_2]^2}{N_{\mathrm st}^2}\sum_{\alpha,\beta}^{N_{\mathrm st}}\mathbb{E} \left[  c_i^{(\alpha)}(t_1)c_j^{(\alpha)*}(t_1) c_k^{(\beta)}(t_2)c_l^{(\beta)*}(t_2) \right] = \delta_{ij}\delta_{kl}.
\end{align}
And for the $\Omega_{ij}^{(2)}$, there are
\begin{align}
    & \frac{[\mathcal{L}_2]([\mathcal{L}_2]-1)}{N_{\mathrm st}(N_{\mathrm st}-1)}\sum_{\alpha \neq \beta}^{N_{\mathrm st}}\mathbb{E} \left[  c_i^{(\alpha)}(t_1)c_j^{(\alpha)*}(t_1) c_k^{(\beta)}(t_2)c_l^{(\beta)*}(t_2) \right] 
    + 
    \frac{[\mathcal{L}_2]}{N_{\mathrm st}}\sum_{\alpha=1 }^{N_{\mathrm st}}\mathbb{E} \left[  c_i^{(\alpha)}(t_1)c_j^{(\alpha)*}(t_1) c_k^{(\alpha)}(t_2)c_l^{(\alpha)*}(t_2) \right] \nonumber \\ 
    = 
    & (\frac{\mathcal{L}_2]([\mathcal{L}_2]-1)+[\mathcal{L}_2]}{[\mathcal{L}_2]^2})\delta_{ij}\delta_{kl} = \delta_{ij}\delta_{kl}.
\end{align}
Therefore, we can also use $\Omega_{ij}^{(2)}$ regardless the time coordinates of the elements of the composite operators.

\red{The preceding first- and second-order expressions are special cases of a
single inverse-inclusion-probability construction.  We state the result
for a fixed gauge configuration, so that the Laplacian subspaces and all
propagator kernels are deterministic during the stochastic expectation.
Write
\[
\mathcal L=\mathcal L_1\oplus\mathcal L_2,\qquad
D\equiv[\mathcal L_2]=N_cN_L^3-N_e,\qquad n\equiv N_{\mathrm st},
\]
and let $\{v_\ell\}_{\ell=1}^{N_e}$ be the exact orthonormal basis of
$\mathcal L_1$.  The $n$ stochastic vectors in $\mathcal L_2$ are obtained
from projected isotropic complex Gaussian vectors by complex QR (or
Gram--Schmidt), and hence form a Haar--Stiefel orthonormal frame.

Equivalently, for the purpose of computing the expectation, choose a
complete orthonormal basis $\{u_a\}_{a=1}^{D}$ of $\mathcal L_2$ and let
$S\subset\{1,\ldots,D\}$ be a uniformly distributed subset of size $n$,
independent of the complete basis.  The sampled vectors are the columns
$\{u_a\}_{a\in S}$ (with an arbitrary random ordering used for the
stochastic labels).  Since the estimator sums all ordered tuples and its
weight depends only on label equalities, this coupling has the same law as
the ordered QR frame.  For an ordered $r$-tuple of low- and high-mode labels
$\boldsymbol j=(j_1,\ldots,j_r)$, define
\[
A(\boldsymbol j)\equiv
\{\text{distinct high-mode labels appearing in }\boldsymbol j\},
\qquad k(\boldsymbol j)\equiv |A(\boldsymbol j)|.
\]
The tuple is present in the sample exactly when
$A(\boldsymbol j)\subseteq S$.  Its inclusion probability is
\begin{align}
\pi_{k(\boldsymbol j)}
&\equiv {\rm Pr}\!\left[A(\boldsymbol j)\subseteq S\right]
=\frac{\binom{D-k(\boldsymbol j)}{n-k(\boldsymbol j)}}{\binom{D}{n}}
=\frac{(n)_{k(\boldsymbol j)}}{(D)_{k(\boldsymbol j)}} .
\label{eq:inclusion_probability_general}
\end{align}
Consequently, the general same-frame reweighting factor is
\begin{align}
\Omega^{(r)}_{\boldsymbol j}
&\equiv \frac{1}{\pi_{k(\boldsymbol j)}}
=\frac{(D)_{k(\boldsymbol j)}}{(n)_{k(\boldsymbol j)}}
=\prod_{q=0}^{k(\boldsymbol j)-1}\frac{D-q}{n-q},
\qquad (x)_0\equiv1 .
\label{eq:Omega_general}
\end{align}

To prove unbiasedness, denote by $R_j$ either a low-mode projector
$|v_\ell\rangle\langle v_\ell|$ or a complete high-mode projector
$|u_a\rangle\langle u_a|$.  The sampled estimator can be written as
\begin{align}
\widehat{\mathcal I}^{(r)}
&=
\sum_{\boldsymbol j}
\frac{\mathbf 1_{\{A(\boldsymbol j)\subseteq S\}}}
{\pi_{k(\boldsymbol j)}}
\bigotimes_{s=1}^{r}R_{j_s}.
\label{eq:general_tensor_estimator}
\end{align}
Conditioning on the complete basis and taking the expectation over $S$
gives, term by term,
\begin{align}
\mathbb E_S\!\left[
\widehat{\mathcal I}^{(r)}\mid\{u_a\}\right]
&=
\sum_{\boldsymbol j}
\frac{\mathbb E_S[
\mathbf 1_{\{A(\boldsymbol j)\subseteq S\}}]}
{\pi_{k(\boldsymbol j)}}
\bigotimes_{s=1}^{r}R_{j_s}
\nonumber\\
&=
\sum_{\boldsymbol j}\bigotimes_{s=1}^{r}R_{j_s}
=I_{\mathcal L}^{\otimes r}.
\label{eq:general_tensor_unbiased}
\end{align}
The right-hand side is independent of the complete-basis realization, so
averaging over the Haar completion yields
\[
\boxed{\mathbb E\!\left[\widehat{\mathcal I}^{(r)}\right]
=I_{\mathcal L}^{\otimes r}.}
\]

For independent stochastic frames at different time slices, let
$r_\tau$ denote the number of tensor factors in the frame-sharing block
at time $\tau$, and let $\kappa_\tau$ be the number of distinct
stochastic labels among those factors.  The joint inclusion probability
factorizes, and
\begin{align}
\Omega^{(r)}_{\boldsymbol i,\boldsymbol t}
&=
\prod_{\tau}
\frac{(D)_{\kappa_\tau}}{(n)_{\kappa_\tau}},
\qquad
n\geq\min(r_\tau,D)
\quad\text{for every frame-sharing block } \tau .
\label{eq:general_block_weight}
\end{align}
This covers arbitrary time-coincidence patterns.  The theorem applies to
raw deterministic multilinear contractions of these tensor resolutions.

The condition $n\geq\min(r,D)$ is necessary for this universal
inverse-inclusion construction.  If $n<\min(r,D)$, a full-basis tensor
monomial containing $n+1$ distinct high-mode labels has zero inclusion
probability, so no finite reweighting of sampled diagonal-projector
monomials can reproduce $I_{\mathcal L}^{\otimes r}$ conditional on the
completed basis.  Thus the valid statement is arbitrary finite tensor
order subject to the per-block sample-size bound, rather than arbitrary
order at fixed $N_{\mathrm st}$.
}

\subsubsection{Two point function}

We shall start from the two point function of the point-source and point-sink pion interpolation field $\pi=\bar{u}\gamma^5 d$,
\begin{align}
    c_2^{\pi,{\rm PP}}(\vec{p},t) &=  
    \int  \dd^3 y e^{-i \vec{y} \cdot \vec{p}} 
        \langle \pi(\vec{y},t) \pi^\dagger(\vec{0},0)\rangle  =\int  \dd^3 y e^{-i \vec{y} \cdot \vec{p}}  \tr[S(\vec{y},t,\vec{0},0) \gamma_5 S(\vec{0},0,\vec{y},t)\gamma_5],
    \label{Eq.c2pionPoint}
\end{align}
and also that with the projected operator ${\mathbf O}$,
\begin{align}
c_2({\mathbf O},\vec{p},t)=\frac{1}{L^3}\int  \dd^3 x \dd^3 y e^{-i (\vec{y} - \vec{x})\cdot \vec{p}}  \tr[  \mathcal{O}(\vec{p},t)  S(t,0) \mathcal{O}^{\dagger}(\vec{p},0)  S(0,t) ],
\end{align}
where the projected propagator is $S_{ij}(t_1,t_2)= \int \dd^3 x \dd^3 y \langle \phi_i(\vec{x},t_1)|S(\vec{x},t_1;\vec{y},t_2)|\phi_j(\vec{y},t_2)\rangle$.

Then the $c_2^{\pi,{\rm PP}}$ 
can be approximate by the blending method as
\begin{align}
   c_2^{\pi,{\rm BLD}}(\vec{p},t)  &\equiv c_2({\mathbf O}_{\pi}^{{\rm BLD}},\vec{p},t)_{\overrightarrow{N_{\mathrm st}\rightarrow [{\cal L}_2]}} \frac{1}{L^3}
        \langle \pi(\vec{y},t) \pi^\dagger(\vec{x},0)\rangle  =c_2^{\pi,{\rm PP}}(\vec{p},t),
\end{align}
where the blending operator $\mathbf{O}^{{\rm BLD}}_{\pi}$ with the re-weight tensor $\Omega^{(2)}$ is, 
\begin{align}
   \mathbf{O}^{{\rm BLD}}_{\pi,ij}(\vec{p},t)  =  \gamma^5  \Omega^{(2)}_{ij}  \int \dd x^3  
    \langle \phi_i(\vec{x},t) | e^{-i \vec{p} \cdot \vec{x}}  | \phi_j(\vec{x},t) \rangle,\  \Omega^{(2)}_{ij}=
        \begin{cases}
            1 & \mathrm{for}\ i,j\le N_e\\
            \omega_0\omega_1 &\mathrm{for}\  i,j> N_e, i\neq j\\
            \omega_0 &\text{for the other cases}\\
        \end{cases},\label{eq:operator_sm}
\end{align}
and $\omega_n=([{\cal L}_2]-n)/[N_{\mathrm st}-n]$. 

For the zero momentum case ($\vec{p}=0$), the operator $\mathbf{O}_{\pi}$ becomes diagonal, rendering the off-diagonal \red{weights $\Omega^{(2)}_{i\neq j}$} irrelevant. As demonstrated in the left panels of Fig. ~\ref{fig:C2Pion_rebuild}, both the two-point correlation function (left panel, normalized by $c_2^{\pi{\rm BLD}}(\vec{p},t)$ with $N_{\mathrm st}=200$) and its effective mass (right panel) exhibit saturation behavior at $N_{\mathrm st}<10$. The results show excellent agreement with those obtained from $c_2^{\pi,{\rm PP}}(\vec{p},t)$ which uses one point-source propagator at each time \red{slice} with the same $N_{\rm cfg}=90$ for the unitary light quark mass on the C24P29 ensemble. 

    \begin{figure}[!h]
        \centering
        \includegraphics[width=0.45\linewidth]{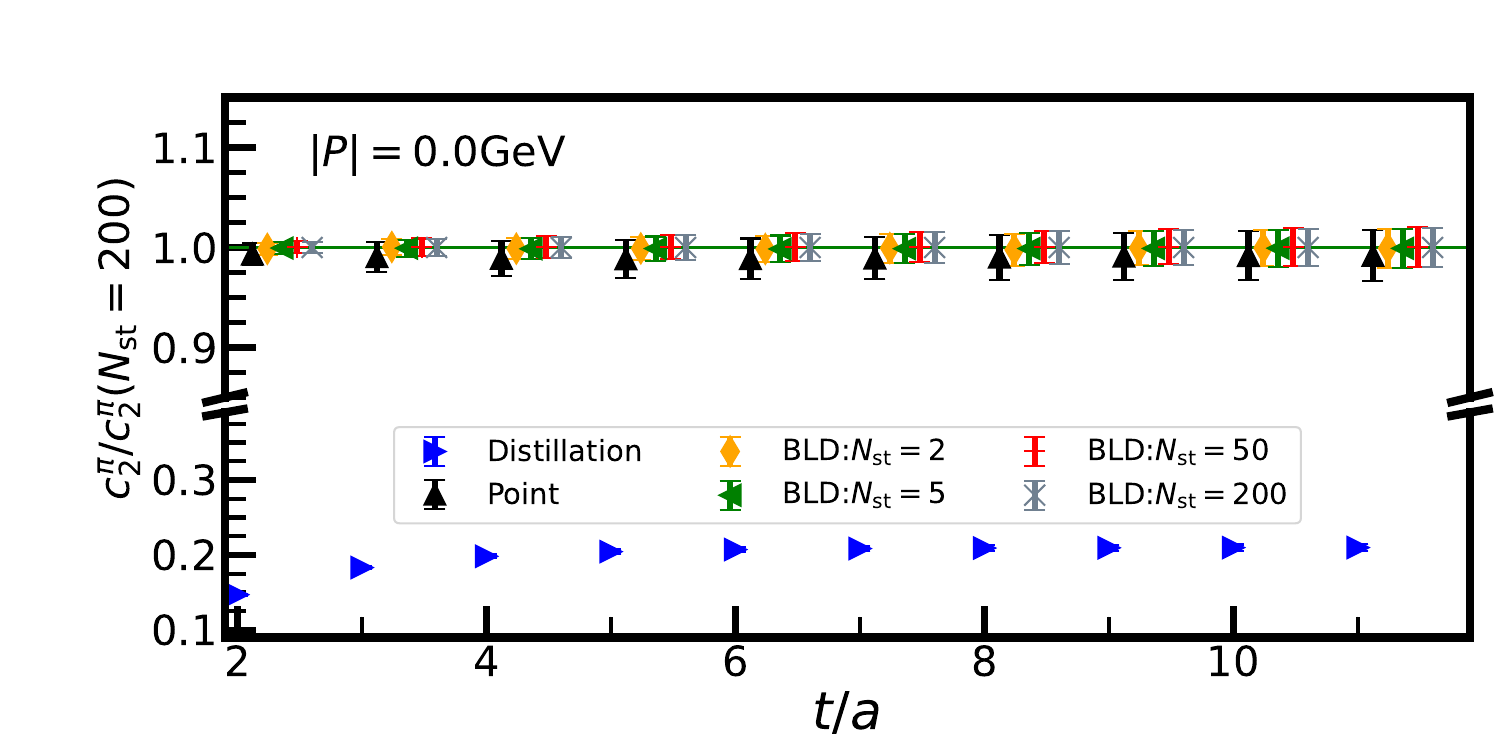}
        \includegraphics[width=0.42\linewidth]{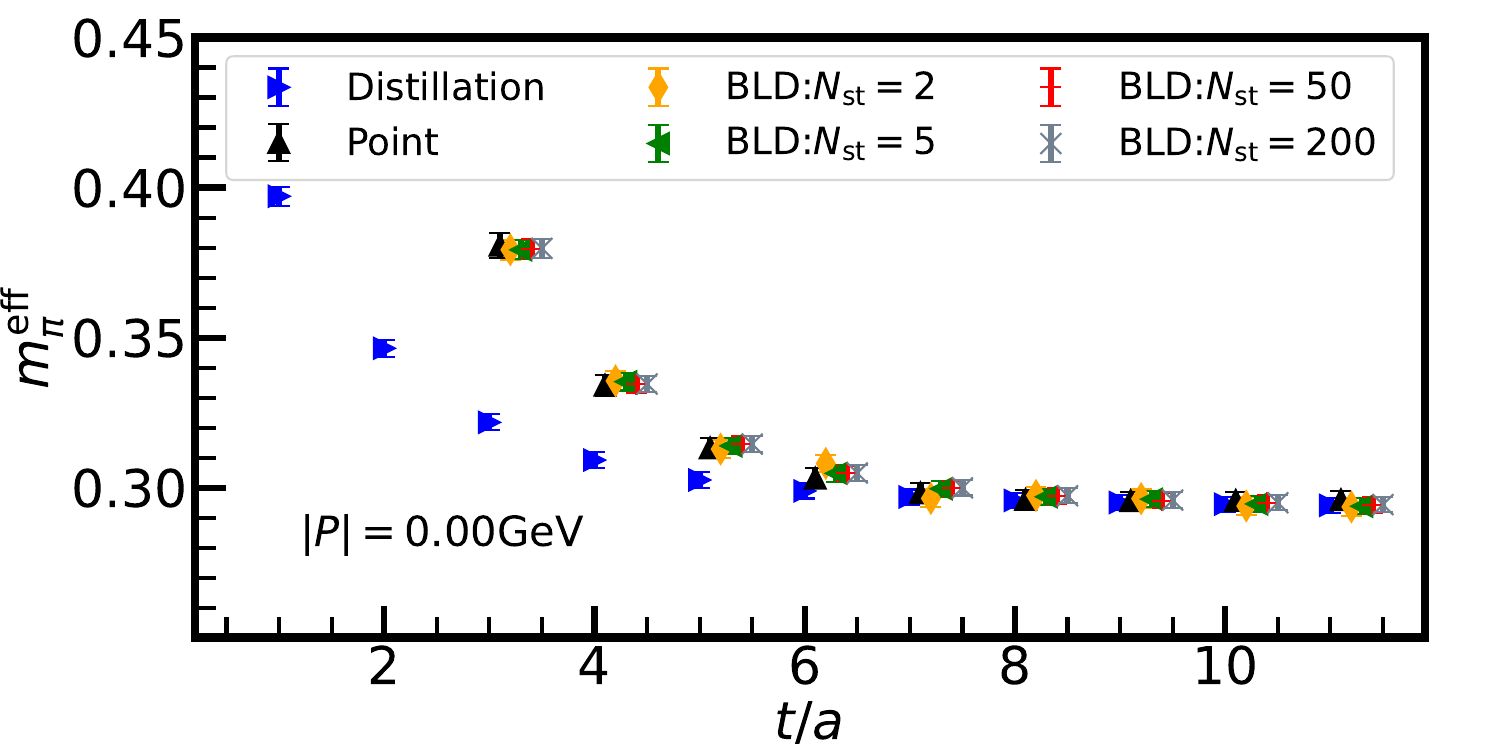}
        \caption{
    The zero momentum two point function (normalized by the saturated value) and the pion effective mass on C24P29 ensemble using the point source (black  dots), blending method with different $N_{\mathrm st}$, and the corresponding distillation results (blue triangles).}
        \label{fig:C2Pion_rebuild}
    \end{figure}

Simultaneously, we define the distilled two-point function $c_2^{\pi{\rm DST}}(\vec{p},t)  \equiv c_2({\mathbf O}_{\pi}^{{\rm DST}},\vec{p},t)$ which excludes all the contributions from the ${\cal L}_2$ subspace with the distilled operator, 
\begin{align}
    \mathbf{O}^{{\rm DST}}_{\pi,ij}(\vec{p},t)  =  \gamma^5  \int \dd x^3 
\langle \phi_i(\vec{x},t) | e^{-i \vec{p} \cdot \vec{x}}  | \phi_j(\vec{x},t) \rangle  \text{\ with\ } i,j\le N_e.
\end{align}
While the effective mass plot of $c_2^{\pi,{\rm DST}}$ demonstrates faster ground-state saturation(blue dots in the right panel of Fig.~\ref{fig:C2Pion_rebuild}), its magnitude never exceeds 16\% of  $c_2^{\pi,{\rm PP}}$ at any time slice (those in the left panel).
This substantial discrepancy reveals that the $\mathcal{L}_2$ subspace contribution to $c_2^{\pi,{\rm BLD}}$ is significant, and the reweighting factor  $\omega_0$ is essential to maintain  $c_2^{\pi,{\rm BLD}}$ as an unbiased estimator of $c_2^{\pi,{\rm PP}}$.

    \begin{figure}[!h]
        \centering
        \includegraphics[width=0.45\linewidth]{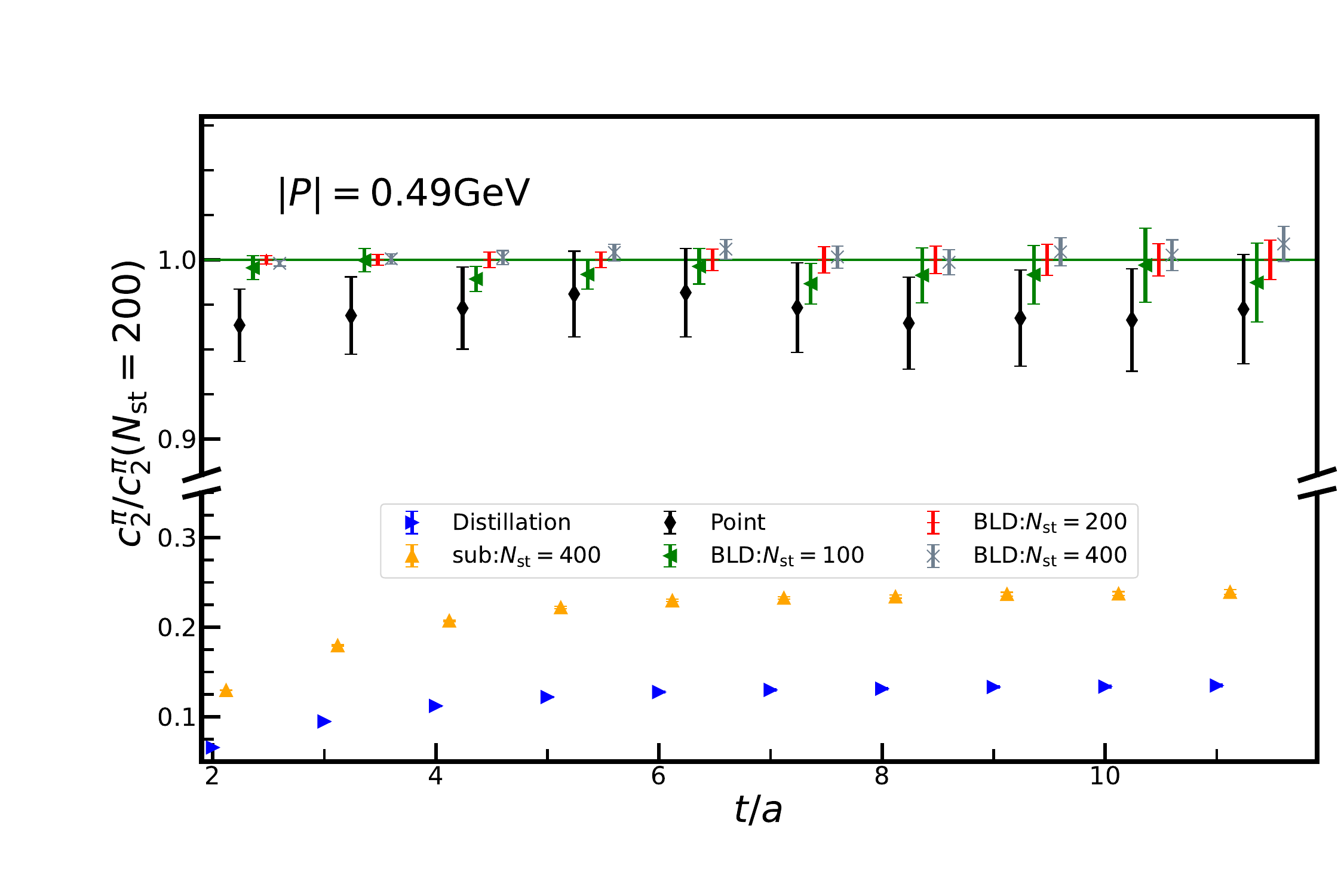}
        \includegraphics[width=0.42\linewidth]{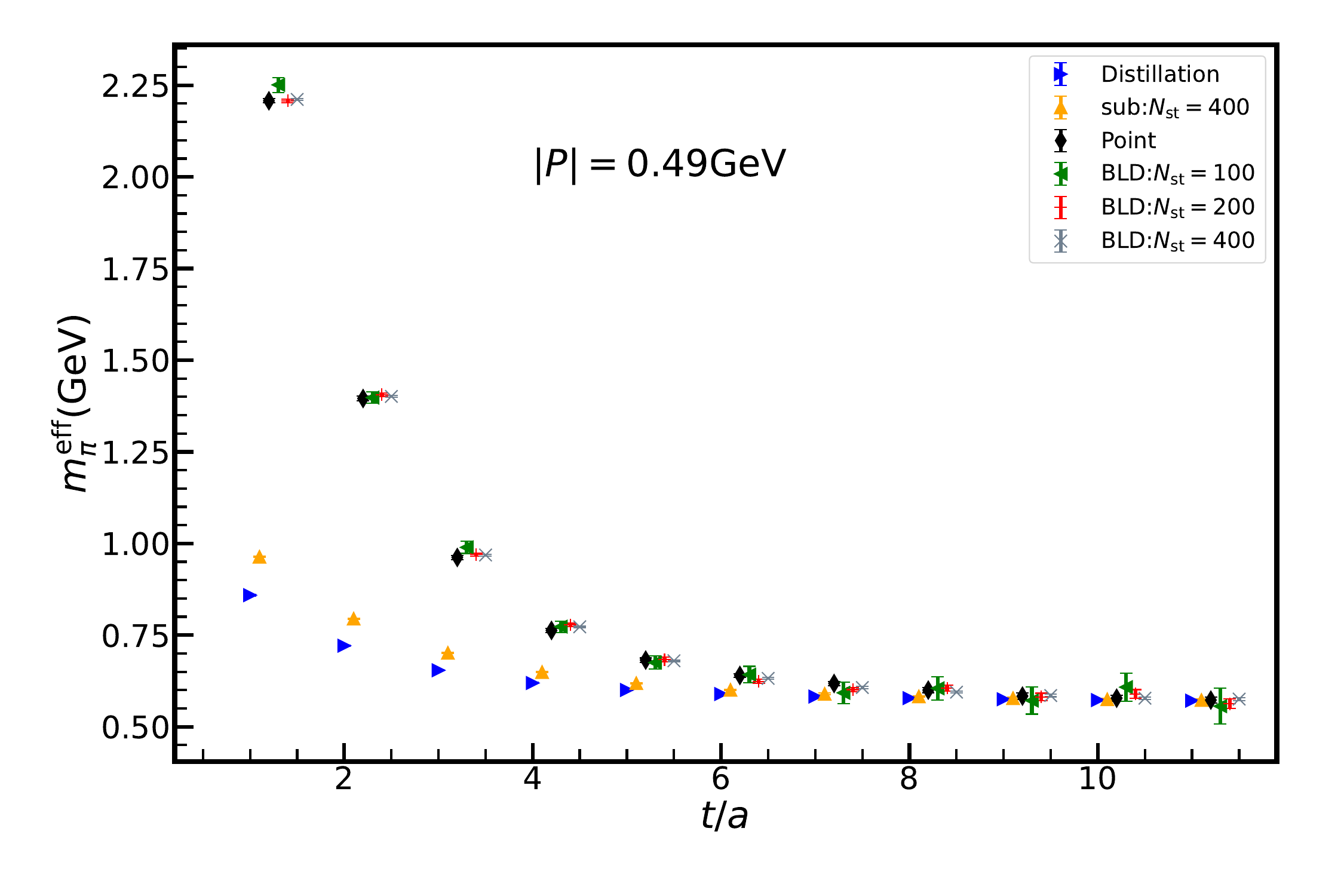}
        \caption{
    Similar to Fig.~\ref{fig:C2Pion_rebuild} but for $|\vec{p}|=2\pi/L$. The orange triangles show the 2-pf and its effective mass without the off-diagonal noise-noise contribution in the interpolation fields.
        }
        \label{fig:C2Pion_rebuild_1}
    \end{figure}

In Fig.~\ref{fig:C2Pion_rebuild_1}, we show similar plots for the case with $|\vec{p}|=2\pi/(N_La)$. We observe that a significantly larger number of stochastic samples $N_{\mathrm st}\sim 200$ is required for the uncertainty of $c_2^{\pi,{\rm BLD}}$ to saturate. 
This requirement arises because the off-diagonal components of $\mathbf{O}_{\pi}$ become non-zero when momentum is introduced, and the total number of such components in the full ${\cal L}_2$ space scales as $[{\cal L}_2]([{\cal L}_2]-1)$. 
However, with only $N_{\mathrm st}$ random vectors, we sample just $N_{\mathrm st}(N_{\mathrm st}-1)$ off-diagonal components. 
Consequently, a substantial reweighting factor $\omega_0\omega_1\sim([{\cal L}_2]/N_{\mathrm st})^2$ is needed to properly account for the unsampled off-diagonal contributions, leading to larger uncertainties at small $N_{\mathrm st}$. But the uncertainty of {$c_2^{\pi,{\rm BLD}}$} exhibits a more favorable $1/N_{\mathrm st}^2$ scaling due to the complete statistical independence between source and sink samples.

To explicitly demonstrate the importance of off-diagonal contributions in the ${\cal L}_2$ space, we define an additional two-point function $c_2^{\pi,{\rm sub}}(\vec{p},t) \equiv c_2(\mathbf{O}_{\pi}^{{\rm sub}},\vec{p},t)$ using a modified operator $\mathbf{O}^{{\rm sub}}_{\pi}$ with specific components subtracted:
\begin{align}
\mathbf{O}^{{\rm sub}}_{\pi,ij}(\vec{p},t) = \gamma^5 \Omega^{(2),{\rm sub}}_{ij} \int \dd x^3 \langle \phi_i(\vec{x},t) | e^{-i \vec{p} \cdot \vec{x}} | \phi_j(\vec{x},t) \rangle,\  \Omega^{(2),{\rm sub}}_{ij}=
\begin{cases}
1 & \mathrm{for}\ i,j\le N_e\\
\mathbf{0} &\mathbf{\text{for}\ i,j> N_e, i\neq j}\\
\omega_0 &\text{for the other cases}\\
\end{cases}.
\end{align}
Fig.~\ref{fig:C2Pion_rebuild_1} reveals that $c_2^{\pi,{\rm sub}}$ contributes no more than 25\% to $c_2^{\pi,{\rm BLD}}$, demonstrating that the omitted off-diagonal components are crucial for obtaining an unbiased approximation of $c_2^{\pi,{\rm PP}}$.

    \begin{figure}[!h]
        \centering
        \includegraphics[width=0.45\linewidth]{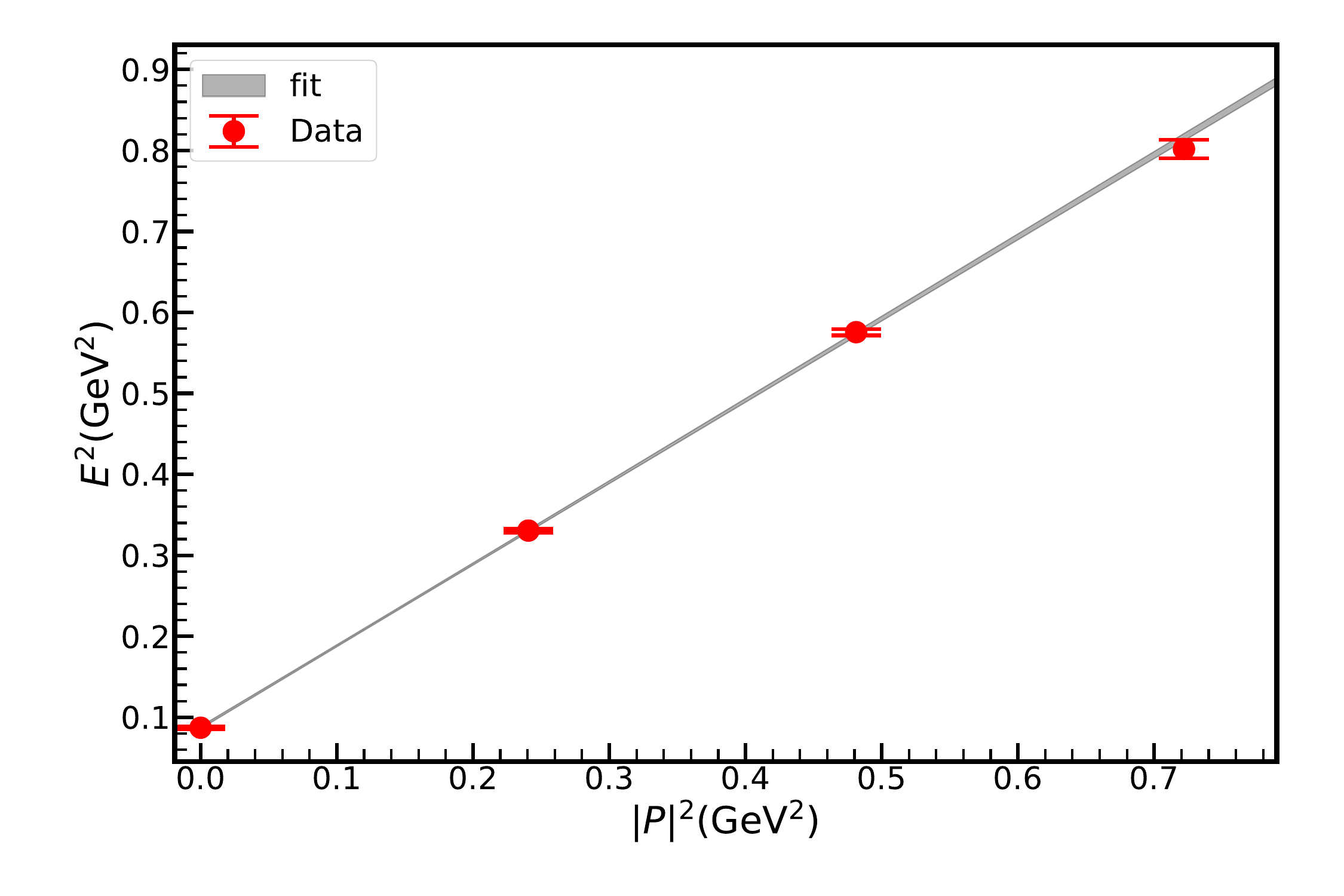}
        \caption{
   $p^2$ dependence of $E_{\pi}(\vec{p})$ obtained using $C^{\pi,{\rm DST}}_{2}$. 
        }
        \label{fig:C2Pion_disp}
    \end{figure}

The pion energy $E_{\pi}(\vec{p})$ with different $\vec{p}$ can be obtained from $C^{\pi,{\rm DST}}_{2}$ using the following parameterization,
\begin{align}
  C^{\pi,{\rm DST}}_{2}(\vec{p},t)&=Z_{\vec{p}} (e^{-E_{\pi}(\vec{p})t}+e^{-E_{\pi}(\vec{p})(T-t)}),
\end{align}
where the second term is introduced introduced to describe the loop around effect in a finite size lattice with temporal length $T$. As shown in Fig~\ref{fig:C2Pion_disp}, a linear fit of $E^2_{\pi}(\vec{p})$ as a function of $p^2$ gives the slope as 1.010(07) which perfectly agrees with the continuum dispersion relation, with $\chi^2$/d.o.f.=0.96. 

\subsubsection{On-shell three point function}

As shown in the right panels of Fig~\ref{fig:C2Pion_rebuild} and \ref{fig:C2Pion_rebuild_1}, $c_2^{\pi,{\rm DST}}$ can saturate to the ground state much faster than $c_2^{\pi,{\rm BLD}}$ with similar uncertainty, it is natural to use the ${\cal O}^{{\rm DST}}$ for the hadron state and ${\cal O}^{{\rm BLD}}$ only for the current operator. Using this setup, the pion matrix element of the conserved vector current 
    \begin{align}
{\mathcal V}\equiv \int \dd^3 x \dd^3 y \bar{q}(x)V^c(x,y)q(y),\ 
V^c(x,y) = \frac{1}{2} \big[
    \delta_{x,y+\hat{n}_4a}(1+\gamma_4) U^\dagger_4(y) - \delta_{x+\hat{n}_4a,y} (1-\gamma_4) U_4(x) 
\big],\nonumber
    \end{align}
can be a good example to show the blending approximation ${\mathcal V}^{/\rm BLD}$ for the case where the quark and anti-quark in the quark bilinear operator are not located at the same time slice, and then the re-weight matrix $\Omega^{(2)}_{ij}=\Omega^{(1)}_i\Omega^{(1)}_j$ with $\Omega^{(1)}_k=1$ for $k \leq N_{\mathrm e}$ and $\omega_0$ for $k > N_{\mathrm e}$.

\begin{figure}[!h]
    \centering
    \includegraphics[width=0.45 \textwidth]{CC_c32_qm-0_2770_pre1e-07_tsep10.pdf}
    \includegraphics[width=0.45 \textwidth]{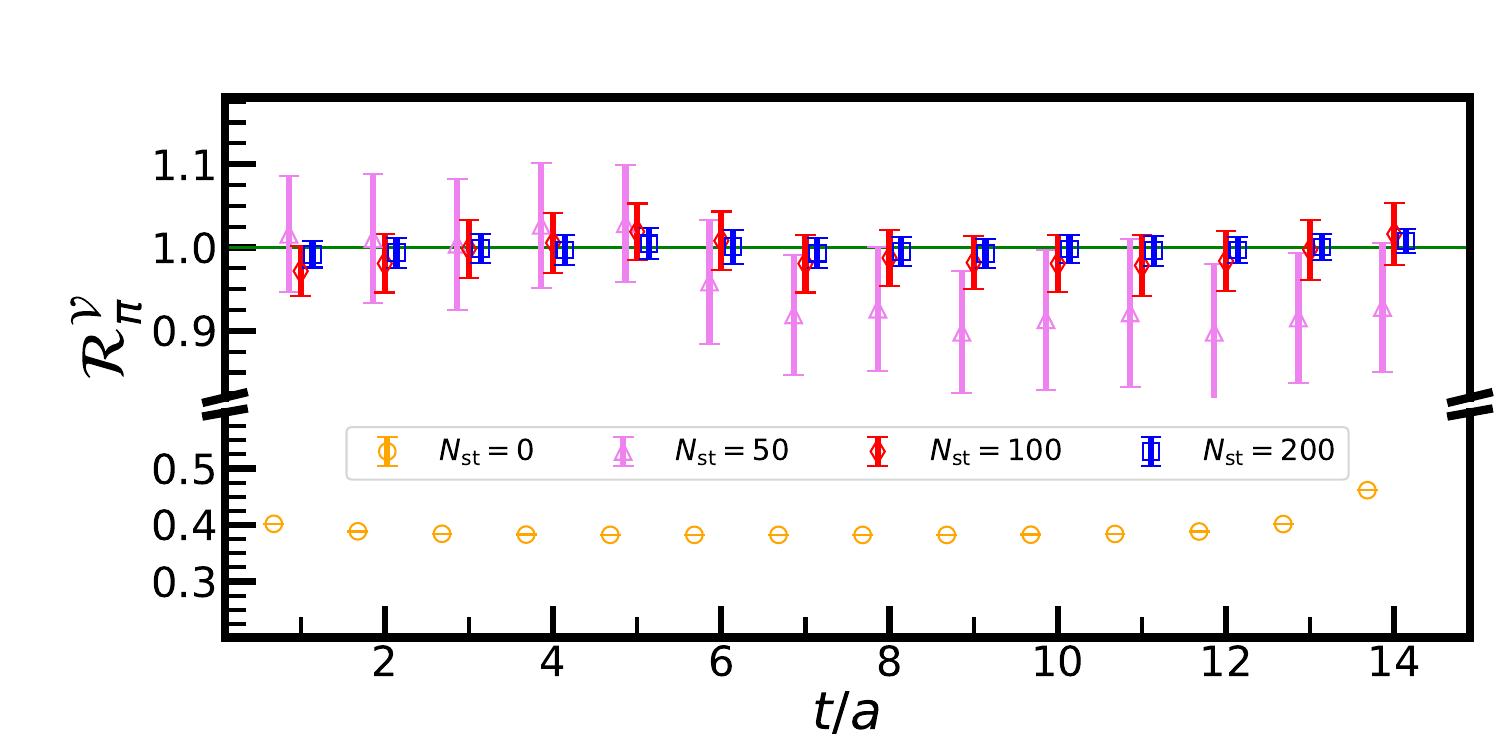}
    \includegraphics[width=0.45 \textwidth]{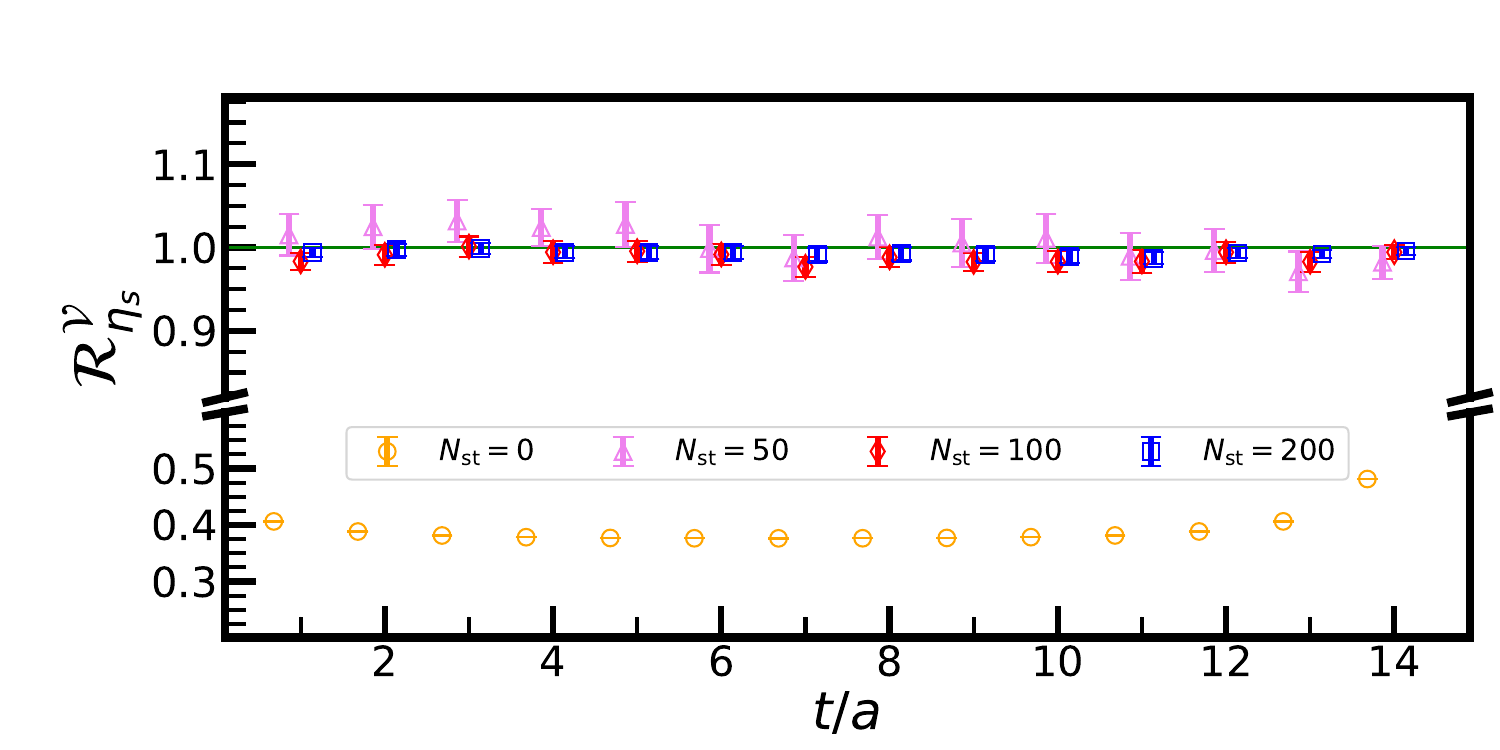}
    \includegraphics[width=0.45 \textwidth]{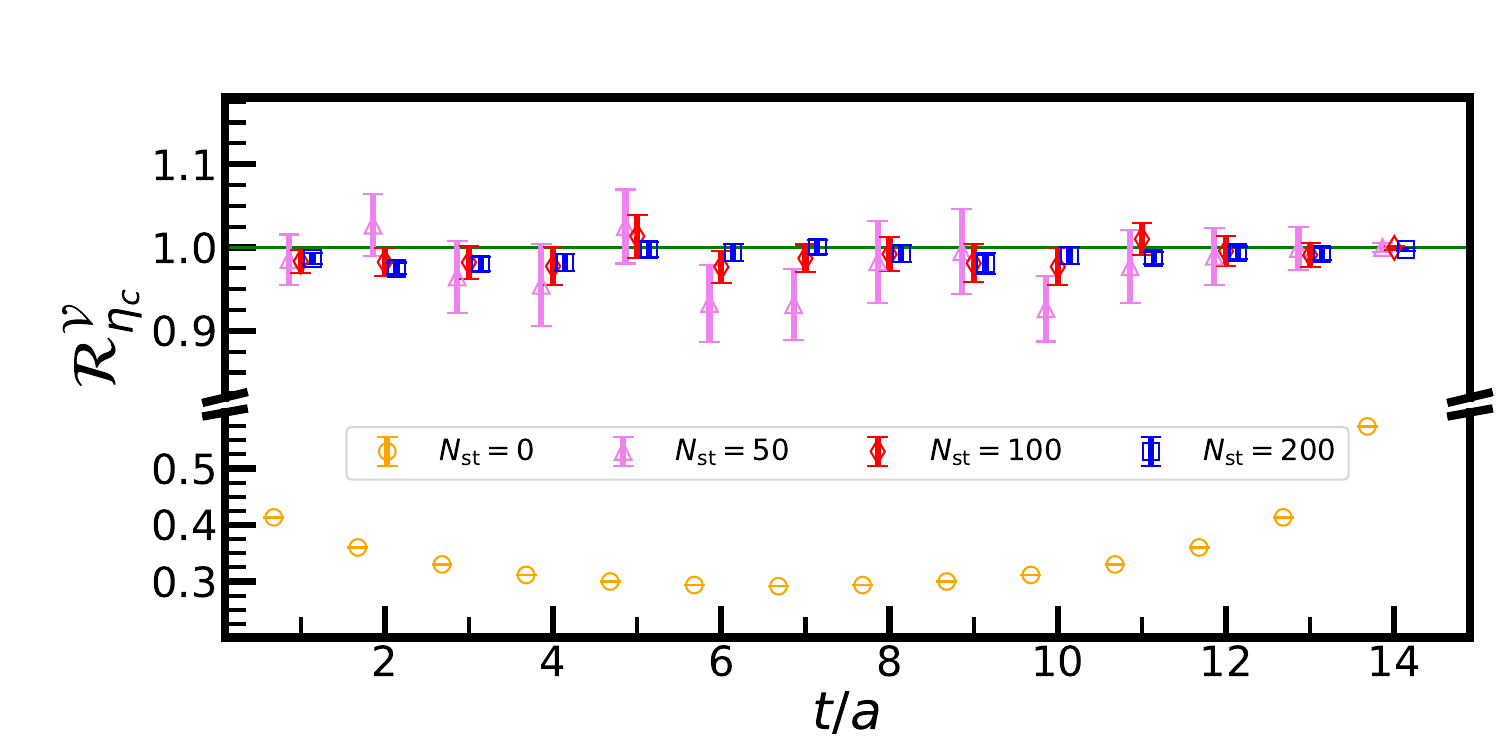}
    \caption{
            The $\pi^+$ matrix element of the conserved vector current operator on C24P29 (left upper panel) and F48P30 ensemble, varies the current-inserted position $t/a$ for different numbers $N_{\mathrm st}$. 
            The lower two panels show the cases of $\eta_s$ (left lower panel) and $\eta_c$ (right lower panel) on the F48P30 ensemble. 
    }
    \label{Fig:CC_C32P32}
\end{figure}

\begin{table}[htbp]
  \centering
    \caption{
    The pseudo-scalar matrix elements of the conserved vector current at $t_{\rm ins}=t_{\rm sep}/2$, on the F48P30 ensemble.
    }
    \begin{tabular}{c|cccc}
    \hline
    \hline
    $N_{\mathrm st}$ & 0     & 50    & 100   & 200 \\
    \hline
    $\pi^{\pm}$ &  0.38250(41) & 0.919(72) & 0.981(35) &  0.993(18) \\
    $\eta_s$ & 0.37640(29) & 0.987(28) &  0.977(12) & 0.991(08) \\
    $\eta_c$ &  0.29183(30) &  0.931(43) &  0.987(17) &  1.001(08) \\
    \hline
    \hline
    \end{tabular}%
  \label{tab:CC}%
\end{table}%

In addition to the pion results obtained on the C24P29 ensemble (left upper panel of Fig.~\ref{Fig:CC_C32P32}), we present corresponding results for the pion (right upper panel), $\eta_s$ (left lower panel), and $\eta_c$ (right lower panel) on the F48P30 ensemble. 

As evidenced by Table~\ref{tab:CC} and Fig.~\ref{Fig:CC_C32P32}, the distillation space contribution to ${\mathcal V}$ remains below 40\% for both pion and $\eta_s$ cases, with an even smaller fraction observed for $\eta_c$. Nevertheless, ${\mathcal V}^{\rm BLD}$ consistently provides unbiased estimates of ${\mathcal V}$, with statistical uncertainties following the expected $1/N_{\mathrm st}$ scaling. 

An interesting observation emerges when comparing statistical uncertainties across different mesons: while $\eta_s$ and $\eta_c$ display comparable error magnitudes, the pion case exhibits significantly larger uncertainties. The origin of this discrepancy is unclear and warrants further investigation in future studies.

\subsubsection{Off-shell three point function}

As an even more complicated example of the blending method, we show its application in the RI/MOM renormalization. Taking the quark self energy under the RI/MOM scheme~\cite{Martinelli:1994ty} as example,
\bal
Z_q(\mu)&=Z_V\frac{1}{48}\text{Tr}[\langle S\rangle^{-1}(p)\langle G_{\Gamma}(p)\rangle\langle S\rangle^{-1}(p)\gamma_\mu],
\eal
where $Z_V$ can be obtained through the pion matrix element of the local vector current, $S$ and $G_{\Gamma}$ can be approximated using the blending method as
\bal
S(p)=&\lim_{N_{\mathrm st}\rightarrow [{\cal L}_2]}S^{\rm BLD}(p;N_{\mathrm st}),\ S^{\rm BLD}(p;N_{\mathrm st})\equiv \sum_{t_1,t_2}\sum^{N_{\mathrm e}+N_{\mathrm st}}_{i=1,j=1}\frac{\Omega^{(2)}_{ij}}{V}| \phi_i(p,t_1)\rangle S_{ij}(t_1,t_2)\langle \phi_j(p,t_2)|,\nonumber\\
G_{\Gamma}(p)=&\lim_{N_{\mathrm st}\rightarrow [{\cal L}_2]}G_{\Gamma}^{\rm BLD}(p;N_{\mathrm st}),\ G_{\Gamma}^{\rm BLD}(p;N_{\mathrm st})\equiv \sum_{t_1,t_2,t_3}\sum^{N_{\mathrm e}+N_{\mathrm st}}_{i=1,j=1,k=1}\frac{\Omega^{(3)}_{ijk}}{V}|\phi_i(p,t_1)\rangle S_{ij}(t_1,t_2)\Gamma S_{jk}(t_2,t_3)\langle \phi_k(p,t_3)|,
\eal
where $|\phi_i(p,t)\rangle =\sum_{\vec{x}}e^{-{\rm i}(p_4t+\vec{p}\cdot \vec{x})} |\phi_i(t;\vec{x})\rangle$, $V=L^3\times T$. $\Omega^{(3)}_{ijk}$ takes different values in different cases:

\begin{align}
\Omega^{(3)}_{ijk}=
    \begin{cases}
        1 & \text{for} \ i,j,k \leq N_{\mathrm e},\\
        \omega_0\omega_1\omega_2 & \text{for} \ i \neq j \neq k>N_{\mathrm e},\\
        \omega_0\omega_1 & \text{for} \ i\neq j>N_{\mathrm e},k \leq N_{\mathrm e}\ \mathrm{or}\ k=i, \text{and all the permutation of } i,j,k,\\
        \omega_0 & \text{for the other cases}.
    \end{cases}
\end{align}
One can also only use the above expression for the case of $t_1=t_2=t_3$, and use $\Omega^{(1)}_i\Omega^{(1)}_j\Omega^{(1)}_k$ when $t_1\neq t_2\neq t_3$, and also the product of $\Omega^{(2)}_{ij}$ defined in Eq.~(\ref{eq:operator_sm}) and $\Omega^{(1)}_k$ when $t_1=t_2\neq t_3$. The left panel of Fig.~\ref{fig:RIMomZq} shows the $Z_q/Z_V$ (on the C24P29 ensemble) from the blending method (red data points) is consistent with that from the exact point source propagators (green band) well, which justifies the correctness of the reweighting factor $\Omega^{(3)}$.

Based on the above procedure, $Z_A/Z_V$ for the SU($n_f$) flavor non-singlet and singlet cases in the RI/MOM scheme can be obtained as  follows:
\begin{align}
\frac{Z^{\rm ns}_A}{Z_V}=\frac{\text{Tr}[\langle S\rangle^{-1}(p)\langle G_{\gamma_{\mu}}(p)\rangle\langle S\rangle^{-1}(p)\gamma_\mu]}{\text{Tr}[\langle S\rangle^{-1}(p)\langle G_{\gamma_{\mu}\gamma_5}(p)\rangle\langle S\rangle^{-1}(p)\gamma_5\gamma_\mu]},\ \frac{Z^{\rm s}_A}{Z_V}=\frac{\text{Tr}[\langle S\rangle^{-1}(p)\langle G_{\gamma_{\mu}}(p)\rangle\langle S\rangle^{-1}(p)\gamma_\mu]}{\text{Tr}[\langle S\rangle^{-1}(p)\langle (G_{\gamma_{\mu}\gamma_5}(p)+N_f S(p)L_{\gamma_{\mu}\gamma_5} )\rangle\langle S\rangle^{-1}(p)\gamma_5\gamma_\mu]},
\end{align}
where the quark loop $L_{\Gamma}$ is defined as $L_{\Gamma}=\sum_{i,t}\mathrm{Tr}[\Omega^{(1)}_i\Gamma S_{ii}(t,t)]$. 
The result for $\frac{Z^{\rm s}_A}{Z_V}$ can be further converted to $\overline{\text{MS}}$ ($\mu_0=2\,$GeV) with the aid of the finite matching coefficient $R_A \equiv  Z_{A}^{{\rm s},\overline{\text{MS}}}(\mu) / Z_{A}^{\rm s,RI}(\mu_{\mathrm{RI}}) $ for converting the renormalization of $J_A^{\mu}$ from the $\RIMOM$ scheme to $\MSbar$ scheme, which will be discussed in detail in Sec.~\ref{sec:Rexpression} of this Supplemental Material.
More specifically, $\frac{Z_{A}^{{\rm s},\overline{\text{MS}}}(\mu_0)}{Z_V} = R_A(\mu_0/\mu_{\mathrm{RI}}) \, \frac{Z_{A}^{\rm s,RI}(\mu_{\mathrm{RI}})}{Z_V}$ with the RG-improved perturbative result for $R_A(\mu_0/\mu_{\mathrm{RI}})$ given in~Eq.~\eqref{eq:RAsRGsolu}.

The renormalized axial charge \( g_A^{q,{\rm R}} \) 
is obtained using the nonsinglet (\( Z_{A}^{\rm ns} \)) and singlet (\( Z_{A}^{\rm s} \)) renormalization constants as:
\begin{align}
g_A^{q,{\rm R}} = Z_A^{\rm ns} g_A^{q,{\rm bare}} + \frac{Z_A^{\rm s} - Z_A^{\rm ns}}{N_f} \sum_q g_A^{q,{\rm bare}}\,,
\end{align}
analogous to the renormalization of the singlet axial-current operator defined with a specific $q$-quark field in Ref.~\cite{Chen:2021rft}.
This ensures that the total renormalized axial charge satisfies $\sum_q g_A^{q,{\rm R}} = Z_A^{\rm s} \sum_q g_A^{q,{\rm bare}}$.

\begin{figure}[!h]
    % \centering
    \includegraphics[width=0.45 \textwidth]{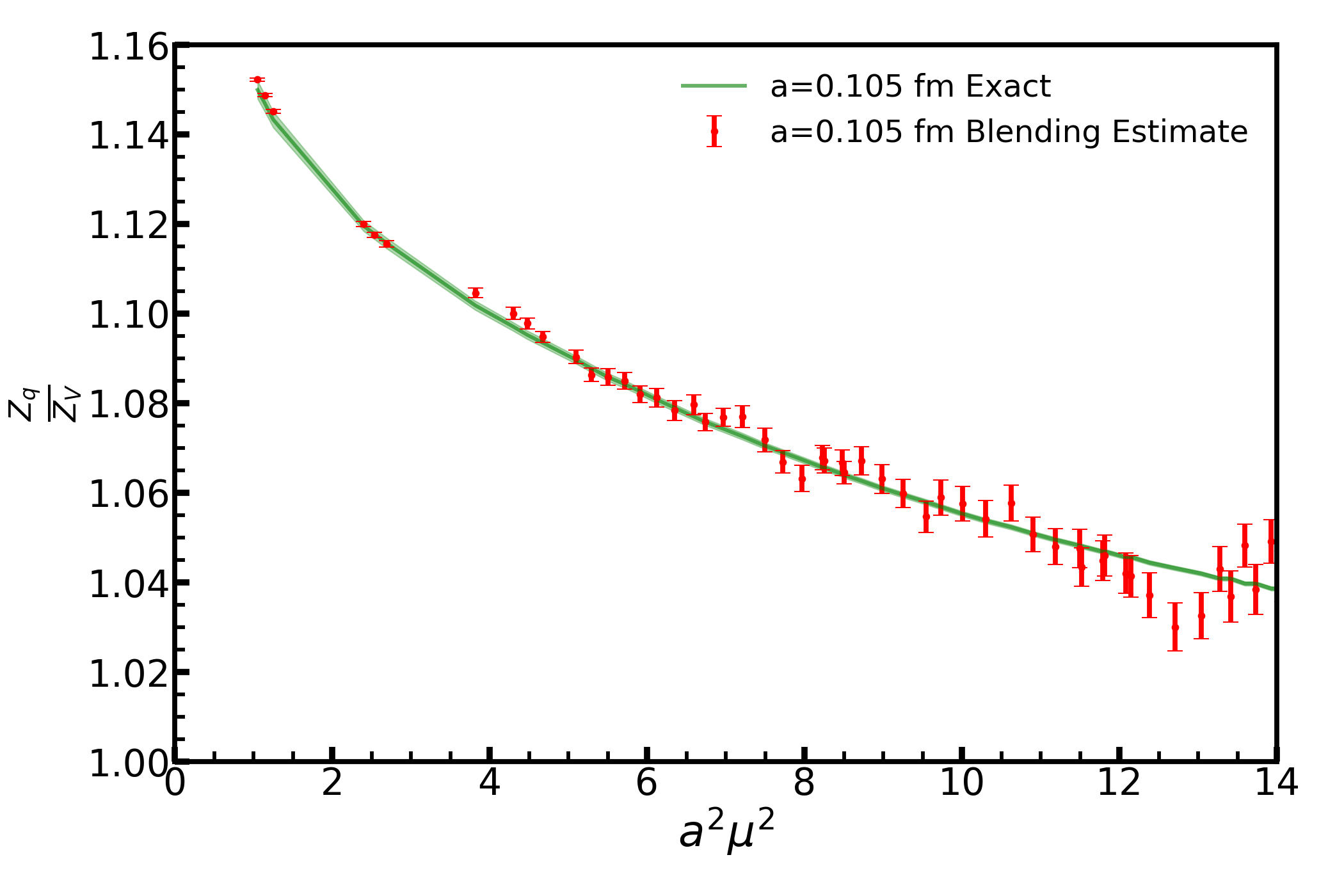}
    \includegraphics[width=0.45 \textwidth]{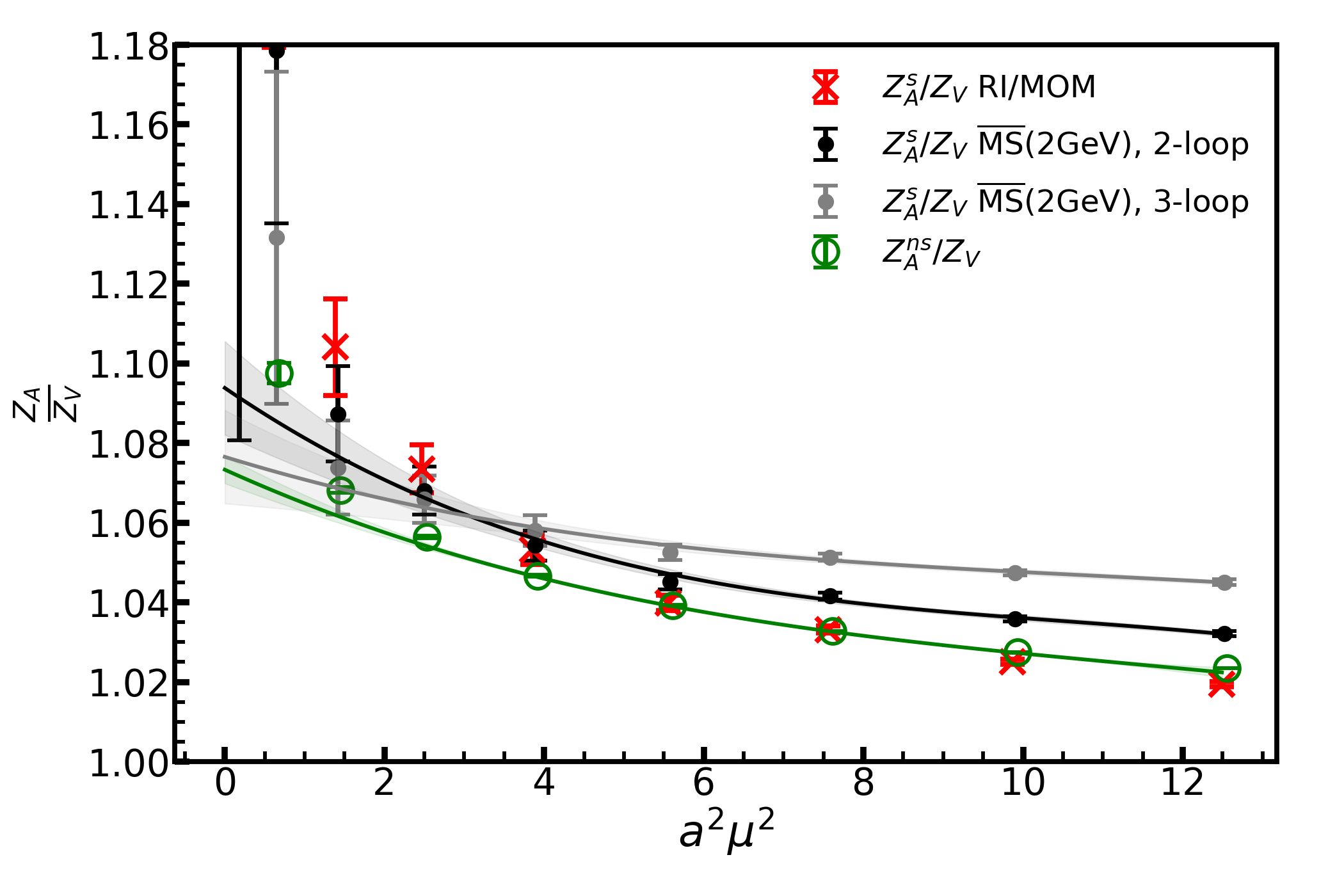}
    \caption{
        Left panel: $Z^{\rm RI/MOM}_q$ as a function of $a^2\mu^2$ on C24P29, using the exact point source propagator (green band) and blending method (red crosses).       
        Right panel: the singlet $Z_A/Z_V$ in the RI/MOM (red crosses) and $\overline{\text{MS}}$ scheme at 2 GeV (black dots) as well as the non-singlet result (green circles) on the F48P30 ensemble. 
    }
    \label{fig:RIMomZq}
\end{figure}

For the F48P30 ensemble used for the $g_A$ calculation, the singlet $Z_A/Z_V$ in the RI/MOM (red crosses) and $\overline{\text{MS}}$ scheme at 2 GeV (black dots), as well as the non-singlet result (green circles) are shown in the right panel of Fig.~\ref{fig:RIMomZq}. 
Both results deviate from 1 and $Z^{\rm s, \overline{\text{MS}}}_A/Z_V$ has a large deviation due to the triangle anomaly. 
Based on the fit in the range of $\mu\in [3~\mathrm{GeV},\sqrt{10/a^2}]$, we can have $Z^{\rm ns}_A/Z_V=1.05794(29)$ and $Z^{{\rm s},\overline{\text{MS}}}_A(2~\mathrm{GeV})/Z_V=\red{1.064(10)}$ using the polynomial form $\sum_{i=0}^3c_0a^{2i}p^{2i}$. We can also see that the $Z^{\rm s,RI}_A/Z_V$ has a stronger $\mu^2$ dependence than $Z^{\rm ns}_A/Z_V$, while such a dependence is majorly canceled by the perturbative matching defined in Eq.~\eqref{eq:RAsRGsolu}. A similar calculation on the C32P23 ensemble gives $Z^{\rm ns}_A/Z_V=1.0733(34)$ and $Z^{{\rm s},\overline{\text{MS}}}_A(2~\mathrm{GeV})/Z_V=\red{1.077(12)}$.

\subsection{The Pion Form Factor: 3pt and 4pt}\label{sec:pionff}

    \begin{figure}[!h]
        \centering
        \includegraphics[width=0.60 \textwidth]{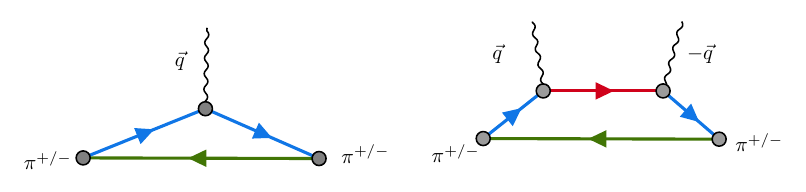}
        \caption{
            Illustration of the 3-pf and 4-pf needed to determine the pion electric form factor. 
            The quark propagators with different color correspond to those with different source-sink combinations, namely green for smeared-to-smeared, blue for smeared-to-point, and red for point-to-point. The summations on the spatial indices are performed for all the operator locations (gray points).
        }
        \label{Fig:Pion34Dig}
    \end{figure}

The pion electric form factor and mean square charge radius are defined as 
\begin{align}
    \langle \pi^+(p)|V_{\mu}|\pi^+(k)\rangle =(p_{\mu}+k_{\mu})f_{\pi\pi}(Q^2),\  \langle r_{\pi}^2\rangle =-6\frac{\partial f_{\pi\pi}(Q^2)}{\partial Q^2}|_{Q^2\rightarrow 0}.
\end{align}
where $Q^2=-(p-k)^2$ and $V_{\mu}=\frac{1}{2}(\bar{u}\gamma_{\mu}u-\bar{d}\gamma_{\mu}d)$. 
And then $f_{\pi\pi}(Q^2)$ can be obtained from either the 3-pf or 4-pf shown in Fig.~\ref{Fig:Pion34Dig},
\begin{align}
    \frac{2\sqrt{m_{\pi}E_{\pi}}}{m_{\pi}+E_{\pi}}\frac{C^{\pi}_{3}(\vec{0},\vec{p},t,t_f)}{\sqrt{C^{\pi}_{2}(\vec{0},t_f)C^{\pi}_{2}(\vec{p},t_f)}}&=f_{\pi\pi}(Q^2)+{\cal O}(e^{-\Delta(t_f-t)},e^{-\Delta 't}),\label{eq:pion_ff_3pt}\\
    \frac{2\sqrt{m_{\pi}E_{\pi}}}{m_{\pi}+E_{\pi}}\sqrt{\frac{C^{\pi}_{4}(\vec{p},t_1,t_2,t_f)}{C^{\pi}_{2}(\vec{0},t_f)}e^{(E_{\pi}-m_{\pi})(t_2-t_1)}}&=f_{\pi\pi}(Q^2)+{\cal O}(e^{-\Delta t_1},e^{-\Delta' (t_2-t_1)},e^{-\Delta (t_f-t_2)}),\label{eq:pion_ff_4pt}
\end{align}
where $Q^2=|\vec{p}|^2-(E_{\pi}-m_{\pi})^2$ in the kinetics used in Eqs.~(\ref{eq:pion_ff_3pt}-\ref{eq:pion_ff_4pt}), $E_{\pi}=\sqrt{m_{\pi}^2+|\vec{p}|^2}$, $\Delta$ is the energy difference between the pion and its first excited state in the rest frame and $\Delta'$ is that with momentum $\vec{p}$, and the correlation function $C_{2,3,4}^{\pi}$ are defined as,
\begin{align}
 C^{\pi}_{2}(\vec{p},t)&=\frac{1}{L^3}\int \dd^3 x \dd^3 y e^{-{\rm i}\vec{p}\cdot (\vec{x}-\vec{y})}\langle H_\pi(\vec{x},t)H_\pi^{\dagger}(\vec{y},0)\rangle,\nonumber\\
C^{\pi}_{3}(\vec{p},\vec{k},t,t_f)&=\frac{1}{L^3}\int \dd^3 x \dd^3 y \dd^3 z e^{-{\rm i}\vec{p}\cdot (\vec{x}-\vec{z})+{\rm i}\vec{k}\cdot (\vec{y}-\vec{z})}\langle H_\pi(\vec{x},t_f)V_4(\vec{z},t)H_\pi^{\dagger}(\vec{y},0)\rangle,\nonumber\\
C^{\pi}_{4}(\vec{p},t_1,t_2,t_f)&=\frac{1}{L^3}\int \dd^3 x \dd^3 y\dd^3 z \dd^3 w e^{{\rm i}\vec{p}\cdot(\vec{w}-\vec{z})}\langle H_\pi(\vec{x},t_f)V_4(\vec{z},t_2)V_4(\vec{w},t_1)H_\pi^{\dagger}(\vec{y},0)\rangle.
\end{align}

Although both Eqs.~(\ref{eq:pion_ff_3pt}-\ref{eq:pion_ff_4pt}) can be employed to extract $f_{\pi\pi}(Q^2)$, their computational demands differ substantially. In the standard sequential method, Eq.~(\ref{eq:pion_ff_3pt}) requires calculation of at least 1 (standard source at $t_i=0$) + $N$ (sequential sources at different $t_f$) quark propagators. By contrast, Eq.~(\ref{eq:pion_ff_4pt}) demands significantly more: 1 (standard source at $t_i=0$) + $N$ (sequential sources at different $t_f$) $\times\ N_2$ (sequential sources at different $t_2$) $\times\ N_{Q^2}$ (different $Q^2$ values) quark propagators. This increased complexity arises because the four-point function $C^{\pi}_{4}$ requires an all-to-all propagator from $(\vec{z},t_2)$ to $(\vec{w},t_1)$ to perform the Fourier transform and control the excited-state contamination. This computational challenge is characteristic of similar four-point calculations needed for various important quantities, including QED corrections to hadron masses, $W-\gamma$ box diagrams, and hadronic tensor evaluations.

The blending method enables extraction of $f_{\pi\pi}(Q^2)$ from both Eqs.~(\ref{eq:pion_ff_3pt}-\ref{eq:pion_ff_4pt}) using the same propagators projected into the blending space. For those calculations on the C24P29 ensemble,  we use 100 eigenvectors for $H_{\pi}$ and 100 (eigen) +400 (noise) blending vectors for $V_4$.

After symmetrizing the three-point function $C^{\pi}_{3}$ in Eq.~(\ref{eq:pion_ff_3pt}), we extract $f_{\pi\pi}(Q^2)$ through the parameterization:
\begin{align}
 {\cal R}_V^{\rm 3pt}(t,t_f;\vec{p})\equiv \frac{\sqrt{C^{\pi}_{3}(\vec{0},\vec{p},t,t_f)C^{\pi}_{3}(\vec{p},\vec{0},t,t_f)}}{\sqrt{C^{\pi}_{2}(\vec{0},t_f)C^{\pi}_{2}(\vec{p},t_f)}}&=\frac{m_{\pi}+E_{\pi}}{2\sqrt{m_{\pi}E_{\pi}}}f_{\pi\pi}(Q^2)+c_1(e^{-\Delta'(t_f-t)}+e^{-\Delta' t})+c_2e^{-\Delta't_f},
\end{align}
where $f_{\pi\pi}(Q^2)$, $c_{1,2}$ and $\Delta'$ are free parameters, and $E_{\pi}(\vec{p})$ and $m_{\pi}=E_{\pi}(\vec{p}=0)$ are determined from independent fits to the two-point function $C^{\pi}_{2}$, as detailed in Sec.~\ref{sec:Proof}.

  \begin{figure}[!h]
      \includegraphics[width=0.49 \textwidth]{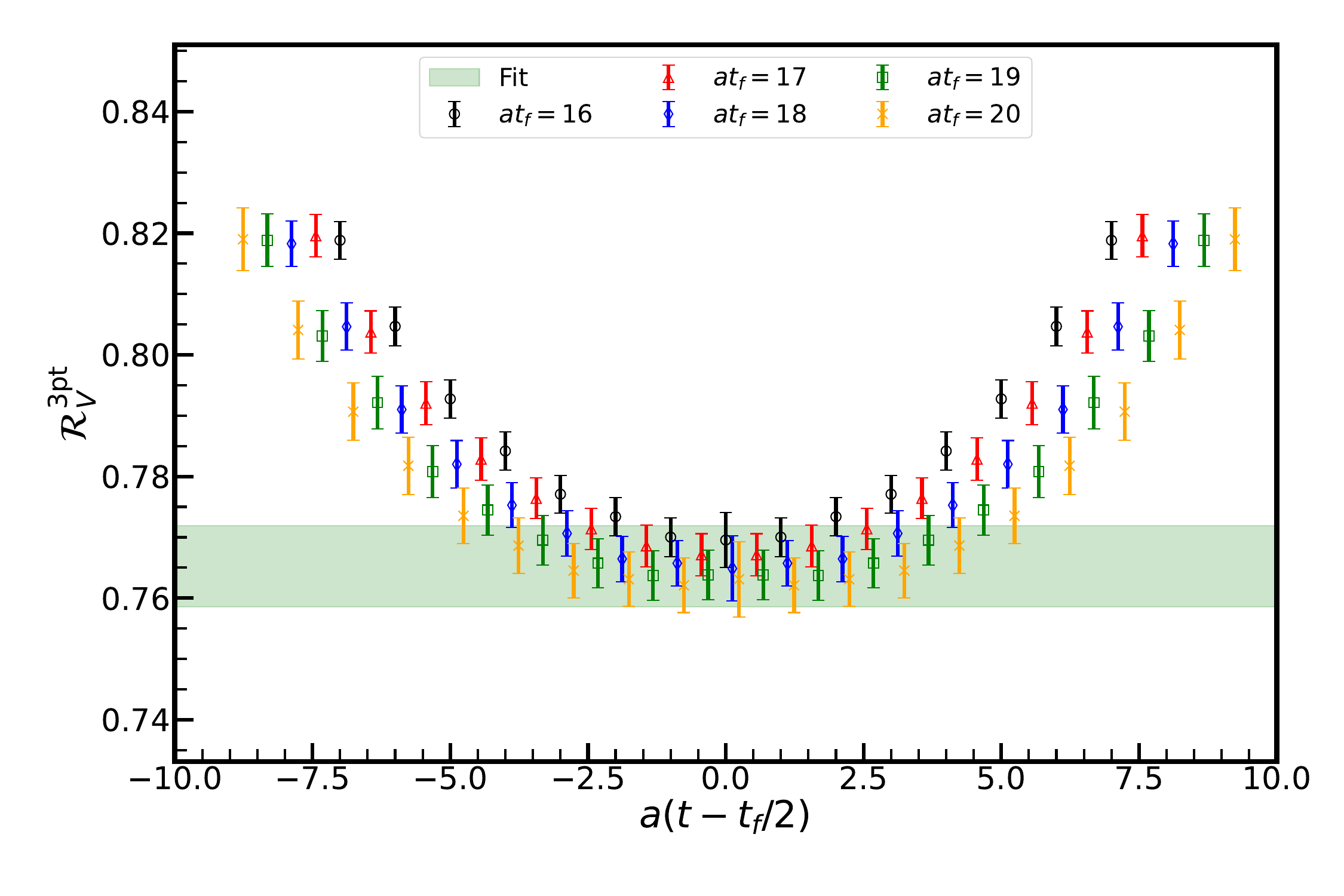}
      \includegraphics[width=0.49 \textwidth]{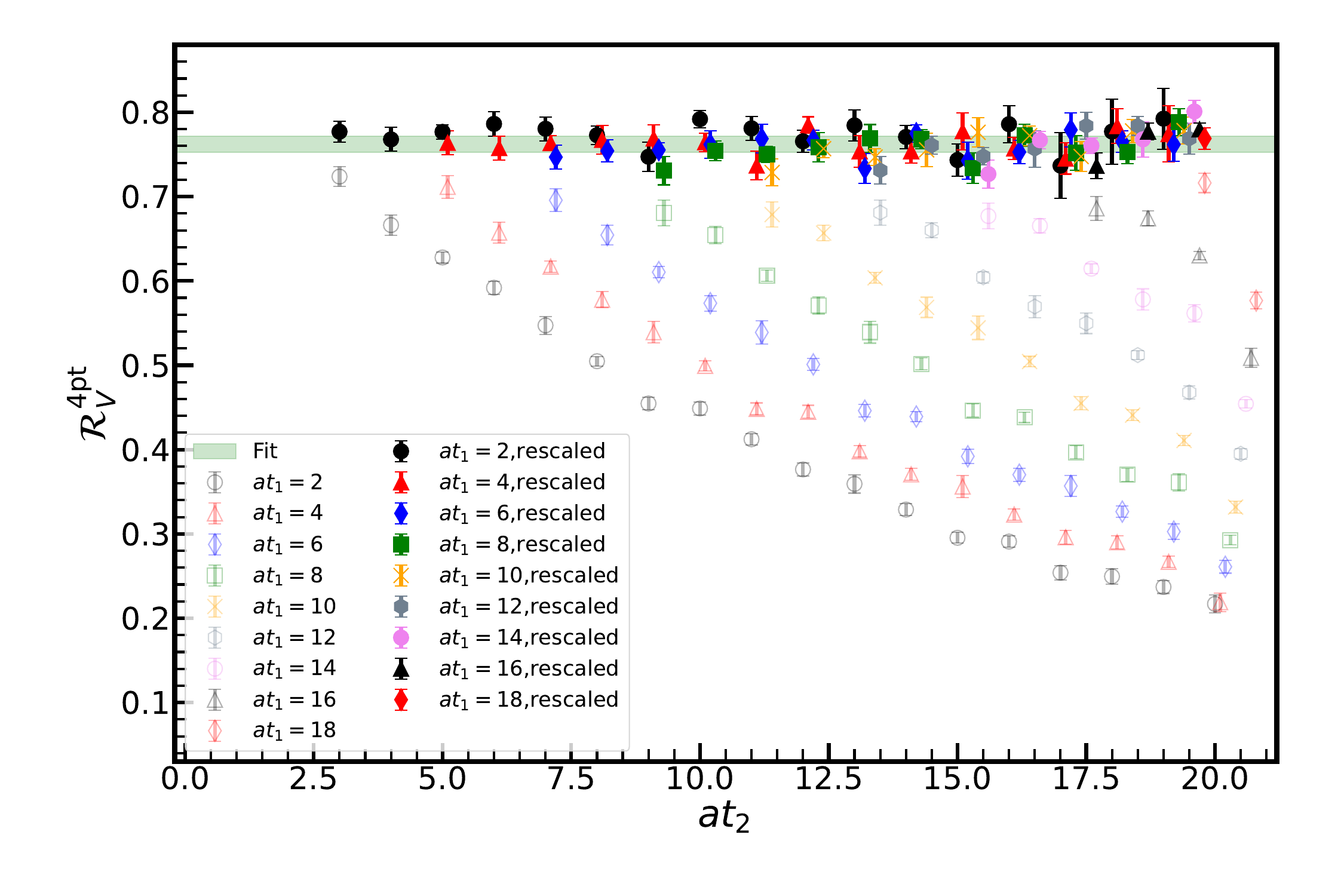}
      \caption{ 
      The left panel shows the symmetrized ratio ${\cal R}_V^{\rm 3pt}(t,t_f;\vec{p})$ for $0<t<t_f$, $t_f\in[1.68,2.11]$ fm and $|\vec{p}|=2\pi/L$;
      The right panel shows the original ratio ${\cal R}_V^{{\rm 4pt}}(t_1,t_2)$ (hollowed point with different color for different $at_1$) of 4-pf and 2-pf, as a function of $at_2$, and the rescaled ratio ${\cal R}_V^{\rm 4pt,rescaled}={\cal R}_V^{\rm 4pt}e^{(E_{\pi}-m_{\pi})(t_2-t_1)/2}$ (solid points). 
        The green bands in both the panels represent the ground state matrix element from the fits.
      }
      \label{fig:PionFF_OriginDate}
  \end{figure}

The left panel of Fig.~\ref{fig:PionFF_OriginDate} displays the ratio ${\cal R}_V^{\rm 3pt}(t,t_f;\vec{p})$ for the smallest non-zero momentum $|\vec{p}|=2\pi/L$, with five source-sink separations $t_f$ spanning 1.68-2.11 fm. As evident from the figure, all ${\cal R}_V^{\rm 3pt}$ curves exhibit plateaus centered around $t\sim t_f/2$, demonstrating satisfactory convergence on the infinite $t_f-t$ and $t$ limits. A two-state fit to these data yields $f_{\pi\pi}(Q^2=0.163~\mathrm{GeV}^2)=0.765(7)$ with excellent fit quality ($\chi^2/{\rm d.o.f}=0.53$).

On the other hand, the four-point function $C^{\pi}_{4}$ exhibits significantly smaller excited-state contamination for the same momentum $|\vec{p}|$. Fixing $t_f=20a$, one can fit the all the data points with $t_1\ge 2a$, $t_2\le t_f-2a$ and $t_2-t_1\ge 2a$ using the following parameterization,
\begin{align}
    {\cal R}_V^{{\rm 4pt}}(t_1,t_2)\equiv \sqrt{\frac{C^{\pi}_{4}(\vec{p},t_1,t_2,t_f)}{C^{\pi}_{2}(\vec{0},t_f)}}&=\frac{m_{\pi}+E_{\pi}}{2\sqrt{m_{\pi}E_{\pi}}}f_{\pi\pi}(Q^2)e^{-(E_{\pi}-m_{\pi})(t_2-t_1)/2},
\end{align}
where $f_{\pi\pi}(Q^2)$ is the only free parameter.  The fit yields $\chi^2$/d.o.f. = 0.8, demonstrating excellent agreement with the data. The right panel of Fig.~\ref{fig:PionFF_OriginDate} provides further validation through the rescaled ratio ${\cal R}_V^{\rm 4pt,rescaled} = {\cal R}_V^{\rm 4pt}e^{(E_{\pi}-m_{\pi})(t_2-t_1)/2}$, which shows perfect consistency with the green fit band. The observed suppression of excited-state contamination likely relates to the exclusion of $t_1>t_2$ contribution, a feature that merits dedicated future study.

\begin{table}[htbp]
  \centering
  \caption{The renormalized $f_{\pi\pi}(Q^2)$ from three-point and four-point functions.}
    \begin{tabular}{c|cccc|cccccc|c}
     & \multicolumn{4}{c|}{$Q^2({\rm GeV}^2)$} 
     & \multicolumn{6}{c|}{Z-expansion} & \multirow{2}{*}{$\langle r^2_{\pi}\rangle~({\rm fm}^2)$ }\\
                           & 0            & 0.163      & 0.269     & 0.354    & $a_0$       & $a_1$      & $a_2$    & $\sum_k a_k$ & $\sum_k ka_k$ & $\chi^2$/d.o.f. &\\
    \hline
    $f_{\pi\pi}$ from 3-pf & 1.00000(42) & 0.7652(67) & 0.673(10) & 0.623(57) &  0.7831(71) & -2.15(12)  & 3.4(1.9) &  2.0(2.0)    &  4.7(4.0)     & 0.01   &  0.332(29)\\
    \hline
    $f_{\pi\pi}$ from 4-pf &1.00001(38) &  0.7623(99) & 0.670(06) & 0.605(23) & 0.7823(96)  & -2.190(79)  & 3.0(1.9) & 1.6(2.0)    & 3.8(3.9)      & 0.15    & 0.329(33) \\  
    \end{tabular}%
  \label{Tab.C4C3PionFF}%
\end{table}%

Using a similar fit strategy, we obtain $f_{\pi\pi}(Q^2)$ for 4 $Q^2$ and list the results in Table~\ref{Tab.C4C3PionFF}, and the values from 4-pf perfectly agree with those from 3-pf, and provide a precise verification of the correctness of our 4-pf calculation. Then the charge radius can be obtained through the z-expansion fit~\cite{zExpansion},
\begin{align}
    f_{\pi\pi}(Q^2) & = \sum_{k=1}^{k_{\rm max}} a_k z^k,\ 
    z(t,t_{\rm cut},t_0) = \frac{ \sqrt{t_{\rm cut}-t} -\sqrt{t_{\rm cut} -t_0}}{ \sqrt{t_{\rm cut}-t} +\sqrt{t_{\rm cut} -t_0}},
\end{align}
where $t=-Q^2$, $t_{\rm cut}=4 m_\pi^2$, $t_0=t_{\rm cut}(1-\sqrt{1+Q_{\rm max}^2/t_{\rm cut}})$ and $Q_{\rm max}^2=0.354$~GeV$^2$. As shown in Table~\ref{Tab.C4C3PionFF}, both the constraints $\sum_ka_k=0$ and $\sum_kka_k=0$ are satisfied within the uncertainties.

\subsection{High-precision $g_A$ from 3-pf}\label{sec:gAFit}

\red{
\subsubsection{Perturbative result for the matching coefficient of $\JA$ from the $\RIMOM$ to the $\overline{\mathrm{MS}}$ scheme}\label{sec:Rexpression}

As a consequence of the famous Adler-Bell-Jackiw (ABJ) anomaly~\cite{Adler:1969gk,Adler:1969er,Bell:1969ts}, the singlet axial-vector current operator $\JA \equiv \Js = \sum_{f} \bar{\psi}_f \, \gamma^{\mu}\gamma_5 \,  \psi_f$ requires a non-trivial renormalization~\cite{Adler:1969gk}.
Converting the renormalization of $\JA$ from the intermediate $\RIMOM$ scheme used in Lattice calculation to the $\overline{\mathrm{MS}}$ scheme typically employed in perturbative calculations in Dimensional Regularization (DR) requires a finite renormalization (matching) coefficient $R_{A}$. 
This section details the perturbative calculation of $R_{A}$ in continuum QCD with DR. 
As in the $\MSbar$ renormalization of $\JA$~\cite{Chanowitz:1979zu,Gottlieb:1979ix,Larin:1991tj,Larin:1993tq},  the renormalization of $J_A^\mu$ in $\mathrm{RI/MOM}$ scheme takes a multiplicative form when the amputated Green function of $\JA$ between a pair of quark states is concerned.
Owing to the Ward-Takahashi identity for the conserved vector current $J_V^{\mu}$, the quark field renormalization constant defined in $\psi_B = \sqrt{Z_{q}}\, \psi_R$ in $\mathrm{RI/MOM}$ can be conveniently determined in perturbative QCD with DR via computing the amputated Green function $\Gamma_V$ of $J_V^{\mu}$ between a pair of quark states with the same external momentum $p$ (with the two external spinors stripped off), denoted by $\big(\Gamma_V\big)^{\mu}_{\alpha\, \beta}$. More specifically, the following $\mathrm{RI/MOM}$ renormalization condition is imposed: 
\begin{equation} \label{eq:fieldRCMOM}
Z_{V} \, Z_{q}\, \big(\Gamma_V\big)^{\mu}_{\alpha\, \beta}\, \hat{\mathcal{P}}^{\alpha\, \beta}_{\mu} \Big|_{p^2 = \mu_s^2} = \big[\Gamma^{\mathrm{tree}}_V\big]^{\mu}_{\alpha\, \beta}\, \hat{\mathcal{P}}^{\alpha\, \beta}_{\mu} \Big|_{p^2 = \mu_s^2} \, 
\end{equation}
where $Z_V=1$ in the continuum, and the pairs of repeated Lorentz/Dirac indices are contracted by default.
The $\RIMOM$ subtraction scale is set at $\mu_s^2 = p^2$ (positive-valued in Euclidean space).  
Depending on the choice of the Lorentz-structure projector, such as $\hat{\mathcal{P}}_{\alpha\, \beta}^{\mu}  = \gamma^{\mu}_{\alpha\, \beta}$ (in the standard $\RIMOM$) or the transverse one $\gamma^{\mu}_{\alpha\, \beta} -  \slashed{p}_{\alpha\, \beta}\,p^{\mu}/p^2$ (in $\RIpMOM$), or even $p^{\mu} \slashed{p}_{\alpha\, \beta}$ (i.e.~projecting out the component along $p$), different expressions for $Z_{q}$ will follow. 
However, they all share the same $\MSbar$ factor, namely the $\MSbar$ wavefunction renormalization constant of the external quark state is independent of the projector $\hat{\mathcal{P}}$ in use.

To determine the $\RIMOM$ operator renormalization of $\JA$, we compute its amputated Green function $\Gamma_A$  between a pair of quark fields with the same external momentum $p$ 
and require 
\begin{equation}\label{eq:ACRCMOM}
Z_{A} \, Z_{q}\, \big(\Gamma_A\big)^{\mu}_{\alpha\, \beta}\, \hat{\mathcal{P}}^{\alpha\, \beta}_{A,\mu} \Big|_{p^2 = \mu_s^2} = \big[\Gamma^{\mathrm{tree}}_A\big]^{\mu}_{\alpha\, \beta}\, \hat{\mathcal{P}}^{\alpha\, \beta}_{A, \mu} \Big|_{p^2 = \mu_s^2}
\end{equation}
where $\hat{\mathcal{P}}^{\alpha\, \beta}_{A, \mu}  = \big(\gamma^{\mu}\,\gamma_5\big)_{\alpha\, \beta}$ and $Z_{q}$ is determined in Eq.~\eqref{eq:fieldRCMOM} using $\hat{\mathcal{P}}_{\alpha\, \beta}^{\mu}  = \gamma^{\mu}_{\alpha\, \beta}$ in the standard $\RIMOM$.
Unlike $Z_V$ in Eq.~\eqref{eq:fieldRCMOM}, $Z_{A}$ differs from unity beginning at two-loop order due to the ABJ anomaly~\cite{Adler:1969gk,Adler:1969er,Bell:1969ts}.
The finite matching coefficient $R_{A}$ for $J_A^\mu$ from the intermediate $\RIMOM$ scheme to the $\overline{\mathrm{MS}}$ scheme can be defined in terms of the operator renormalization constants as follows:
\begin{eqnarray}\label{eq:ACRCmatch}
R_{A} &\equiv& \frac{\overline{Z}_A}{Z_{A}} \Big|_{\epsilon \rightarrow 0}\, 
= \frac{\overline{Z}_A \, Z_{q}\, \big(\Gamma_A\big)^{\mu}_{\alpha\, \beta}\, \hat{\mathcal{P}}^{\alpha\, \beta}_{A,\mu} }{Z_{A} \, Z_{q}\, \big(\Gamma_A\big)^{\mu}_{\alpha\, \beta}\, \hat{\mathcal{P}}^{\alpha\, \beta}_{A,\mu}}  \Big|_{p^2 = \mu_s^2} \nonumber \\
&=&  \frac{\overline{Z}_A \, Z_{q}\, \big(\Gamma_A\big)^{\mu}_{\alpha\, \beta}\, \hat{\mathcal{P}}^{\alpha\, \beta}_{A,\mu} }{\big[\Gamma^{\mathrm{tree}}_A\big]^{\mu}_{\alpha\, \beta}\, \hat{\mathcal{P}}^{\alpha\, \beta}_{A, \mu}}  \Big|_{p^2 = \mu_s^2}\,, 
\end{eqnarray}
where $\overline{Z}_A$ is the $\MSbar$-renormalization constant, and  Eq.~\eqref{eq:ACRCMOM} is employed to arrive at the form given in the last line.
Therefore, $R_{A}$ is essentially the properly normalized matrix element of the $\MSbar$-renormalized $\JA$ but with $\RIMOM$-renormalized external quark fields or states, by the virtue of $\RIMOM$ renormalization condition Eq.~\eqref{eq:ACRCMOM}.
According to the idea of $\RIMOM$, the $R_{A}$ determined above using DR can be employed to obtain the value of $\overline{Z}_A^{\mathrm{lat}} \equiv R_{A}\, Z_{A}^{\mathrm{lat}}$ for $J_A^\mu$ but with all UV divergences regularized in $Z_{A}^{\mathrm{lat}}$ computed in the $\RIMOM$ implementation in Lattice QCD (which can be applied directly to the bare operator matrix elements evaluated in Lattice QCD to obtain the $\MSbar$-renormalized results).
Before we proceed further, let us emphasize that the $\MSbar$-renormalization constant $\overline{Z}_A$ is known to be independent of QCD gauge-fixing~\cite{Breitenlohner:1983pi,Larin:1993tq,Luscher:2021bog}, verified explicitly up to four, and partially five, loops~\cite{Chen:2021gxv,Chen:2022lun}, while the $Z_{A}$, as well as its Lattice counterpart $Z_{A}^{\mathrm{lat}}$, defined in $\RIMOM$ does. 
Furthermore, unlike the $\RIMOM$ $Z_A$ which depends on the particular choice of projector $\hat{\mathcal{P}}$ in  Eq.~\eqref{eq:ACRCMOM}, $\overline{Z}_A$ is independent of the $\hat{\mathcal{P}}$ in use, which we have verified explicitly by considering $\RIpMOM$ projector and by projecting onto the component along $p^{\mu}$. 
In other words, both the gauge dependence and $\RIMOM$-projector dependence of the matching factor $R_{A}$ arise solely from those of $Z_A$ defined in $\RIMOM$, and will drop in the $\overline{Z}_A$ reconstructed using the two.

Due to the notorious issue in treating $\gamma_5$ in DR, extra care is required in the calculation of the dimensionally-regularized bare matrix elements of $\JA$, and the subsequent extraction of its renormalization constant involved in  Eq.~\eqref{eq:ACRCmatch}, especially so in the present case due to the involvement of ABJ anomaly~\cite{Adler:1969gk,Adler:1969er,Bell:1969ts}. 
In view of the applications related to the study of the spin structures and axial-vector charges of nucleons, we choose a (modified) $\MSbar$ prescription where the gauge invariance and ABJ equation~\cite{Adler:1969gk,Adler:1969er} are preserved (hence without chiral symmetry for the anomalous $J_{A,s}^{\mu}$). 
So far, there is no practical prescription known to the authors to apply consistently an anticommuting $\gamma_5$ in DR for computing loop amplitudes with (closed) fermion chains with an odd number of $\gamma_5$, especially when the amplitudes are anomalous~\cite{Chen:2023lus,Chen:2024hlv}.
On the other hand, the Larin prescription~\cite{Larin:1993tq} for $\gamma_5$, in which $\gamma_5$ is treated as non-anticommuting in DR~\cite{tHooft:1972tcz,Breitenlohner:1977hr,Breitenlohner:1975hg,Breitenlohner:1976te} while the spacetime-metric tensors resulting from contracting a pair of $\epsilon^{\mu\nu\rho\sigma}$ are evaluated in D-dimensions, is particularly suited for dealing with the QCD corrections at hand: 
both gauge invariance and the ABJ equation (i.e.~the anomalous axial Ward identity) are maintained and it is technically convenient to apply at high loop orders in QCD.
In particular, both the pure $\MSbar$ factor and the additional finite renormalization constant determined in this prescription  --- with the latter needed to maintain the anomalous axial Ward identity ---  for $J_{A,s}^{\mu}$ are independent of QCD gauge fixing~\cite{Breitenlohner:1983pi,Larin:1993tq,Luscher:2021bog,Ahmed:2021spj,Chen:2021gxv,Chen:2022lun}.
~\\

Before embarking on the actual perturbative-QCD calculations, let us first derive an alternative formula for $R_{A}$ (see Eq.~\eqref{eq:ACRCmatch_rw} below) based on the anomalous axial Ward identity and the $\RIMOM$ renormalization condition, that would allow us to more efficiently and conveniently extract its value from high-order loop corrections.  
The all-order axial-anomaly equation~\cite{Adler:1969gk,Adler:1969er}, which in terms of properly ($\MSbar$) renormalized local composite operators (indicated by the subscript $R$) reads 
\begin{eqnarray} 
\label{eq:ABJanomalyEQ}
\big[\partial_{\mu} \Js \big]_{R} = a_s\, n_f\, \TR \,  \big[F \tilde{F} \big]_{R}\,,
\end{eqnarray}
where $\TR=1/2$, $F \tilde{F} \equiv  - \epsilon^{\mu\nu\rho\sigma} F^a_{\mu\nu} F^a_{\rho\sigma} = \epsilon_{\mu\nu\rho\sigma} F^a_{\mu\nu} F^a_{\rho\sigma}$ denotes the contraction of the field strength tensor $F^a_{\mu\nu} = \partial_{\mu} A_{\nu}^{a} - \partial_{\nu} A_{\mu}^{a} + g_s \,  f^{abc} A_{\mu}^{b} A_{\nu}^{c}$ of the gluon field $A_\mu^a$ with its \textit{dual} form. (We use the convention $\epsilon^{0123} = -\epsilon_{0123} = +1$.)
We use the shorthand notation $a_s \equiv \frac{\alpha_s}{4 \pi} = \frac{g_s^2}{16 \pi^2}$ for the QCD coupling, and $f^{abc}$ denotes the structure constants of the non-Abelian color group of QCD. 
It is well-known that the axial anomaly operator assumes an equivalent form in terms of the divergence of the Chern-Simons current, also known as the topological current $K^{\mu}$, 
\begin{eqnarray} 
\label{eq:Kcurrent}
F \tilde{F}  &=& \partial_{\mu} K^{\mu} = \partial_{\mu} \Big(-4 \,\epsilon^{\mu\nu\rho\sigma} \,\big(A_{\nu}^{a} \partial_{\rho} A_{\sigma}^{a} \,+\, g_s\,\frac{1}{3} f^{abc} A_{\nu}^{a} A_{\rho}^{b} A_{\sigma}^{c} \big) \Big)\,,
\end{eqnarray}
where the current $K^{\mu} \equiv -4 \, \epsilon^{\mu\nu\rho\sigma} \,\left(A_{\nu}^{a} \partial_{\rho} A_{\sigma}^{a} \,+\, g_s\,\frac{1}{3} f^{abc} A_{\nu}^{a} A_{\rho}^{b} A_{\sigma}^{c} \right)$ is not exactly gauge-invariant, however, its infinitesimal local-gauge variation takes a linear form in the total derivatives of the gauge field~\cite{Espriu:1982bw,Breitenlohner:1983pi,Lam:1982ii,Breitenlohner:1983pi,Altarelli:1988nr,Cheng:1996jr,Luscher:2021bog,Pang:2024sdl}.
This alternative equivalent form is particularly convenient both for computing the matrix elements of $F \tilde{F}$ and for the subsequent determination of its renormalization constant to be shown below.
Given the operator-level axial-anomaly equation  Eq.~\eqref{eq:ABJanomalyEQ} in the canonical quantization formalism, one can derive the following anomalous Ward-Takahashi identity~\cite{Adler:1969gk} in momentum space: %
\begin{eqnarray} 
\label{eq:WTI_s}
q_{\mu}\, \Gamma_{A,s}^{\mu}(p',p) = a_s\, n_f\, \TR \, q_{\mu}\,  \Gamma_{K}^{\mu}(p',p)  \,+\, \gamma_5\, \hat{S}^{-1}(p) \,+\, \hat{S}^{-1}(p')\, \gamma_5\, ,
\end{eqnarray} %
with $q_{\mu}$ the momentum flow through the singlet axial current operator $\Js$ defined with massless quark fields. 
$\hat{S}(p)$ in the contact terms on the r.h.s. denotes the full propagator of a massless quark with an off-shell momentum $p$ up to an overall $i$ factor. 
$\Gamma_{K}^{\mu}(p',p)$ denotes the amputated Green function of the topological current $K^{\mu}$ between a pair of external quark fields.  
From Eq.~\eqref{eq:WTI_s}, 
we can derive the following alternative expression for the finite matching coefficient:  
\begin{eqnarray}\label{eq:ACRCmatch_rw}
R_{A} &\equiv& \frac{\overline{Z}_A}{Z_{A}} \Big|_{\epsilon \rightarrow 0}\, 
= \frac{ Z_{q}\, \big(\overline{Z}_A \Gamma_A\big)^{\mu}_{\alpha\, \beta}\, \hat{\mathcal{P}}^{\alpha\, \beta}_{A,\mu} }{\big[\Gamma^{\mathrm{tree}}_A\big]^{\mu}_{\alpha\, \beta}\, \hat{\mathcal{P}}^{\alpha\, \beta}_{A, \mu}}  \Big|_{p^2 = \mu_s^2}  \nonumber \\
&=& \frac{Z_{q}\, \big(\overline{Z}_{A,ns} \, \mathrm{\Gamma}_{A,ns} + a_s\, n_f\, \TR \,  \big[\Gamma_{K}\big]_R \big)^{\mu}_{\alpha\, \beta}\, \hat{\mathcal{P}}^{\alpha\, \beta}_{A,\mu} }{\big[\Gamma^{\mathrm{tree}}_A\big]^{\mu}_{\alpha\, \beta}\, \hat{\mathcal{P}}^{\alpha\, \beta}_{A, \mu}}  \Big|_{p^2 = \mu_s^2}  \nonumber \\
&=& 1 \,+\, \frac{  \big( a_s\, n_f\, \TR \, Z_{q}\, \big[\Gamma_{K}\big]_R \big)^{\mu}_{\alpha\, \beta}\, \hat{\mathcal{P}}^{\alpha\, \beta}_{A,\mu} }{\big[\Gamma^{\mathrm{tree}}_A\big]^{\mu}_{\alpha\, \beta}\, \hat{\mathcal{P}}^{\alpha\, \beta}_{A, \mu}}  \Big|_{p^2 = \mu_s^2}\,, 
\end{eqnarray}
where, for the final equality, we have employed the chiral-symmetry property of the non-singlet axial-current operator $\Jns$ (with massless quark fields), $Z_{A,ns} = Z_{V} = 1$, and consequently $Z_{q}\, \big(\overline{Z}_{A,ns} \, \Gamma_{A, ns}\big)^{\mu}_{\alpha\, \beta}\, \hat{\mathcal{P}}^{\alpha\, \beta}_{A,\mu} \Big|_{p^2 = \mu_s^2} = \big[\Gamma^{\mathrm{tree}}_A\big]^{\mu}_{\alpha\, \beta}\, \hat{\mathcal{P}}^{\alpha\, \beta}_{A, \mu} \Big|_{p^2 = \mu_s^2}$ holding in $\RIMOM$, just as in  Eq.~\eqref{eq:fieldRCMOM} for the conserved vector-current case.
We refrain here from repeating the $\MSbar$ renormalization in $\big[\Gamma_{K}\big]_R$ discussed in detail in Refs. \cite{Breitenlohner:1983pi,Larin:1993tq,Ahmed:2021spj,Luscher:2021bog}.
With the aid of the formula~\eqref{eq:ACRCmatch_rw} for $R_{A}$, we are then able to conveniently derive its perturbative result at $\mathcal{O}(\alpha_s^{N+1})$ based solely on the $N$-loop computation of $ \big[\Gamma_{K}\big]_R$ in DR. 
(A direct determination $R_{A}$ at $\mathcal{O}(\alpha_s^{N+1})$ using  Eq.~\eqref{eq:ACRCmatch} will necessarily entail a $N+1$-loop computation.) 
We note that a similar shortcut formula can also be derived and applied to the finite matching coefficient $R_{22}$ defined in Ref.~\cite{Zhao:2025piw}. 

With this set-up, we have derived the explicit perturbative expressions for the bare matrix elements involved in  Eq.~\eqref{eq:ACRCmatch_rw} up to three loops, truncated to high orders in $\epsilon$ sufficient for obtaining their renormalized counterparts to $\mathcal{O}(\epsilon^0)$. 
They are subsequently employed to extract the perturbative result for the finite $R_{A}$ according to  Eq.~\eqref{eq:ACRCmatch}.  
For reference, we document below the perturbative expression for $R_{A}$ in Landau gauge. 
(The gauge dependence in $R_{A}$ arises from that of $Z_{A}$ in  Eq.~\eqref{eq:ACRCmatch}, and will cancel against that of $Z_{A}^{\mathrm{lat}}$ in the $\RIMOM$-renormalized matrix element.)
The perturbative corrections to $R_{A}$ start only at two loops and read:   
\begin{eqnarray} \label{eq:RAexp}
R_{A} &=& 1 \,+\, 
a_s^2 \, 
\Big(
C_F n_f \, \left(-6 L_{\mu }-9\right)
\Big) \nonumber\\
&+&
a_s^3 \, 
\Big(
C_A C_F n_f \, \left(54 \zeta _3-22 L_{\mu }^2-\frac{340 L_{\mu }}{3}-\frac{3461}{18}\right)
\nonumber\\ &+&
C_F^2 n_f \, \left(-48 \zeta _3+18 L_{\mu }+\frac{83}{2}\right)
\,+\, 
C_F n_f^2 \, \left(4 L_{\mu }^2+\frac{40 L_{\mu }}{3}+\frac{211}{9}\right)
\Big) 
\,+\, \mathcal{O}(a_s^4)\,,
\end{eqnarray}
where the shorthand notations $a_s \equiv \frac{\alpha_s}{4\, \pi}$ and $L_{\mu } \equiv \ln \big(\frac{{\mu}^2}{\mu_s^2} \big)$ are introduced.
The definitions of the quadratic Casimir color constants involved above are as usual: $C_A = N_c \,, \, C_F = (N_c^2 - 1)/(2 N_c) \,$ with $N_c =3$ in QCD and we have set the color-trace normalization factor to its value ${1}/{2}$.
($\zeta_n$ is the usual Riemann $\zeta(n)$ function.) 
Note that this expression differs from that of $R_{22}$ reported in Ref.~\cite{Zhao:2025piw}, due to the choice of Lorentz-structure projectors used in the standard $\RIMOM$ renormalization condition  Eq.~\eqref{eq:fieldRCMOM} and \eqref{eq:ACRCMOM}.
However, the value of the $\MSbar$ operator renormalization constant $\overline{Z}_A^{\mathrm{lat}} \equiv R_{A}\, Z_{A}^{\mathrm{lat}}$ --- reconstructed from the $\RIMOM$ operator renormalization constant and the corresponding consistently determined matching factor --- shall remain the same, i.e.~independent of the particular choice of the projectors in the intermediate $\RIMOM$ scheme in use. 
Moreover, the differences in the wavefunction renormalization constants $Z_q$ for quark fields in different $\RIMOM$ schemes shall cancel in properly defined ratios, such as the ratio between the matrix elements of the $\MSbar$-renormalized $\JA$ and $J_V^{\mu}$ (over the same external quark states), and also in physical observables with properly normalized external states.

The dependence of $R_{A}$ on ${\mu}^2$ exhibited in the above perturbative expression is governed by its RG equation 
\begin{equation} \label{eq:RAsRGE}
{\mu}^2\, \frac{\mathrm{d}}{\mathrm{d}\, {\mu}^2 }\, R_{A} = \gamma_{s}\, R_{A}  \,,
\end{equation} 
with the anomalous dimension $\gamma_{s}$ fully determined by the (proper or modified) $\MSbar$ renormalization constant $\overline{Z}_A$ in Larin’s prescription~\cite{Larin:1993tq}.
Up to four-loop order, it reads  
\begin{eqnarray} \label{eq:RAsAMD}
\gamma_{s} &=& 
a_s^2 \,
 \left(-8 n_f\right)
\,+\, 
a_s^3 \, \Big(
n_f^2 \, \frac{16}{9}\,-\,  n_f \, \frac{472}{3}
\Big)
\,+\, 
a_s^4 \, \Big(
n_f^3 \, \frac{104}{9} \,+\, n_f^2 \, \big(320 \zeta _3+\frac{728}{9}\big)
\,-\, 2134 n_f \,
\Big)\, + \mathcal{O}(a_s^5)\,.
\end{eqnarray}
The solution to the RG equation  Eq.~\eqref{eq:RAsRGE} of $R_A$ can be parameterized as 
\begin{equation} \label{eq:RAsRGsolu}
R_{A}(\mu_t/\mu_{\mathrm{RI}}) =  R_{A}(\mu_i/\mu_{\mathrm{RI}})\, \mathrm{Exp}\Big(\int_{\alpha_s(\mu_i)}^{\alpha_s(\mu_t)}\, \frac{\gamma_s}{\beta\, \alpha_s}\, \mathrm{d}\, \alpha_s \Big)
\end{equation} 
where the $\beta$ function of QCD is defined through $\frac{\mathrm{d}}{\mathrm{d}\, \ln {\mu}^2 }\, \alpha_s(\mu) = \beta\, \alpha_s(\mu)$. 
(Note that the RG-evolution factor in Eq.~\eqref{eq:RAsRGsolu} remains invariant under the redefinition $a_s = \alpha_s/(4\pi)$.)
The quantity $R_{A}(\mu_i/\mu_{\mathrm{RI}})$ may be approximated by evaluating the fixed-order expression Eq.~\eqref{eq:RAexp} with the common choice $\mu_i = \mu_{\mathrm{RI}}$,~i.e.~$L_{\mu } = 0$, 
or alternatively but slightly better $\mu_i = e^{-3/4}\, \mu_{\mathrm{RI}}$ (motivated by the $L_{\mu }$-structure of $R_{A}$) for sufficiently large $\mu_{\mathrm{RI}}$. }

\subsubsection{Parameterization}

By inserting the complete sets of states, the ratio ${\cal R}_H^{{\cal O}}$ can be parameterized as:
\begin{align}
        {\cal R}_H^{{\cal O}}(t_f,t;\bar{N}_{\rm e}) &=\frac{\int \dd^3 x \dd^3 y\dd^3 z\langle H(\vec{x},t_f){\cal O}(\vec{y},t)H^{\dagger}(\vec{z},0)\rangle}{\int \dd^3 x\dd^3 z\langle H(\vec{x},t_f)H^{\dagger}(\vec{z},0)\rangle}\nonumber\\
        &= \frac{\sum_{i}c_{ii}d_i^2(\bar{N}_{\rm e}) e^{-\Delta_{i} t_f} +\sum_{i<j}c_{ij}d_i(\bar{N}_{\rm e})d_j(\bar{N}_{\rm e}) (e^{-\Delta_{i} (t_f-t)-\Delta_{j} t} +e^{-\Delta_{j}(t_f-t)-\Delta_{i} t})}{\sum_i d_i^2(\bar{N}_{\rm e})e^{-\Delta_i t_f}}\label{eq:3pt_fit}
\end{align}
where $c_{ij}=\frac{1}{2\sqrt{m_im_j}}\langle H_i|{\cal O}|H_j\rangle$  is the normalized matrix element between the $i$-th state $H_i$ and $j$-th state $H_j$ (thus $c_{ii}=\langle {\cal O}\rangle_{H_i}\equiv \langle H_i|{\cal O}|H_i\rangle/(2m_i)$), $d_i=\sqrt{\frac{m_0}{m_i}}\frac{\langle 0|H|H_i\rangle}{\langle 0|H|H_0\rangle}$ is the rescaled weight of the $i$-th state in the two-point function which is parameterized as (with $d_0=1$)
\begin{align}
   C_2(t_f;\bar{N}_{\rm e})=Z(\bar{N}_{\rm e})e^{-m_0t_f}\times \left[\sum_i d_i^2(\bar{N}_{\rm e})e^{-\Delta_i t_f} \right],\label{eq:2pt_fit}
\end{align}
 $\Delta_{i}=m_i-m_0$ is the mass gap between the n-th state and ground state (thus $\Delta_0=0$). 

The Feynman-Hellmann inspired method provides an efficient approach for suppressing excited-state contamination and extracting ground-state matrix elements from data at relatively small source-sink separations $t_f$. First one defines the summed ratio with cutoff $t_c$,
\begin{align}
\sum_{t=t_c}^{t_f-t_c}{\cal R}_H^{\cal O}(t_f,t)
&=\frac{\sum_{i}c_{ii}d_i^2 e^{-\Delta_{i} t_f}(\tilde{t}_f+1-2\tilde{t}_c) +2\sum_{i<j}c_{ij}d_id_j e^{-\Delta_{i}t_f}(e^{-\Delta_{ij}t_c}-e^{-\Delta_{ij}(t_f-t_c+a)})/(1-e^{-\Delta_{ij}})}{\sum_i d_i^2e^{-\Delta_i t_f}}\nonumber\\
&=\big\{\langle {\cal O}\rangle_H(\tilde{t}_f+1-2\tilde{t}_c)+2\sum_{i=1}c_{0i}d_ie^{-\Delta_{i}t_c}/(1-e^{-\Delta_{i}a})+\sum_{i=1}e^{-\Delta_{i} t_f}[c_{ii}d_i^2(\tilde{t}_f+1-2\tilde{t}_c)\nonumber\\
&\quad \quad \quad +\sum_{j\neq i}c_{ij}d_id_je^{-\Delta_{ij}t_c}/(1-e^{-\Delta_{ij}a})]\big\}/(\sum_i d_i^2e^{-\Delta_i t_f}),
\end{align}
where $\Delta_{ij}=m_j-m_i=\Delta_j-\Delta_i$, $\tilde{t}=t/a$. Then
 the Feynman-Hellmann ratio~\cite{chang2018per,PhysRevD.96.014504} is defined as:
\begin{align}
{\cal R}_H^{{\cal O},{\rm FH}}(t)&\equiv \sum_{t=t_c}^{t_f+a-t_c}{\cal R}_H^{\cal O}(t_f+a,t)-\sum_{t=t_c}^{t_f-t_c}{\cal R}_H^{\cal O}(t_f,t)=\langle {\cal O}\rangle_H +\sum_{i=1}e^{-\Delta_{i} t_f}(\bar{c}_{i0}+\bar{c}_{i1}t_f)+{\cal O}(e^{-2\Delta_1t_f}),
\end{align}
with $\bar{c}_{i0(1)}$ representing redefined expansion coefficients.

Although the Feynman-Hellmann inspired method suppresses excited-state contamination from ${\cal O}(e^{-\Delta_1 t_f/2})$ in ${\cal R}_H^{\cal O}(t_f,t_f/2)$ to ${\cal O}(e^{-\Delta_1 t_f})$ in ${\cal R}_H^{{\cal O},{\rm FH}}(t_f)$, the statistical uncertainty at the same $t_f$ increases by $\sqrt{t_f}$, and information from different insertion times $t$ is lost. 

 \begin{figure}
      \includegraphics[width=0.49 \textwidth]{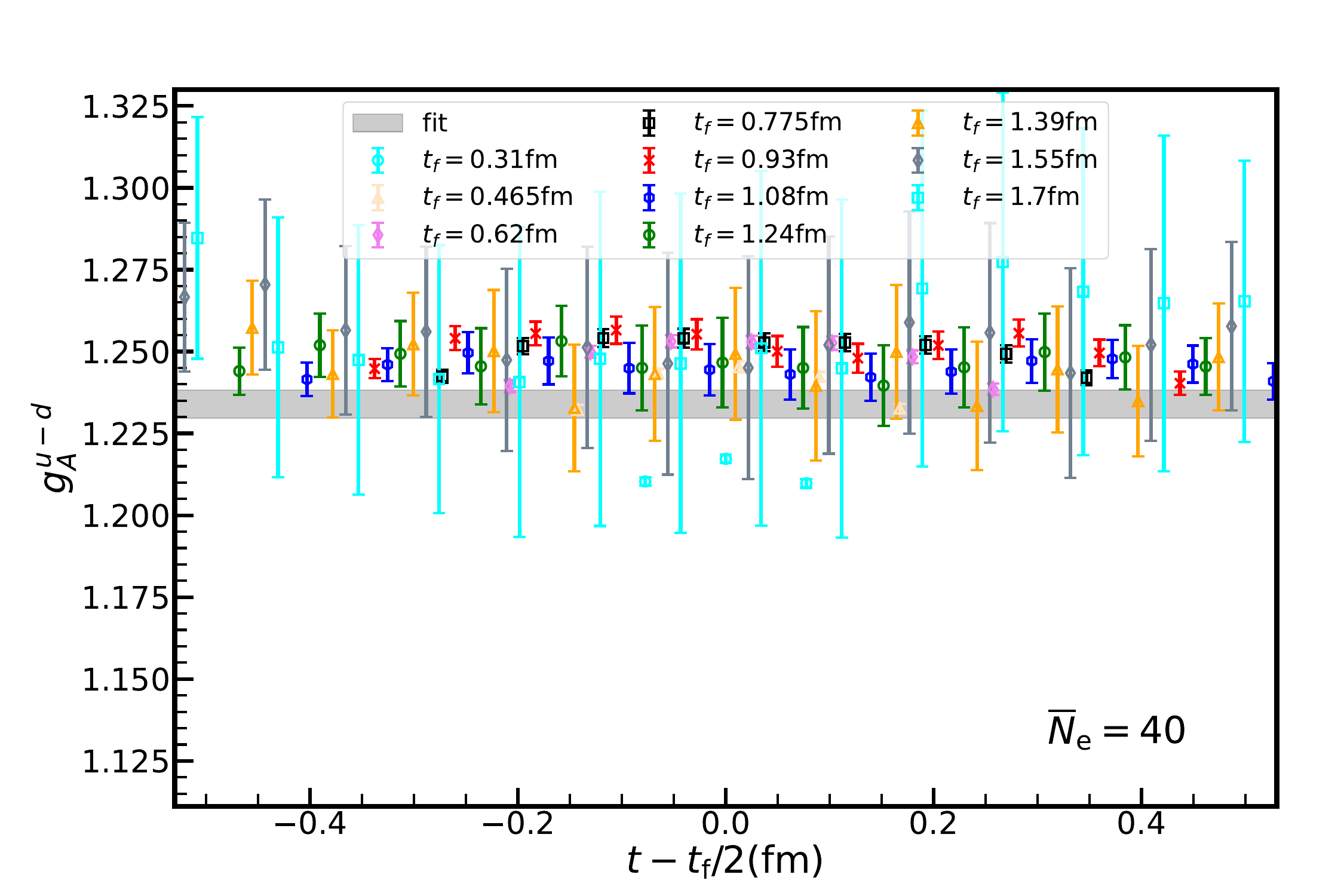}
      \includegraphics[width=0.49 \textwidth]{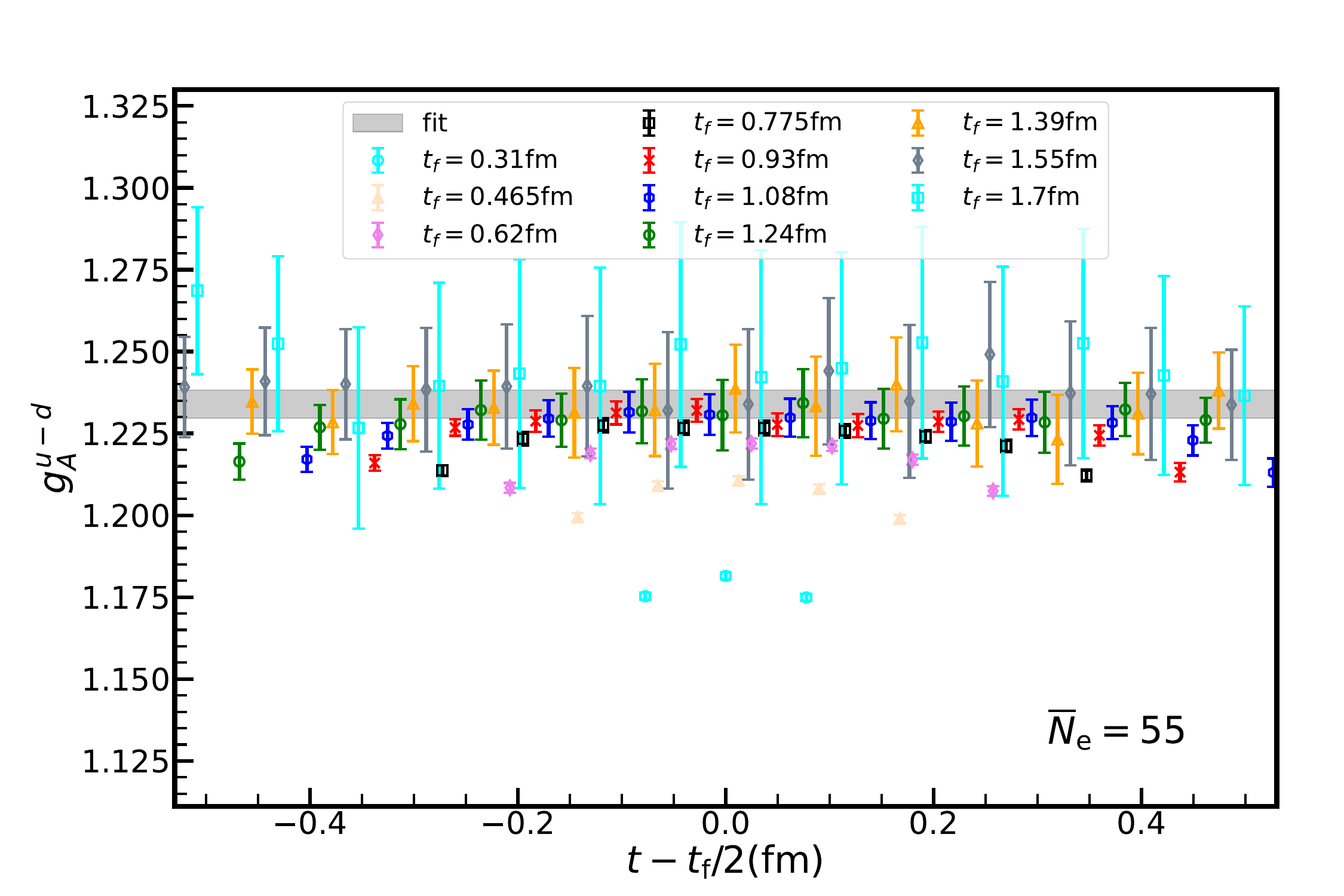}
      \includegraphics[width=0.49 \textwidth]{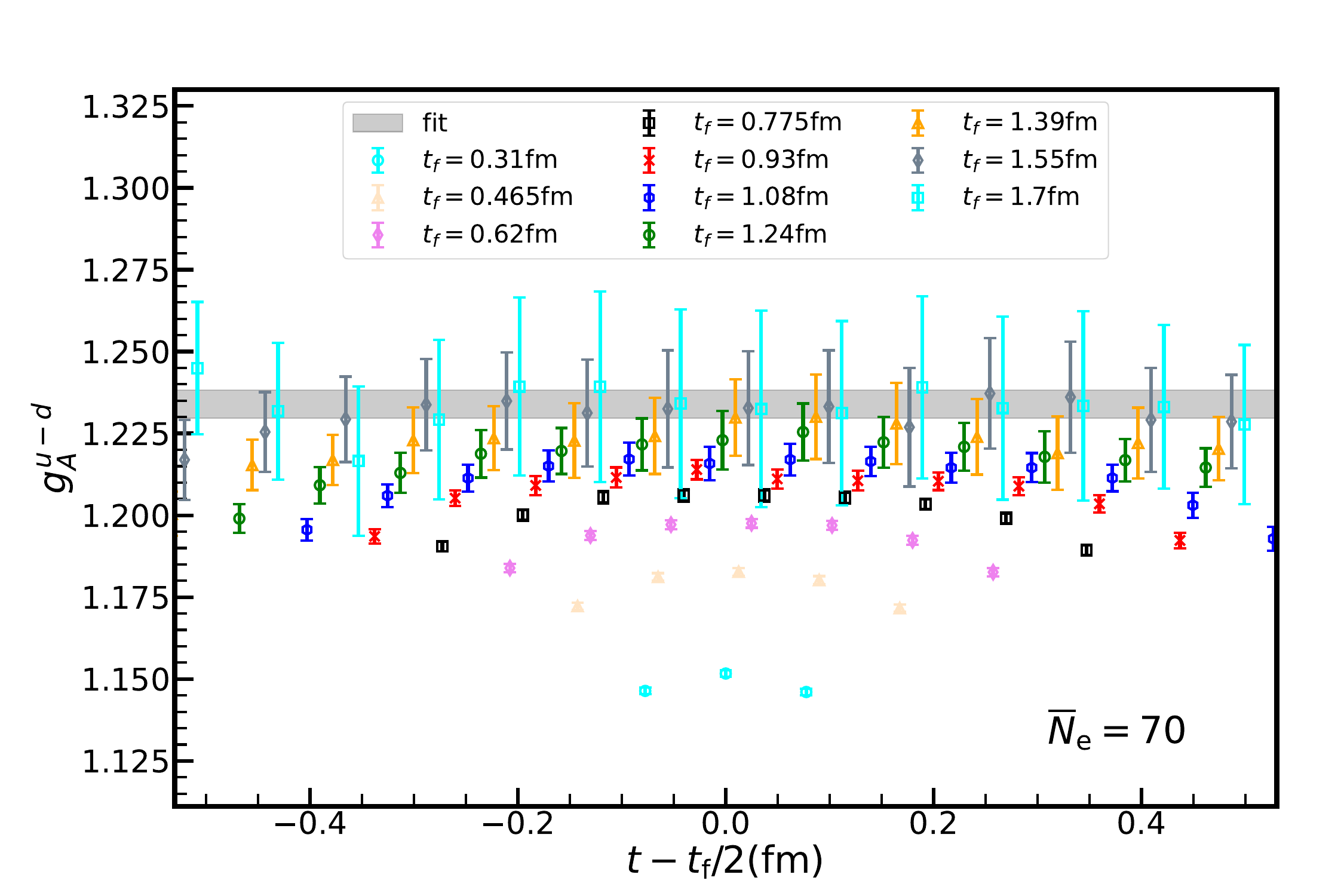}
      \includegraphics[width=0.49 \textwidth]{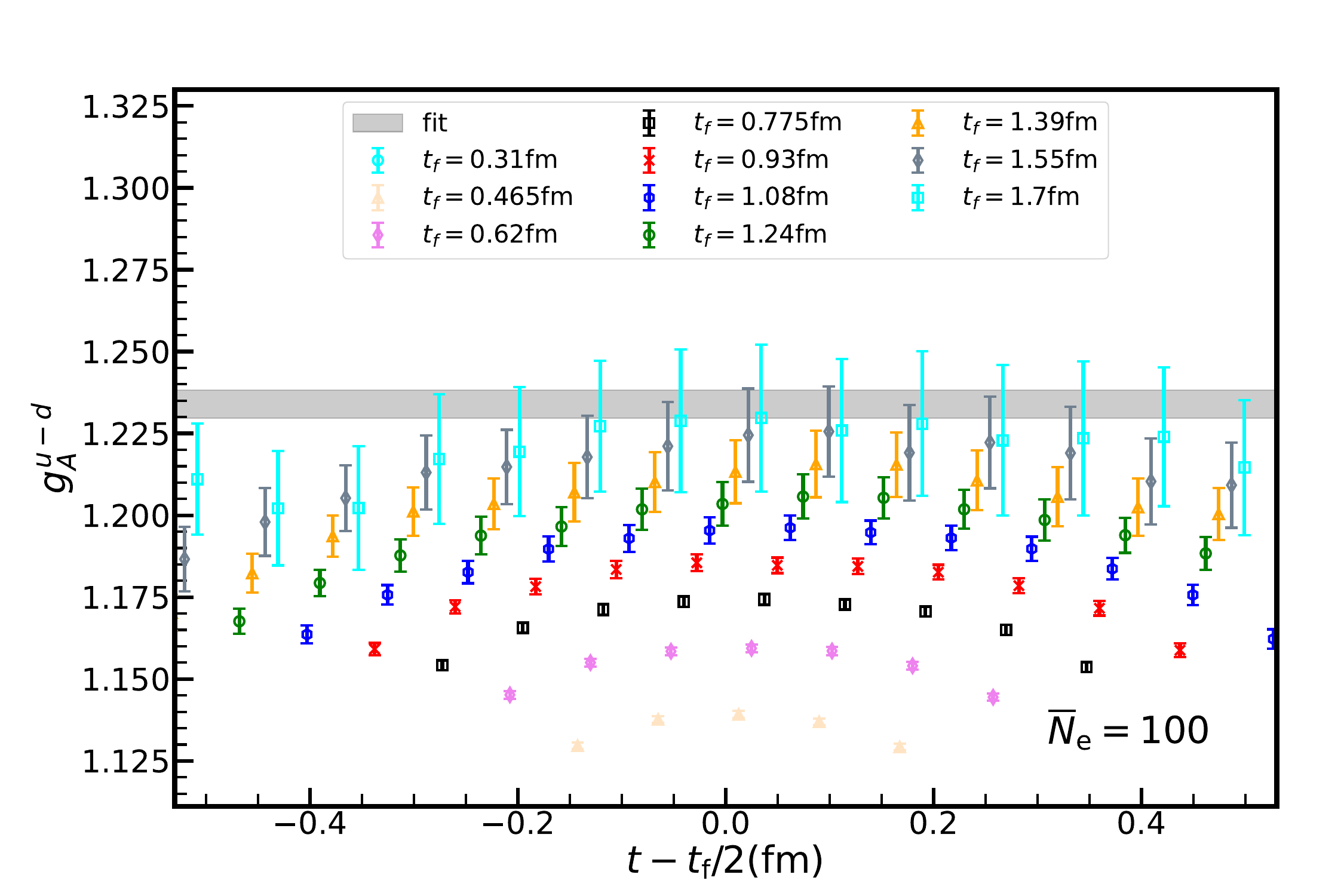}
      \caption{ 
            ${\cal R}_N^{A_{u-d}}(t_f,t)$ as a function of $t-t_f/2$ with different $t_f$, $\bar{N}_{\rm e}=40,55,70,100$ and the same $N_{\mathrm st}=200$ on the F48P30 ensemble. 
      }
      \label{Fig.gA_3pt}
  \end{figure}

\subsubsection{$g_A^{u,d,s}$ using 3-state fit}

\red{We compute \(g_A\) on two ensembles (F48P30 and F64P13) with the same lattice spacing \(a = 0.0775(5)\) fm but different pion masses (300 MeV and 134 MeV).} 

In Fig.~\ref{Fig.gA_3pt}, we present \( \mathcal{R}_N^{A_{u-d}}(t_f,t) \) \red{on the F48P30 ensemble} for various \( t \) and \( t_f \) values, using \( \bar{N}_{\rm e} = 40 \) (left upper panel), 55 (right upper panel), 80 (left lower panel), and 100 (right lower panel).
All panels display the same gray bands representing the ground-state matrix element {\color{black}\( g_A^{u-d} =1.2339(43) \)} obtained from the joint 3-state fit of 2-pf and 3-pf across all \( \bar{N}_e \) values using Eq.~(\ref{eq:3pt_fit}-\ref{eq:2pt_fit}), with $\chi^2/{\rm d.o.f} = 1.2$.
Due to the limited number of configurations (only 40) in this calculation, we account only for correlations between different $t$ at fixed $t_f$ and $\bar{N}_{\rm e}$, estimating the uncertainties of the fit parameters via jackknife resampling.
The other fit parameters for the excited state contamination will be discussed in detail later.

The results show that \( \mathcal{R}_N^{A_{u-d}}(t_f,t) \) with \( \bar{N}_{\rm e} = 55 \) agrees well with \( g_A^{u-d} \) for \( t_f \geq 0.9 \) fm and \( t \in (0.15~\mathrm{fm}, t_f - 0.15~\mathrm{fm}) \). 
As expected from the behavior of \( d_{1} \), larger \( \bar{N}_{\rm e} \) values delay the saturation to later \( t_f \).
The \( \bar{N}_{\rm e} = 40 \) result, however, requires caution: \( \mathcal{R}_N^{A_{u-d}}(t_f,t) \) exhibits an apparent plateau at \( t_f \sim 0.78 \) fm with \( \sim 1.245 \) (\( \sim 2\sigma \) above the gray band), but this trend reverses at larger $t_g$ (e.g., \( t_f \sim 1.1 \) fm) and ultimately converges to the band.
We will see that such a weird situation majorly comes from the competition of the contamination from the first and second excited states.

 \begin{figure}[!h]
      \includegraphics[width=0.49 \textwidth]{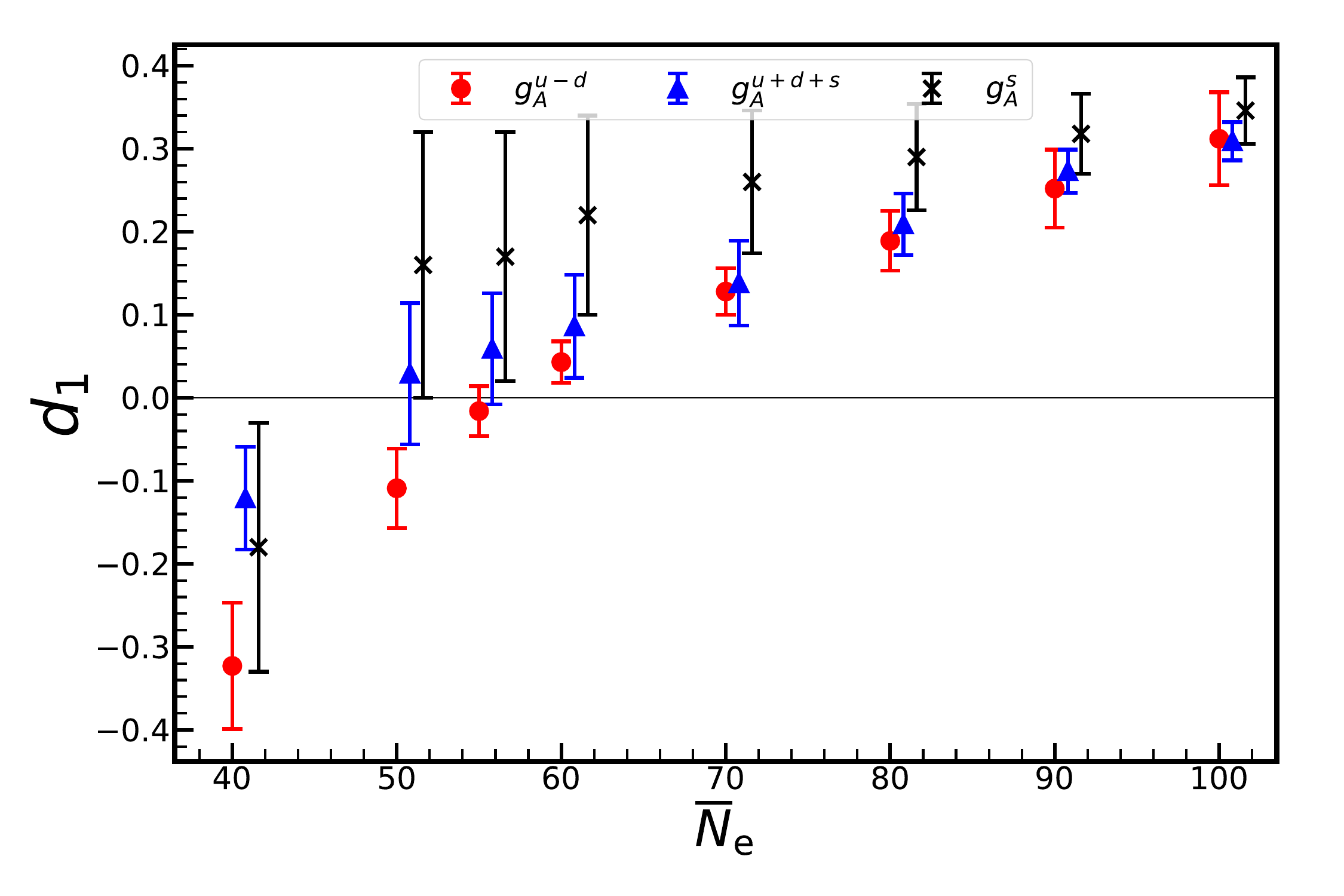}
      \includegraphics[width=0.49 \textwidth]{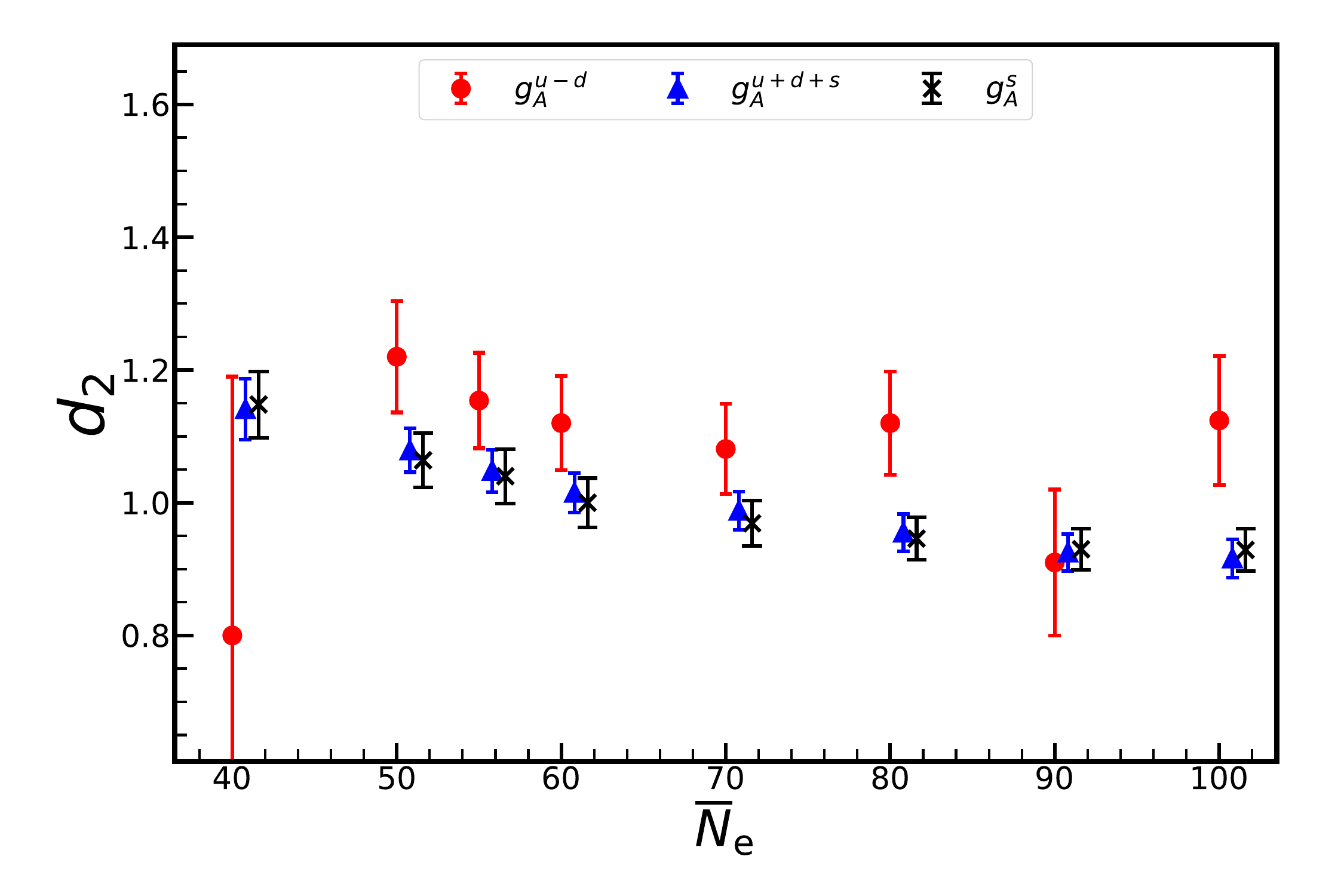}
      \caption{ 
             The rescaled weights of the effective first ($d_1$) and second ($d_2$) excited states obtained from joint 2-pf and 3-pf fit, \red{on the F48P30 ensemble}.
             The weight from iso-vector (u-d, red dot), flavor singlet (u+d+s, blue triangles), and strange (black crosses) channels agree with each other. 
      }
      \label{Fig:gA_ex_strengh}
  \end{figure}

In the left panel of Fig.~\ref{Fig:gA_ex_strengh}, we plot the rescaled weight \( d_{1}(\bar{N}_{\rm e}) \) of the first excited state (red dots), with an energy gap \( \Delta_1 = 0.484(20) \) GeV.
This value is consistent with the expected \(\pi N\) state energy (\(E_{\pi N} - m_N = 0.499(1)\) GeV).
The second excited state, however, has a significantly larger mass gap \( \Delta_2 \sim 1.5 \) GeV, suggesting an effective contribution from higher excitations.
Its rescaled weight \( d_{2}(\bar{N}_{\rm e}) \) is similarly displayed as red dots in the right panel of Fig.~\ref{Fig:gA_ex_strengh}.

Since the rescaled matrix elements \( c_{10} = -0.172(35) \) and \( c_{20} = -0.0208(17) \) share the same sign, the opposite signs of \( d_1 \) and \( d_2 \) at small \( \bar{N}_{\rm e} \) lead to a partial cancellation of the excited-state contamination at relatively small \( t_f \).
This results in a false plateau, as seen in the upper left panel of Fig.~\ref{Fig.gA_3pt}.
Notably, \(d_1 \) undergoes a sign change at \( \bar{N}_{\rm e} = 55 \), indicating that for this specific value, the ratio \( \mathcal{R}_N^{A_{u-d}}(t_f,t) \) at finite \( t \) and \( t_f \) should exhibit reduced excited-state contamination, bringing it closer to the ground-state prediction for \( g_A^{u-d} \) compared to other choices of \( \bar{N}_{\rm e} \).
At larger \( \bar{N}_{\rm e} \), however, \( d_1 \) and \( d_2 \) take on the same sign, causing their excited-state contributions to add constructively and enhancing the overall contamination.

A key finding is the hierarchy \( d_1 = \mathcal{O}(0.1) \ll d_2 \sim 1 = d_0 \), consistent with recent GEVP analyses of nucleon 2-point functions employing \( \pi N \) and \( \pi\pi N \) interpolators ~\cite{Hackl:2024whw}. This suggests that while the \( \pi N \) state is heavily suppressed in 2-point functions (scaling as \( d_1^2 \)), its influence on 3-point functions—suppressed only by \( d_1 \)—could remain significant.

\begin{table}[htbp]
  \centering
  \caption{Parameters of the joint fit of \( \mathcal{R}_N^{A_{u-d}}(t_f,t) \), with the data in the range of $t_f\in [t_{\rm min}(\bar{N}_{\rm e}),20]a$ and $t\in[\tau(\bar{N}_{\rm e})a,t_f-\tau(\bar{N}_{\rm e})a]$.}
    \begin{tabular}{ccccccccc}
    \hline
    \hline
    $\bar{N}_{\rm e}$ & 40    & 50    & 55    & 60    & 70    & 80    & 90    & 100 \\
    \hline
    $a\tau(\bar{N}_{\rm e})$ & 5     & 3     & 2     & 2     & 2     & 3     & 4     & 4 \\
    $at_{\rm min}$ & 11    & 7     & 6     & 6     & 6     & 7     & 9     & 13 \\
    \hline
    $Z(\bar{N}_{\rm e})$ & 0.0008812(95) & 0.001883(16) & 0.002572(21) & 0.003373(25) & 0.005317(36) & 0.007622(53) & 0.010425(84) & 0.01333(13) \\
    $d_1(\bar{N}_{\rm e})$ & -0.323(76) & -0.109(48) & -0.016(30) & 0.043(25) & 0.128(28) & 0.189(36) & 0.252(47) & 0.312(56) \\
    $d_2(\bar{N}_{\rm e})$ & 0.80(39) & 1.220(84) & 1.154(72) & 1.120(71) & 1.081(68) & 1.120(78) & 0.91(11) & 1.124(97) \\
    \hline
           $c_{00}$ & 1.2339(43) &       &       & \boldmath{}\textbf{$\chi^2/{\rm d.o.f}$}\unboldmath{} & \textbf{1.2} & & $m_0({\rm GeV})$ & 1.0574(22) \\
           $c_{10}$ & -0.172(35) & $c_{11}$ & -0.08(53) &       &      &  & $\Delta_1({\rm GeV})$ & 0.493(27) \\
           $c_{20}$ & -0.0208(17) & $c_{21}$ & 0.088(84) & $c_{22}$ & 0.892(29) & & $\Delta_2({\rm GeV})$ & 1.420(74) \\
    \hline
    \hline
    \end{tabular}%
  \label{tab:param_u-d}%
\end{table}%

All the fit parameters for \( \mathcal{R}_N^{A_{u-d}}(t_f,t) \) are collected in Table.~\ref{tab:param_u-d}, and we can see that $c_{22}$ is ${\cal O}(1)$ which is consistent with our expectation since they correspond to $g_A$ of the excited states. 

    \begin{figure*}
        \includegraphics[width=0.90 \textwidth]{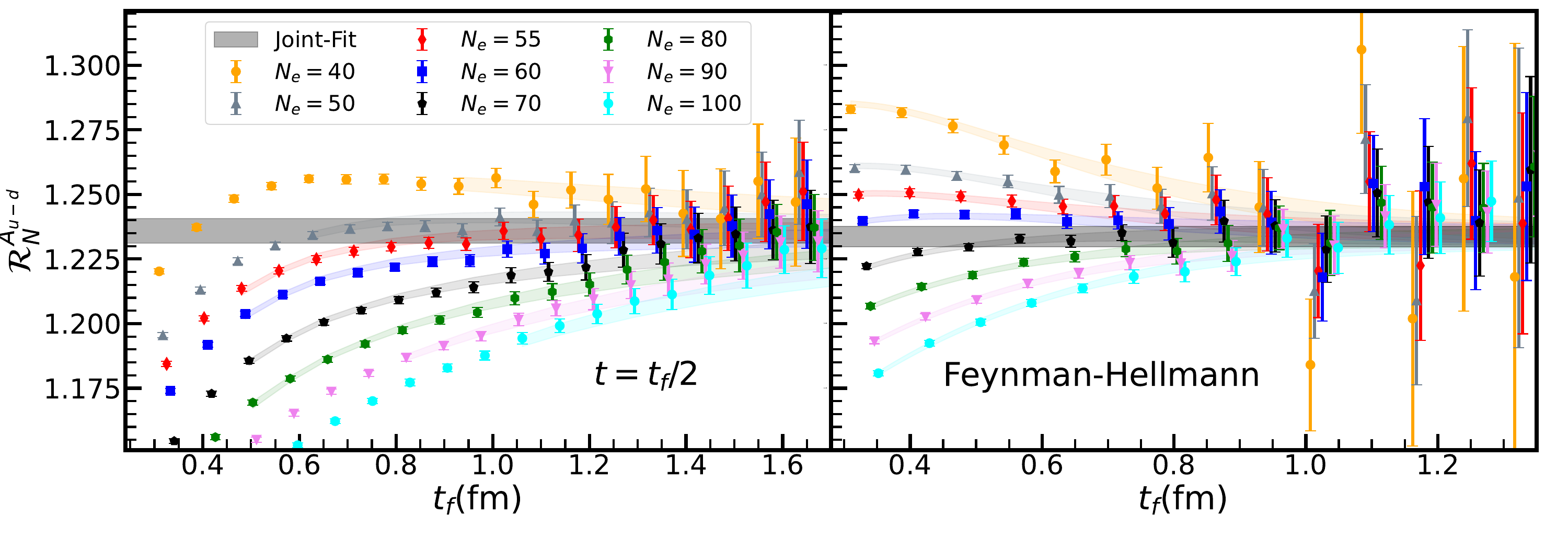}\\
        \includegraphics[width=0.90 \textwidth]{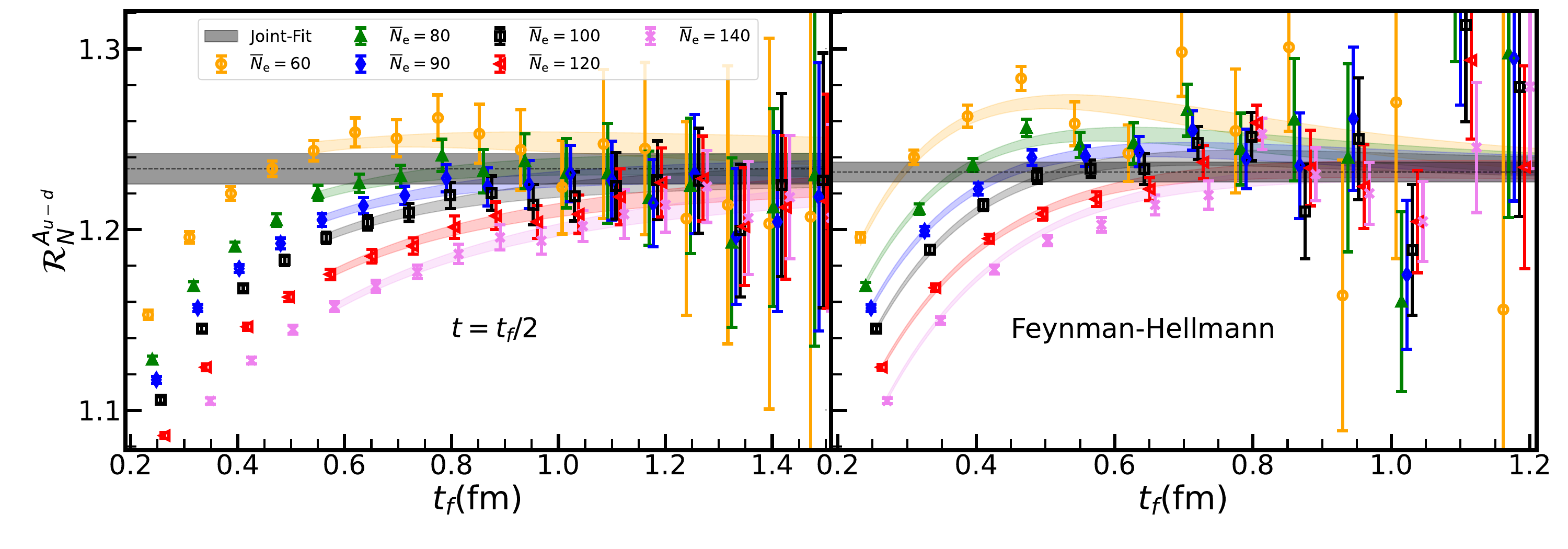}
        \caption{
            ${\cal R}_N^{A_{u-d}}(t_f,t_f/2)$ with different $\bar{N}_e$ used for the nucleon interpolator (left panel) and ${\cal R}_N^{A_{u-d},{\mathrm FH}}(t_f)$ for $\bar{N}_e=80$ which has good suppression of the excited state contamination (right panel) on the CLQCD F48P30 ensemble with 40 configurations (upper panels) and F64P13 ensemble with 41 configurations (lower panels).
            The gray bands correspond to that from the joint fit of the data with different $\bar{N}_{\rm e}$. 
        }
        \label{Fig:GA_sm}
    \end{figure*}

\red{
We extract \(g_A\) using two methods. The first is a joint three-state fit incorporating data across various \(\bar{N}_e\), \(t\), and \(t_f\) values. The second employs the Feynman–Hellmann-inspired differential summed ratio \({\cal R}_H^{{\cal O},{\rm FH}}(t)\) followed by a joint two-state fit with various \(\bar{N}_e\). On the F48P30 ensemble, the two methods yield:
\begin{align}
    g_A^{u-d, \mathrm{3-state}}(\mathrm{F48P30})=1.2339(43),\qquad g_A^{u-d, \mathrm{FH, 2-state}}(\mathrm{F48P30})=1.2337(40),
\end{align}
The original data and fitted curves are shown in the upper two panels of Fig.~\ref{Fig:GA_sm}: the left panel displays the three-state fit of \({\cal R}_N^{A_{u-d}}(t_f,t)\), and the right panel shows the two-state fit of \({\cal R}_N^{A_{u-d},{\rm FH}}(t_f)\). Both fits give consistent results with comparable uncertainties.

A similar analysis on the F64P13 ensemble gives:
\begin{align}
g_A^{u-d,\;\mathrm{3-state}}(\mathrm{F64P13}) = 1.2337(84),\qquad 
g_A^{u-d,\;\mathrm{FH,\;2-state}}(\mathrm{F64P13}) = 1.2318(55),
\end{align}
with the corresponding figures shown in the lower two panels of Fig.~\ref{Fig:GA_sm}. The uncertainties are larger, as expected due to the lighter pion mass. As a more conservative estimate, we take the results from the three-state fit (which has the larger uncertainty) as the main result used in the text. The inversions cost efficiency at the physical pion mass is discussed in the next section of this supplemental material.

Note that the consistency between the results on the two ensembles likely arises from a combination of pion-mass dependence and finite-volume effects, warranting further investigation in future studies.
}

  \begin{figure*}
      \includegraphics[width=0.49 \textwidth]{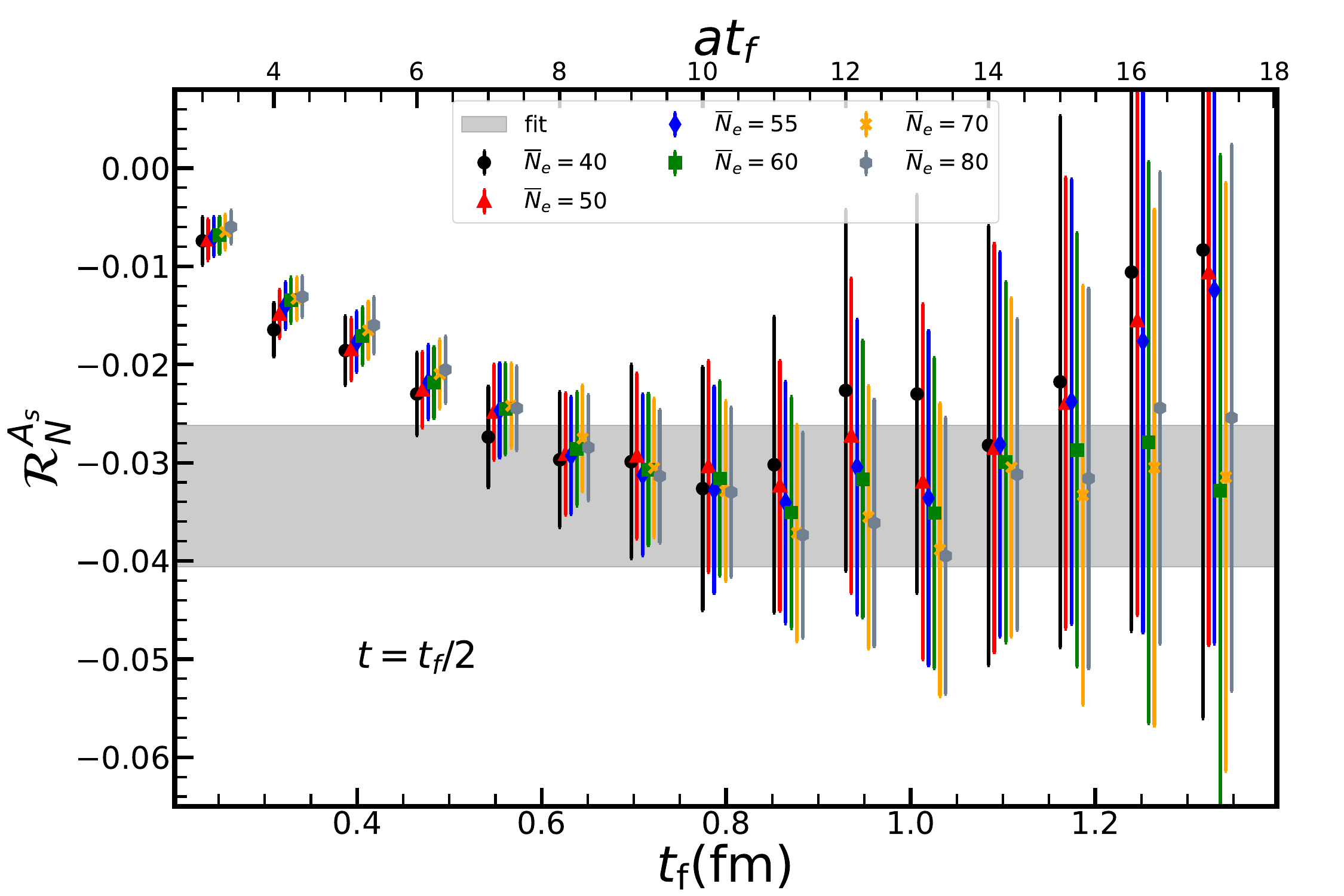}
      \includegraphics[width=0.49 \textwidth]{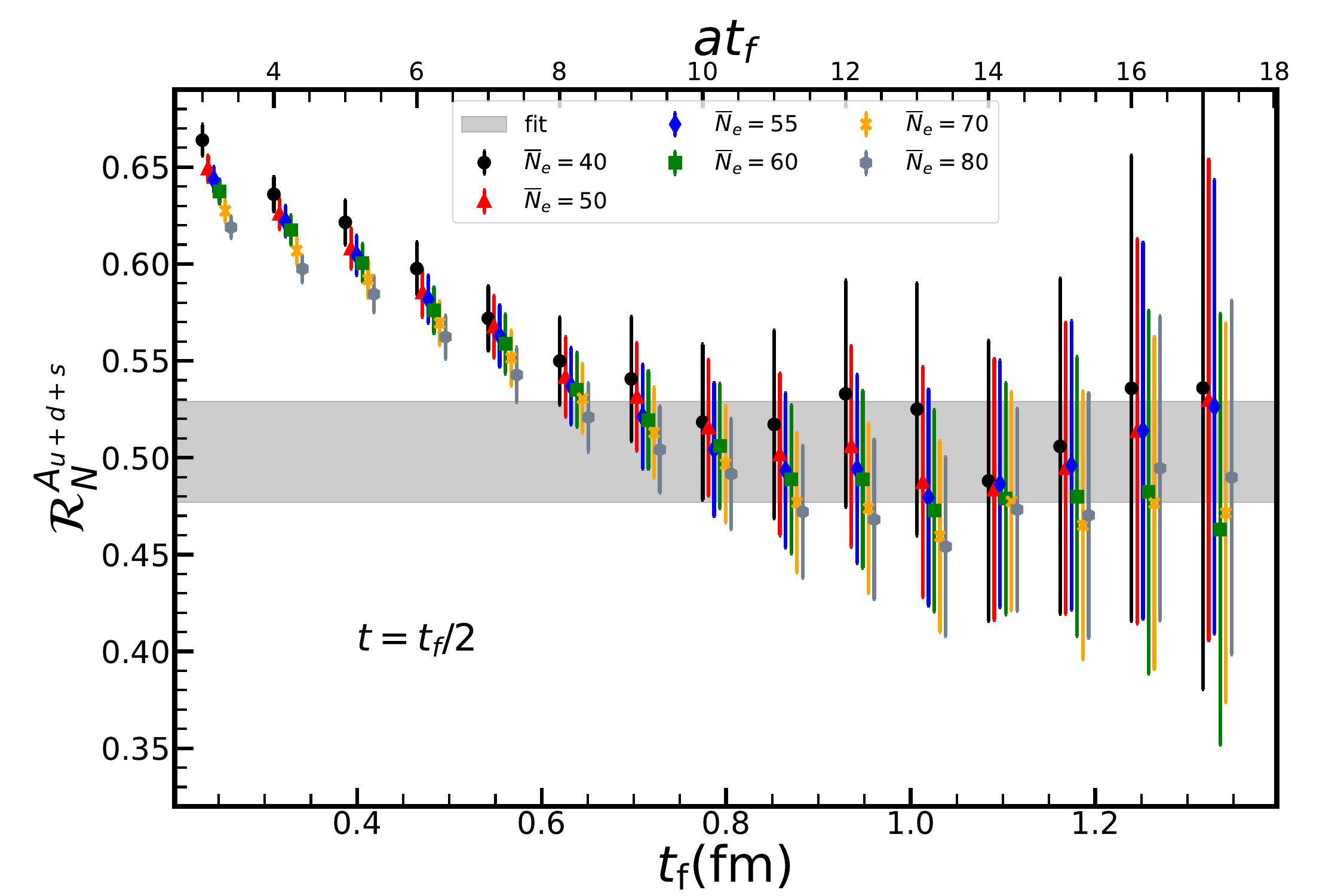}
      \caption{ 
            ${\cal R}_N^{A_{s}}(t_f,t_f/2)$ (left panel) and ${\cal R}_N^{A_{u+d+s}}(t_f,t_f/2)$ as a function of $t_f$ with different $\bar{N}_{\rm e}$ on the F48P30 ensemble. We can see that the $\bar{N}_{\rm e}$ dependencies are much weaker than those of ${\cal R}_N^{A_{u-d}}$.
      }
      \label{fig:ga_s_sum}
  \end{figure*} 

Our calculations on the F48P30 ensemble reveal that the \( \bar{N}_{\rm e} \)-dependence of both \( \mathcal{R}_N^{A_{s}} \) and \( \mathcal{R}_N^{A_{u+d+s}} \) is significantly weaker than expected across various \( t \) and \( t_f \) values, as demonstrated in Fig.~\ref{fig:ga_s_sum}. These results incorporate the disconnected contributions from light and strange quark loops, and then have relatively larger uncertainties.

Using a similar 3-state joint fit with priors on $\Delta_1$ and $d_1(\bar{N}_{\rm e})$ using the central values from the u-d case with 5 times of the uncertainties, we obtain {$g_A^{u+d+s}=0.503(26)$} with $\chi^2$/d.o.f.=1.0, and {$g_A^s=-0.0335(72)$} with $\chi^2/{\rm d.o.f.}$=0.8.
The resulting $d_{1,2}$ are consistent with those in the $g_A^{u-d}$ case, as shown in Fig.~\ref{Fig:gA_ex_strengh} with blue triangles (u+d+s) and black crosses (s).

\begin{table}
  \centering
  \caption{Parameters of the joint fit of \( \mathcal{R}_N^{A_{u+d+s}}(t_f,t) \), with the data in the range of $t_f\in [t_{\rm min}(\bar{N}_{\rm e}),20]a$ and $t\in[\tau(\bar{N}_{\rm e})a,t_f-\tau(\bar{N}_{\rm e})a]$.}
    \begin{tabular}{ccccccccc}
    \hline
    \hline
    $\bar{N}_{\rm e}$ & 40    & 50    & 55    & 60    & 70    & 80    & 90    & 100 \\
    \hline
    $a\tau(\bar{N}_{\rm e})$ & 2     & 2     & 2     & 2     & 2     & 2     & 2     & 2 \\
    $at_{\rm min}$ & 5     & 5     & 5     & 5     & 5     & 5     & 5     & 5 \\
    \hline
    $Z(\bar{N}_{\rm e})$ & 0.0008736(99) & 0.001870(17) & 0.002541(23) & 0.003329(31) & 0.005249(54) & 0.007523(85) & 0.01019(14) & 0.01317(21) \\
    $d_1(\bar{N}_{\rm e})$ & -0.120(61) & 0.028(84) & 0.058(67) & 0.085(62) & 0.137(51) & 0.208(38) & 0.272(26) & 0.308(23) \\
    $d_2(\bar{N}_{\rm e})$ & 1.141(46) & 1.079(33) & 1.048(32) & 1.015(30) & 0.988(29) & 0.955(28) & 0.925(28) & 0.916(29) \\
    \hline
         $c_{00}$ & 0.503(26)  &       &       & \boldmath{}\textbf{$\chi^2/{\rm d.o.f}$}\unboldmath{} & \textbf{1.0} & & $m_0({\rm GeV})$ & 1.0556(24) \\
         $c_{10}$ & -0.135(63) & $c_{11}$ & 0.66(22) &       &       & & $\Delta_1({\rm GeV})$ & 0.39(13) \\
         $c_{20}$ & 0.101(40) & $c_{21}$ & 0.37(26) & $c_{22}$ & 1.17(23) & & $\Delta_2({\rm GeV})$ & 1.289(51) \\
    \hline
    \hline
    \end{tabular}%
  \label{tab:param_uds}%
\end{table}%

\hypertarget{target1}{} 
\begin{table}[htbp]
  \caption{ 
      Parameters of the joint fit of \( \mathcal{R}_N^{A_{s}}(t_f,t) \), with the data in the range of $t_f\in [t_{\rm min}(\bar{N}_{\rm e}),20]a$ and $t\in[\tau(\bar{N}_{\rm e})a,t_f-\tau(\bar{N}_{\rm e})a]$.
      % \hypertarget{target1}{~~} 
    }
  \centering
    \begin{tabular}{ccccccccc}
    \hline
    \hline
    $\bar{N}_{\rm e}$ & 40    & 50    & 55    & 60    & 70    & 80    & 90    & 100 \\
    \hline
    $a\tau(\bar{N}_{\rm e})$ & 2     & 2     & 2     & 2     & 2     & 2     & 2     & 2 \\
    $at_{\rm min}$ & 5     & 5     & 5     & 5     & 5     & 5     & 5     & 5 \\
    \hline
    $Z(\bar{N}_{\rm e})$ & 0.000853(15) & 0.001835(39) & 0.002491(53) & 0.003243(68) & 0.00507(10) & 0.00730(14) & 0.00991(19) & 0.01278(27) \\
    $d_1(\bar{N}_{\rm e})$ & -0.18(15) & 0.16(16) & 0.17(15) & 0.22(12) & 0.260(86) & 0.290(64) & 0.318(48) & 0.346(40) \\
    $d_2(\bar{N}_{\rm e})$ & 1.148(50) & 1.064(41) & 1.040(41) & 1.000(37) & 0.969(34) & 0.946(32) & 0.930(31) & 0.929(32) \\
    \hline
           $c_{00}$ &-0.0335(72)  &       &       & \boldmath{}\textbf{$\chi^2/{\rm d.o.f}$}\unboldmath{} & \textbf{0.8} && $m_0({\rm GeV})$ & 1.0521(33) \\
           $c_{10}$ & -0.015(11) & $c_{11}$ & 0.10(12) &       &       && $\Delta_1({\rm GeV})$ & 0.31(11) \\
           $c_{20}$ & 0.008(11) & $c_{21}$ & 0.076(36) & $c_{22}$ & 0.062(56) && $\Delta_2({\rm GeV})$ & 1.28(10) \\
    \hline
    \hline
    \end{tabular}
   \label{tab:param_s}
\end{table}

From these results and {\color{black} $g_A^{u-d} = 1.2339(43)$}, we derive the flavor-decomposed axial charges as :\red{
\begin{align}
    g_A^{u} &= \frac{1}{2}\left(g_A^{u+d+s} - g_A^s + g_A^{u-d}\right) = 0.886(14),\nonumber\\
    g_A^{d} &= \frac{1}{2}\left(g_A^{u+d+s} - g_A^s - g_A^{u-d}\right) = -0.350(14),\label{eq:decomp}
\end{align}}
where uncertainties are propagated through jackknife resampling. 
Those values are consistent with the direct 3-state fit of $g_A^{u,d}$ using similar priors, as {$g_A^u=0.887(11)$, $g_A^d=-0.364(21)$}.

All the fit parameters for ${\cal R}_N^{A_{u+d+s}}(t_f,t_f/2)$ and ${\cal R}_N^{A_{s}}(t_f,t_f/2)$ over the range of $t$ and $t_f$ are collected in Tables~\ref{tab:param_uds} and  \hyperlink{target1}{\textcolor{blue}{VII}}, respectively.
\red{A crucial difference between \( \mathcal{R}_N^{A_{u-d}} \) and \( \mathcal{R}_N^{A_{u+d+s}} \) is that, for the latter, the coefficient \(c_{20}\) of the contamination from the second excited state is comparable to \(c_{10}\) (the coefficient for the first excited state). Since the spectral weight \(d_2\) of the second excited state is always \(\mathcal{O}(1)\), varying \(\bar{N}_e\) cannot efficiently suppress the contamination from this state, which lies about 1 GeV above the ground state.} Thus identifying these states through a generalized eigenvalue problem (GEVP) analysis with multi-particle interpolation fields can be highly beneficial~\cite{Alexandrou:2024tps,Barca:2024hrl}. Notably, such an analysis can be performed without additional inversion cost using the blending method.

\subsubsection{\red{Autocorrelation, covariance matrix regularization and AIC model averaging}}
\label{sec:grid_aic}

\begin{figure}
    \centering
    \includegraphics[width=0.45\linewidth]{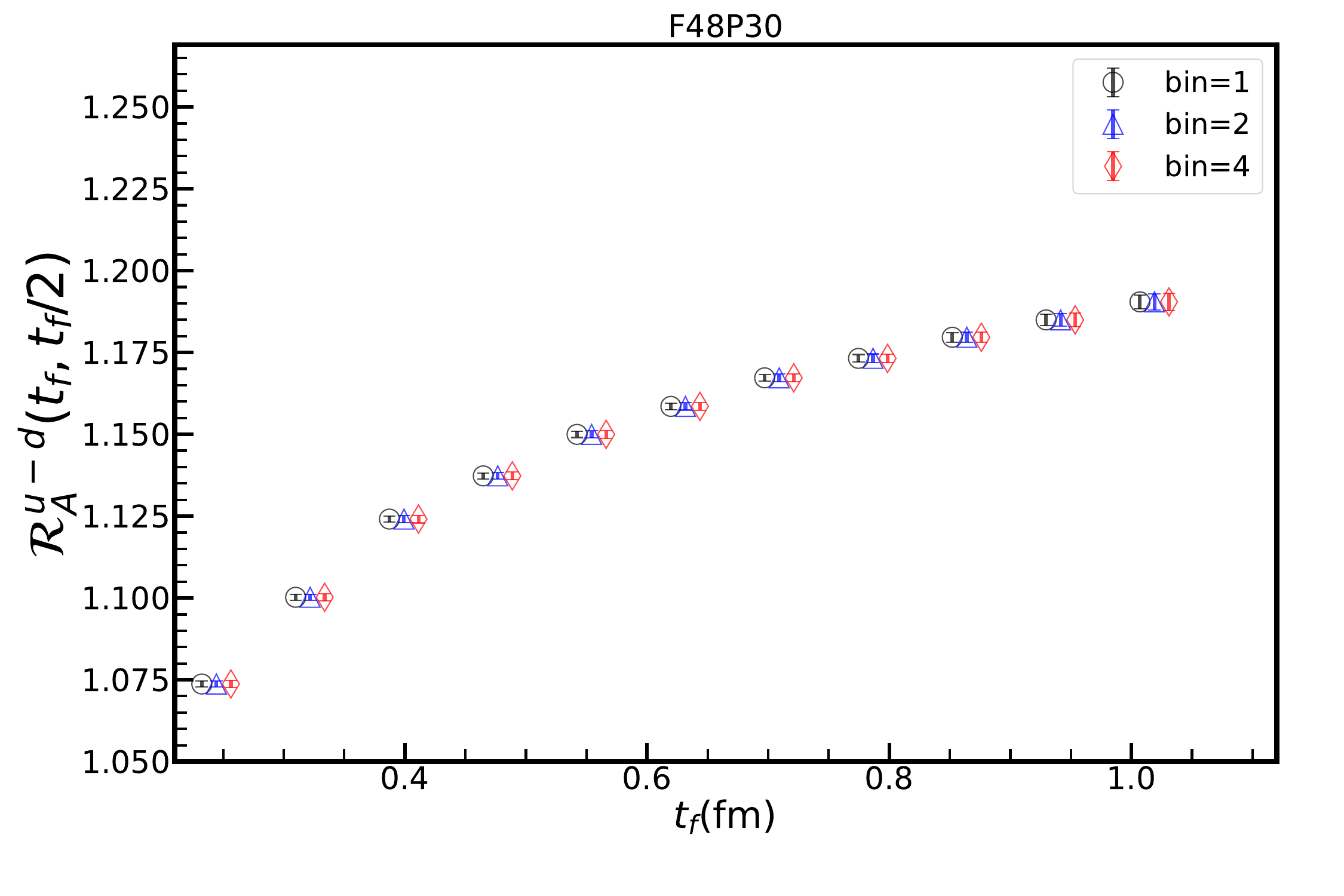}
    \includegraphics[width=0.45 \textwidth]{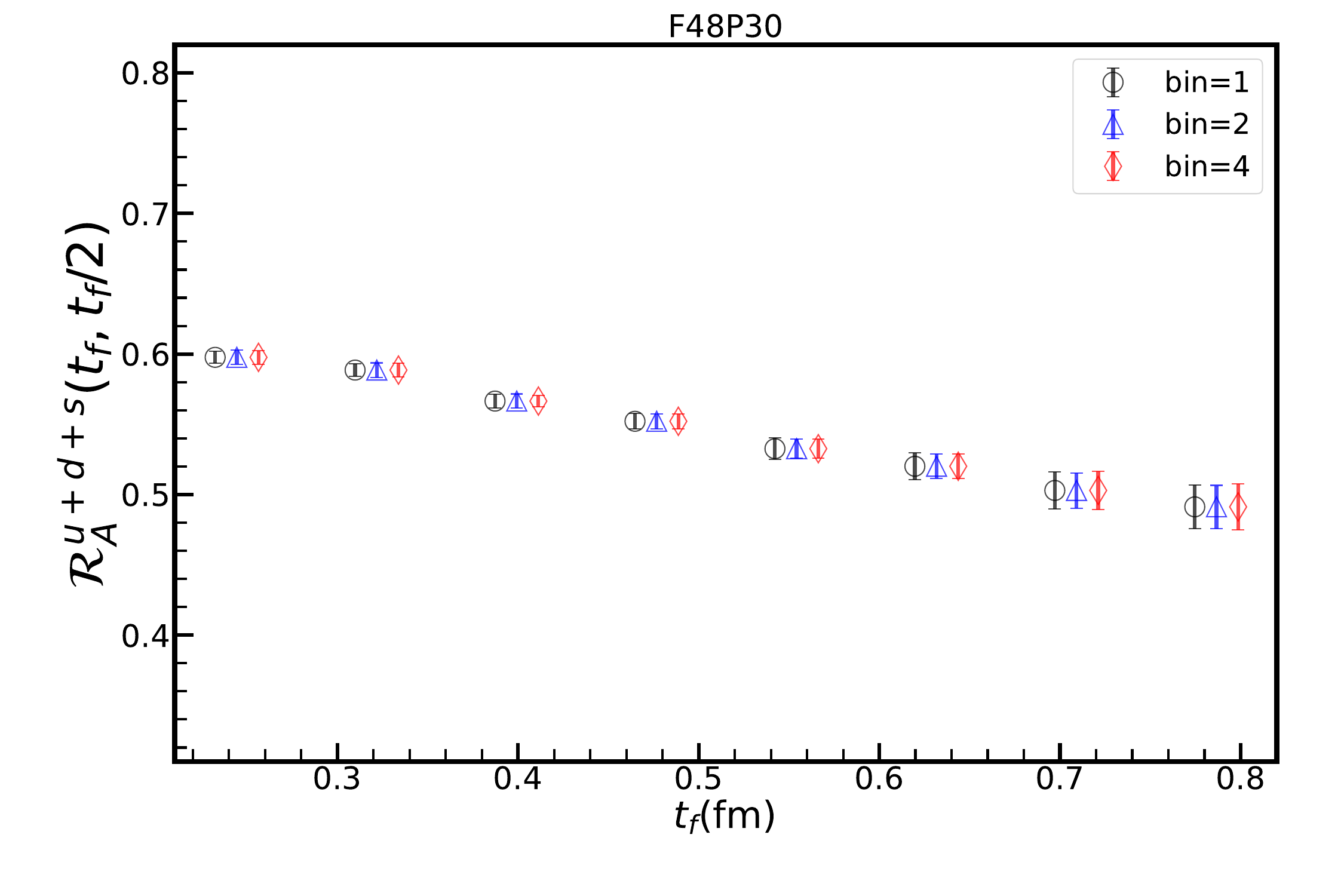}
    \includegraphics[width=0.45\linewidth]{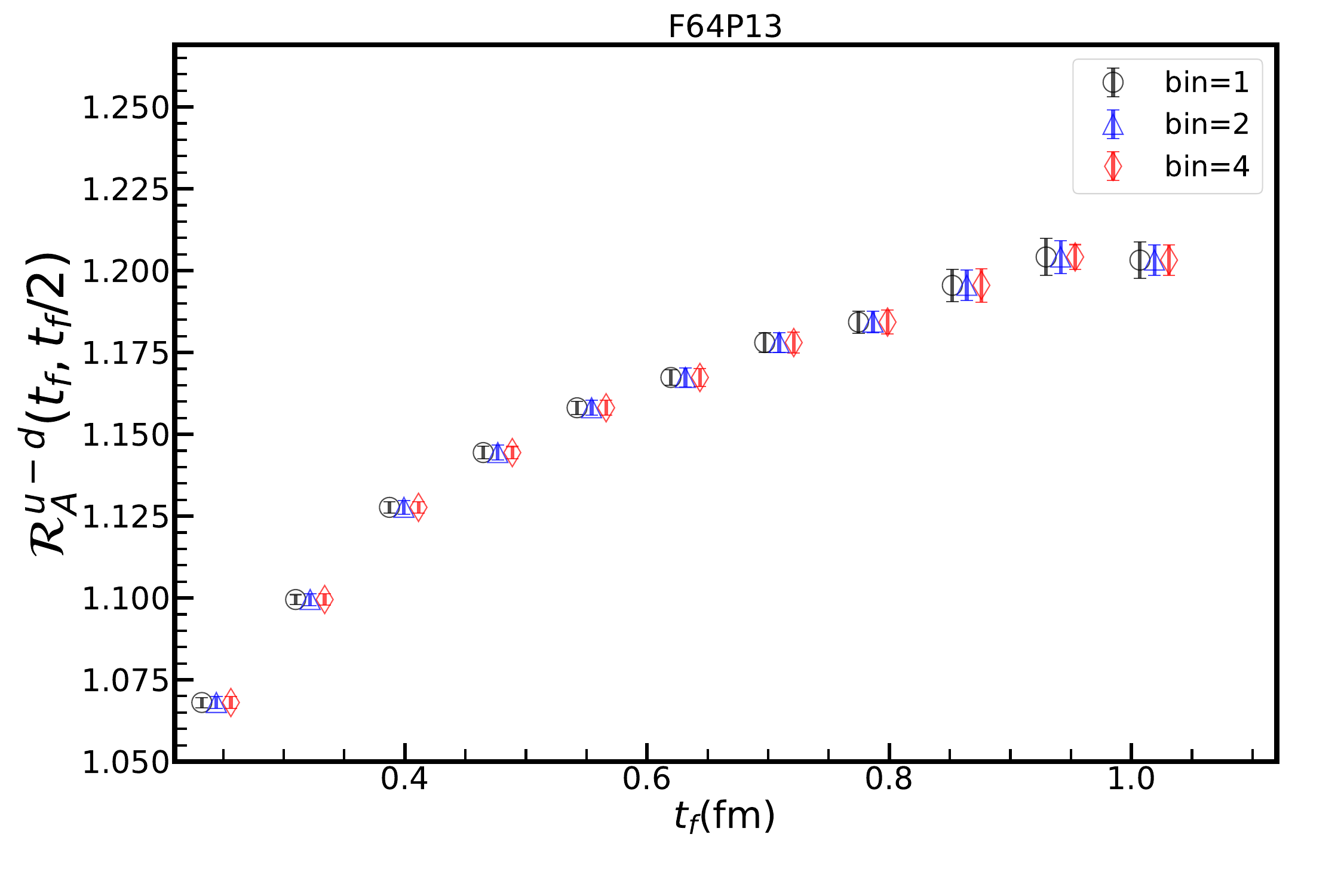}
    \includegraphics[width=0.45 \textwidth]{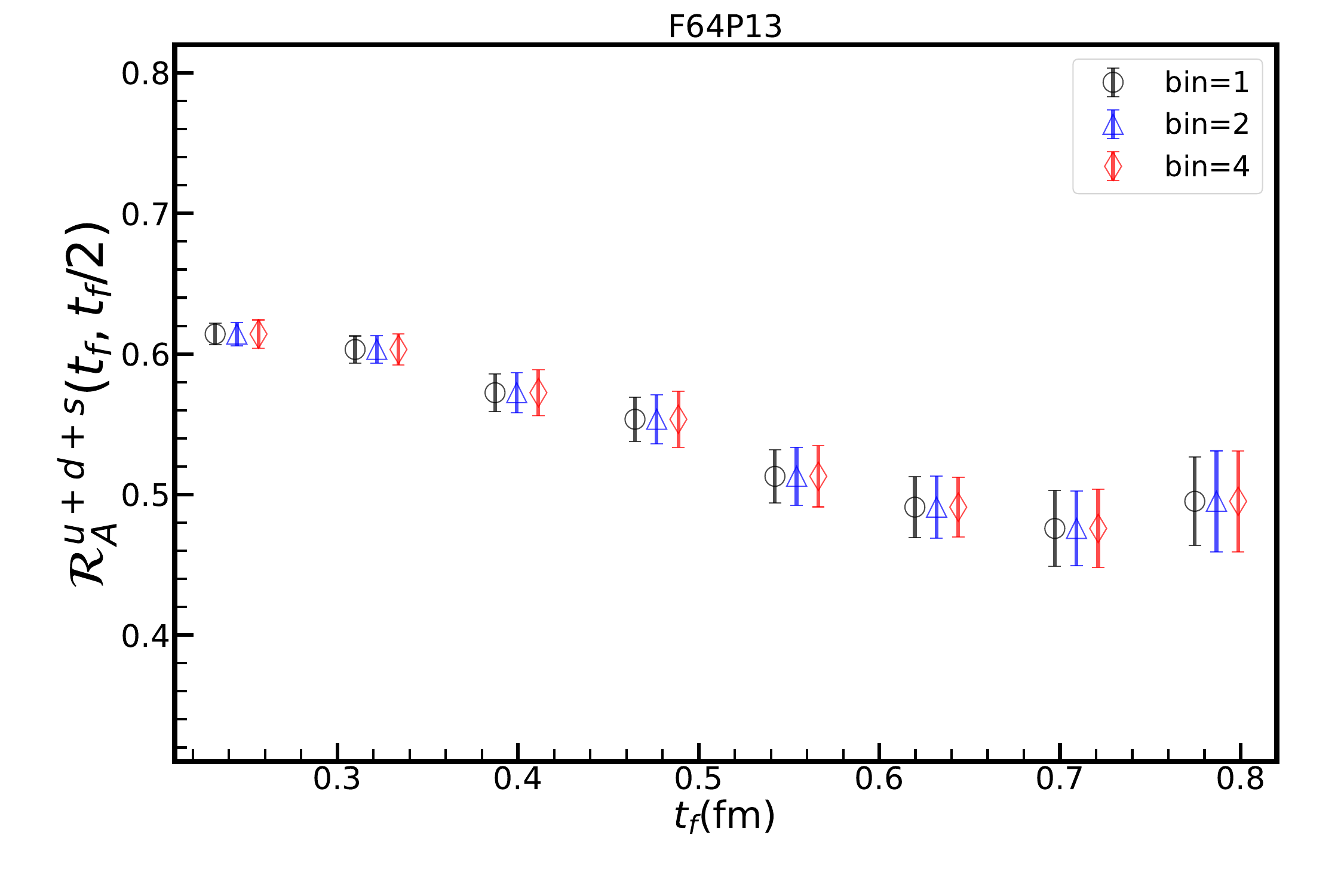}
    \caption{The autocorrelation analysis of $g^{u-d}_{A}$ (left panels) and $g_A^{u+d+s}$ (right panels) are shown by illustrating ${\cal R}_{N}^{A}(t_f,t_f/2)$ with bin sizes is $\{1,2,4\}$(from left to right with colors ranging from black to red ) on the F48P30 ensemble with $\bar{N}_{\rm e}$=100 (upper panels), and F64P13 ensemble with $\bar{N}_{\rm e}$=140 (lower panels).
    }
    \label{fig:autocorr_all}
\end{figure}

\red{To assess the impact of autocorrelations, we perform a dedicated binning analysis. For this purpose, we compute the ratio \( \mathcal{R}^{A,\text{mid}}_{N}(t_f, t_f/2) \) for both the \( u-d \) and \( u+d+s \) cases on the F64P13 ensemble, employing bin sizes of \( n = \{1, 2, 4\} \).

For each bin size, the mean and standard deviation are estimated from a large number of bootstrap resamples. In the absence of significant autocorrelations, the standard deviation is expected to remain stable as \( n \) varies. As shown in Fig.~\ref{fig:autocorr_all}, this is indeed the case across the F48P30 and F64P13 ensembles with different pion masses, for both $g_A^{u-d}$ and $g_A^{u+d+s}$, confirming that autocorrelation effects are negligible in our dataset. The stability of the statistical errors across bin sizes validates their robustness.

We emphasize that this binning procedure is applied to the sequence of gauge configurations after aggregating all measurements per configuration. It therefore probes gauge-field autocorrelations, rather than serving as a rebinning of intra-configuration source measurements.

Given the independence of the configurations, we should further consider the correlation between different quantities from the same configuration. When performing a simultaneous fit of the three-point function ratio $R(t,t_{\rm sep})$ across multiple source-sink separations $t_{\rm sep}$, the conventional approach is to include all time slices $t$ satisfying $t_{\rm cut} \le t \le t_{\rm sep}-t_{\rm cut}$ for each $t_{\rm sep}$ in the fit window.
The data are symmetrized under $t \leftrightarrow t_{\rm sep}-t$ (i.e., $R(t,t_{\rm sep})$ is averaged with $R(t_{\rm sep}-t,t_{\rm sep})$), so only points with $t \le t_{\rm sep}/2$ carry independent information; the half of the window with $t > t_{\rm sep}/2$ is redundant by construction.
However, even after symmetrization, an inclusive selection produces a highly redundant dataset: data points with similar source-insertion separations from adjacent $t_{\rm sep}$ remain strongly correlated, and the total number of data points $N_{\rm data}$ can approach or exceed the number of gauge configurations $N_{\rm cfg}$.
This leads to a near-singular or exactly singular covariance matrix that cannot be reliably inverted in the $\chi^2$ minimization. 

To avoid this problem, our default analysis for \(g_A^{u-d}\) employs bootstrap resampling with uniform seeds to preserve the underlying correlations. We have verified that the uncertainties are stable with respect to bin size, confirming that no statistically resolved bin-size dependence over the tested bins.

As a cross-check, we adopt a manual grid-selection strategy to regularize the covariance matrix while preserving sensitivity to the ground-state matrix element $c_{00}$.
Rather than including all data points within a standard window, we specify a sparse set of $(t_{\rm sep}, t)$ combinations in a configuration file, separately for each ensemble and each $t_{\rm min}$.
The selection is guided by the following principles:

\begin{enumerate}
    \item \textbf{Dimensionality constraint.}
    The total number of selected points $N_{\rm data}$ is kept well below $N_{\rm cfg}$ (typically $ 8\leq N_{\rm data} \leq 16$ for $N_{\rm cfg} \sim 40$ in this work), ensuring the covariance matrix remains positive definite with numerically stable inversion.

    \item \textbf{Source-sink separation spacing.}
    To avoid strong correlations between data at adjacent $t_{\rm sep}$ values, the selected $t_{\rm sep}$ values are not consecutive but spaced by an interval that depends on the available range.
    When many $t_{\rm sep}$ values are available (small $t_{\rm min}$), a larger spacing (e.g., $\Delta t_{\rm sep}=2$ or $3$) can be used; at larger $t_{\rm min}$, where fewer $t_{\rm sep}$ values remain above the cut, the spacing is reduced to maintain $N_{\rm data}$ in the range $8$--$16$.
    This adaptive spacing is not a fundamental principle but a practical consequence of the dimensionality constraint.

    \item \textbf{Window coverage.}
    Within each included $t_{\rm sep}$, a limited set of $t$ values is selected from the symmetrized half-window $t_{\rm cut} \le t \le t_{\rm sep}/2$ .
    The $t$ values are chosen to sample the central region of the available time range, avoiding the source boundary where excited-state contamination is largest. So 
    the starting $t$ value in each $t_{\rm sep}$ equals $t_{\rm cut}$, and $t$ values increase in unit steps.
\end{enumerate}

The grid selection is performed independently for each fit window (i.e., each $t_{\rm min}$), so that the AIC model averaging described below combines fits with different data-point selections in a consistent framework.

\textbf{AIC model averaging for fit-window variation.}
In LQCD, extracting ground-state matrix elements from correlation functions requires suppressing excited-state contamination by fitting data at sufficiently large Euclidean times. 
The choice of the minimum time slice $t_{\rm min}$ included in the fit is a source of systematic uncertainty, as different choices yield different results. 
The Akaike Information Criterion (AIC) provides a data-driven method to average over fit windows, weighting each by its relative fit quality.

Consider a set of $N$ fit windows, each yielding a result $\bar{c}_i\pm\sigma_i$ with chi-squared $\chi^2_i$, $n_i$ data points, and $k_i$ model parameters. The AIC for window $i$ is $\mathrm{AIC}_i = \chi^2_i + 2k_i - n_i$, and the corresponding weight is
\begin{equation}\label{eq:aic}
w_i = \frac{\exp\big[-\tfrac{1}{2}(\chi^2_i + 2k_i - n_i)\big]}{\sum_j\exp\big[-\tfrac{1}{2}(\chi^2_j + 2k_j - n_j)\big]}.
\end{equation}
Since all windows use the same two-state fit function ($k_i\equiv4$), the $2k_i$ term is a constant offset and does not affect the relative weights. Nevertheless we retain it for completeness.
The final estimate is obtained from the weighted probability distribution
\begin{equation}
P(c) = \sum_{i=1}^{N} w_i\,
\frac{1}{\sigma_i\sqrt{2\pi}}
\exp\!\left[-\frac{(c-\bar{c}_i)^2}{2\sigma_i^2}\right],
\end{equation}
with the central value and $1\sigma$ error taken from the 16th and 84th percentiles of $P(c)$. This construction folds the spread among fit windows into the total uncertainty.

Throughout this work, the nucleon matrix element $c_{00}$ is extracted from a two-state fit to the three-point function ratio,
\begin{equation}
R(t,t_{\rm sep}) = c_{00} + c_{10}\left[e^{-E_1(t_{\rm sep}-t)}+e^{-E_1 t}\right] + c_{11}\,e^{-E_1 t_{\rm sep}},
\label{eq:2state_fit}
\end{equation}
with priors $E_1=2.0(1.5)$, $c_{10}=c_{11}=0(3)$. For each ensemble, multiple windows are considered with $t_{\rm max}=12a$ and varying $t_{\rm min}$, where data satisfy $t_{\rm cut}\le t\le t_{\rm sep}-t_{\rm cut}$.

\textbf{Validation on $g_A^{u-d}$.}
We first apply the grid+AIC framework to the $g_A^{u-d}$ channel on the F64P13 ensemble, using the full covariance matrix that would be singular without the grid-based regularization. The fit windows span $t_{\rm min}/a = 8$--$12$, each with a distinct grid-based data-point selection. The results are shown in Fig.~\ref{fig:aic_ud}.

\begin{figure}[htbp]
\centering
\includegraphics[width=0.49\textwidth]{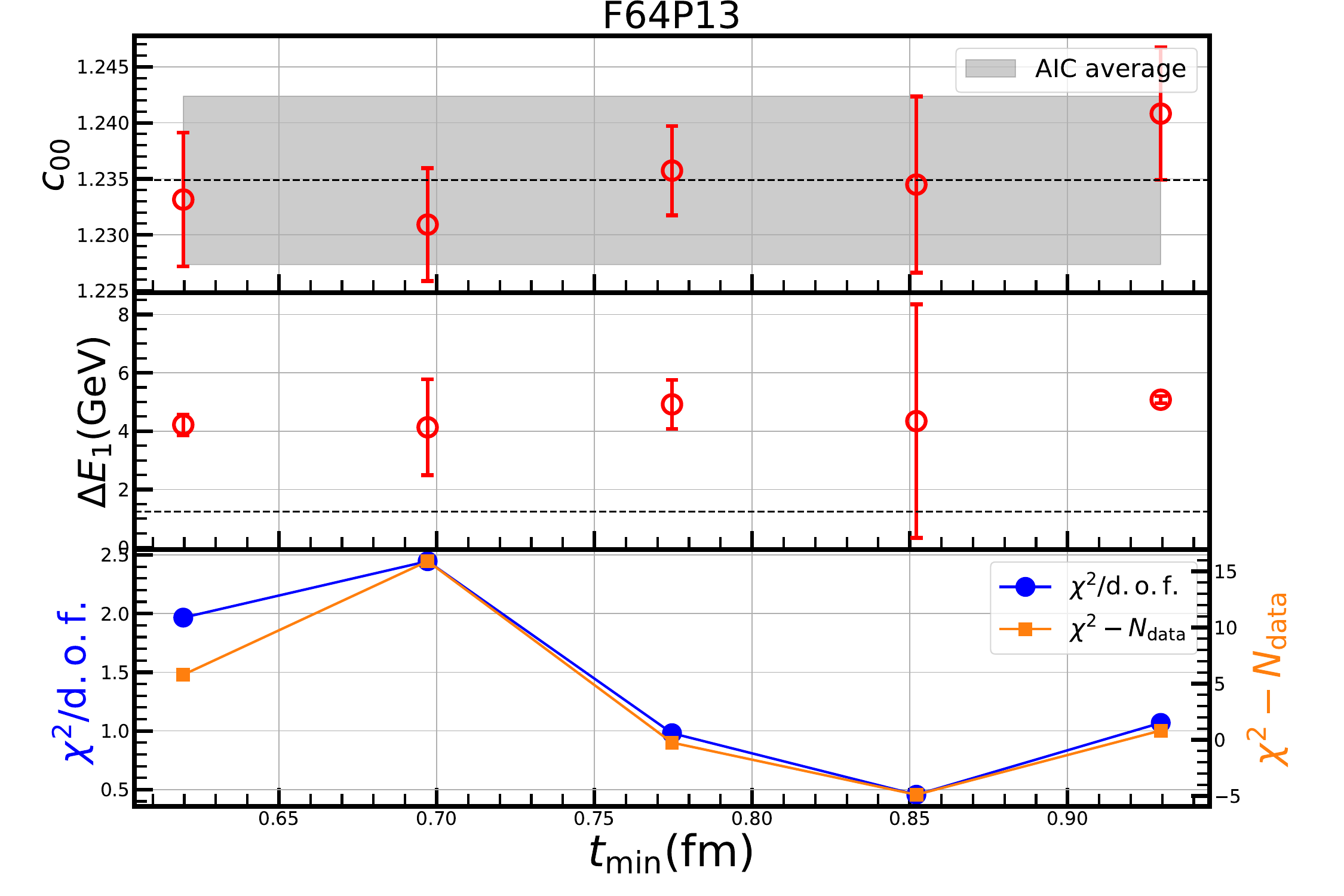}
\caption{
    AIC analysis of the grid-based full-covariance fits for $g_A^{u-d}$ on the F64P13 ensemble.
    Top panel: $c_{00}$ as a function of $t_{\rm min}$, with the AIC-averaged result shown as the gray band.
    Middle panel: the first excited-state energy gap $\Delta E_1$ in GeV.
    Bottom panel: $\chi^2/{\rm dof}$ (blue) and $\chi^2-n_i$ (orange) for each fit window.
    The AIC strongly favors $t_{\rm min}/a=11$ (weight 0.86) where $\chi^2/{\rm dof}=0.46$ with only $n_{\rm data}=9$ grid-selected data points.
}
\label{fig:aic_ud}
\end{figure}

The AIC-averaged result is
\begin{equation}
g_A^{u-d}(\mathrm{F64P13,AIC}) = 1.2349(75),
\label{eq:gA_ud_aic}
\end{equation}
in excellent agreement with the multi-$\bar{N}_e$ three-state fit ($1.2337(84)$) and the Feynman--Hellmann summed-ratio fit ($1.2318(55)$) reported in the previous subsection. This quantitative consistency validates the partial-correlation approximation used in the multi-$\bar{N}_e$ analysis and demonstrates that the grid+AIC framework yields stable, unbiased results even with a fully correlated covariance matrix. Having established the reliability of this methodology, we now apply it to the flavor-singlet channel, where the GEVP is essential for isolating the $\mathcal{N}f_1$ scattering states.}

{
\subsubsection{$g_A^{u,d,s}$ using a GEVP with current-involved operators}

In lattice QCD calculations, the imaginary-time evolution leads to an exponential suppression of excited-state contributions. In principle, the ground-state matrix element can be extracted by taking a sufficiently large time separation between source and sink. In practice, however, the signal-to-noise ratio also decays exponentially with time, making it challenging to reach the asymptotic region where only the ground state survives. The generalized eigenvalue problem (GEVP) provides a powerful tool to overcome this difficulty at the quantum field theory level~\cite{Barca:2022uhi,Barca:2024hrl,Barca:2025det,Barca:2026juc}.

The concept of “current‑enhanced excited states” was introduced in Ref.~\cite{Barca:2025det}. In lattice QCD, the standard nucleon interpolator \( \mathcal{N} = P^+ u (u C \gamma^5 d) \) possesses the quantum numbers of the nucleon. However, quantum field theory allows this operator to couple to all states sharing the same quantum numbers – including five‑quark and higher Fock components (particle number is not conserved). A separate CLQCD work on \(g_S\) and \(g_T\)~\cite{Wang:2025nsd} found that even though the mixing between the three‑quark and five‑quark operators is very small (well below 1\%), the resulting excited‑state contamination can still be significant at small source‑sink separations.

For the axial coupling constant, the $\mathcal{N}f_1$ scattering state is expected to be a major source of excited-state contamination~\cite{Barca:2025det,Wang:2025nsd}, as the $f_1(1^+)$ meson carries the same quantum numbers as the axial current. To identify and systematically remove this contamination at the spectroscopic level, we construct explicit scattering interpolating operators. The standard nucleon interpolator is $\mathcal{N}= P^+u(uC\gamma^5d)$. We define flavor-singlet and flavor-octet operator to create the $f_1(1^+)$ state:
\begin{align}
f_1 &= \frac{1}{\sqrt{3}}\bigl(\bar{u}\gamma_5\gamma^i u + \bar{d}\gamma_5\gamma^i d + \bar{s}\gamma_5\gamma^i s\bigr), \\
f'_1 &= \frac{1}{\sqrt{6}}\bigl(\bar{u}\gamma_5\gamma^i u + \bar{d}\gamma_5\gamma^i d - 2\bar{s}\gamma_5\gamma^i s\bigr).
\end{align}
The scattering operators are constructed as local S-wave products, yielding a three-operator GEVP basis:
\begin{equation}
X \equiv \{\mathcal{N},\ \mathcal{N}f_1,\ \mathcal{N}f'_1\}.
\end{equation}
This basis explicitly includes the $\mathcal{N}f_1$ scattering states in the variational analysis. Based on the $3\times3$ two-point correlation matrix $C_{2,\alpha\beta}(t) \equiv \langle X_\alpha(t) X_\beta^\dagger(0)\rangle$ with proper spin projection, we solve the GEVP via:
\begin{equation}
C_{2,\alpha\beta}(t) V_{\beta}^{n}(t,t_0) = \lambda^n(t,t_0) C_{2,\alpha\beta}(t_0) V_{\beta}^{n}(t,t_0),
\label{Eq:GEVP}
\end{equation}
where $\alpha,\beta=0,1,2$ index the interpolator basis, and the eigenvectors satisfy $(V_{\alpha}^{n})^{\dagger}C_{2,\alpha\beta}V_{\beta}^{m}=\delta^{nm}$. The eigenvalues give the effective masses $m_{\rm eff}^{(n)}(t)=\ln[\lambda^n(t,t_0)/\lambda^n(t+1,t_0)]$, and the eigenvectors $V_\alpha^{(n)}$ encode the operator composition of each state.

To obtain the nucleon--$f_1$ channel with quantum numbers
$I(J^P)=\frac12(\frac12^+)$, we project the nucleon--$f_1$ system onto
total spin $J=\frac12$ as follows.  The nucleon has
$I_{\mathcal{N}}=\frac12$ and $J_{\mathcal{N}}^P=\frac12^+$, whereas the isoscalar axial-vector
meson $f_1$ has $I_{f_1}=0$ and $J_{f_1}^P=1^+$.  Therefore, an
$S$-wave nucleon--$f_1$ system has total isospin $I=\frac12$ and positive
parity,
\begin{equation}
 I_{\mathcal{N}}\otimes I_{f_1}=\frac12\otimes0=\frac12,
 \qquad
 P=P_{\mathcal{N}}P_{f_1}(-1)^L=+1
 \quad (L=0).
\end{equation}
Its constituent spins couple according to  $\frac12 \otimes 1=\frac12\oplus\frac32$.  Using the Condon--Shortley phase convention and ordering the product states as
$|J_{\mathcal{N}}=\frac12,m_{\mathcal{N}}\rangle|J_{f_1}=1,m_{f_1}\rangle$, the state with
$(J,M)=(\frac12,\frac12)$ is
\begin{align}
\left|\mathcal{N}f_1;\frac12,\frac12\right\rangle
&=\sum_{m_{\mathcal{N}},m_{f_1}}
\left\langle \frac12,m_{\mathcal{N}};1,m_{f_1}
\middle|\frac12,\frac12\right\rangle
|\mathcal{N},m_{\mathcal{N}}\rangle|f_1,m_{f_1}\rangle \\\\[3pt]
&=\sqrt{\frac13}\,
|\mathcal{N},\tfrac12\rangle|f_1,0\rangle
-\sqrt{\frac23}\,
|\mathcal{N},-\tfrac12\rangle|f_1,+1\rangle .
\label{eq:spherical-coupling}
\end{align}
Here the two nonzero Clebsch--Gordan coefficients are
\begin{equation}
\left\langle \tfrac12,\tfrac12;1,0
\middle|\tfrac12,\tfrac12\right\rangle
=\sqrt{\frac13},
\qquad
\left\langle \tfrac12,-\tfrac12;1,+1
\middle|\tfrac12,\tfrac12\right\rangle
=-\sqrt{\frac23}.
\end{equation}

For the spherical-to-Cartesian polarization convention
\begin{equation}
|f_1,\pm1\rangle
=\mp\frac{1}{\sqrt2}  ( f_1^x \pm i f_1^y)
\qquad
|f_1,0\rangle= f_1^z,
\end{equation}
Eq.\ref{eq:spherical-coupling} is equivalently written as 
\begin{equation}
\left|\mathcal{N}f_1;\frac12,\frac12\right\rangle
=\frac{1}{\sqrt3}\left[
|\mathcal{N},\tfrac12\rangle|f_1,z\rangle
+|\mathcal{N},-\tfrac12\rangle|f_1,x\rangle
+i|\mathcal{N},-\tfrac12\rangle|f_1,y\rangle
\right].
\label{eq:cartesian-coupling}
\end{equation}
Accordingly, a Cartesian interpolating operator for this channel is
chosen as
\begin{equation}
\mathcal O_{\mathcal{N} f_1}^{\frac12,\frac12}
=\frac{1}{\sqrt3}\left[
\mathcal{N}_{\uparrow}f_{1,z}
+\mathcal{N}_{\downarrow}f_{1,x}
+i\mathcal{N}_{\downarrow}f_{1,y}
\right].
\end{equation}

    \begin{figure}
        \centering
        \includegraphics[width=0.45 \textwidth]{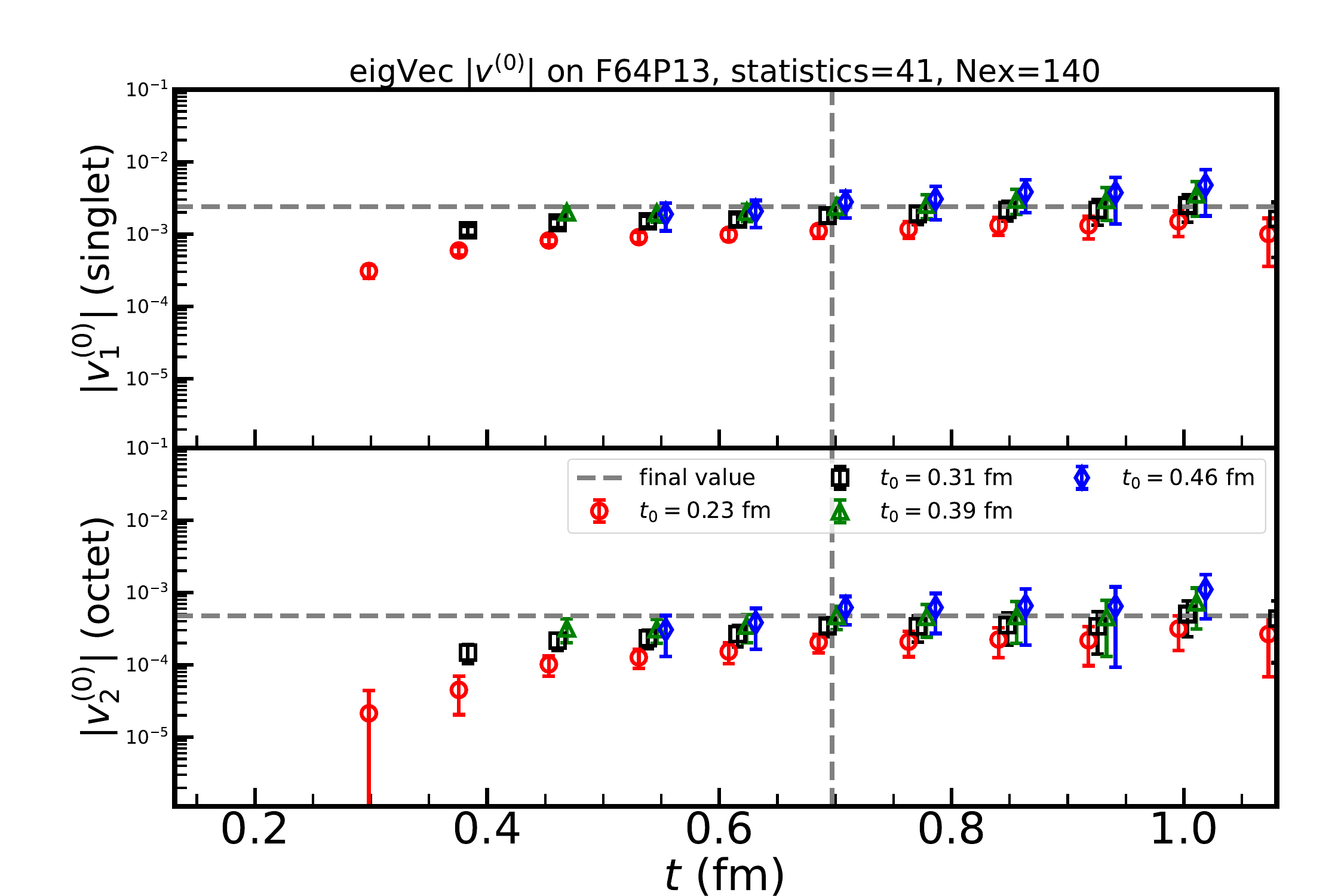}        
        \includegraphics[width=0.45 \textwidth]{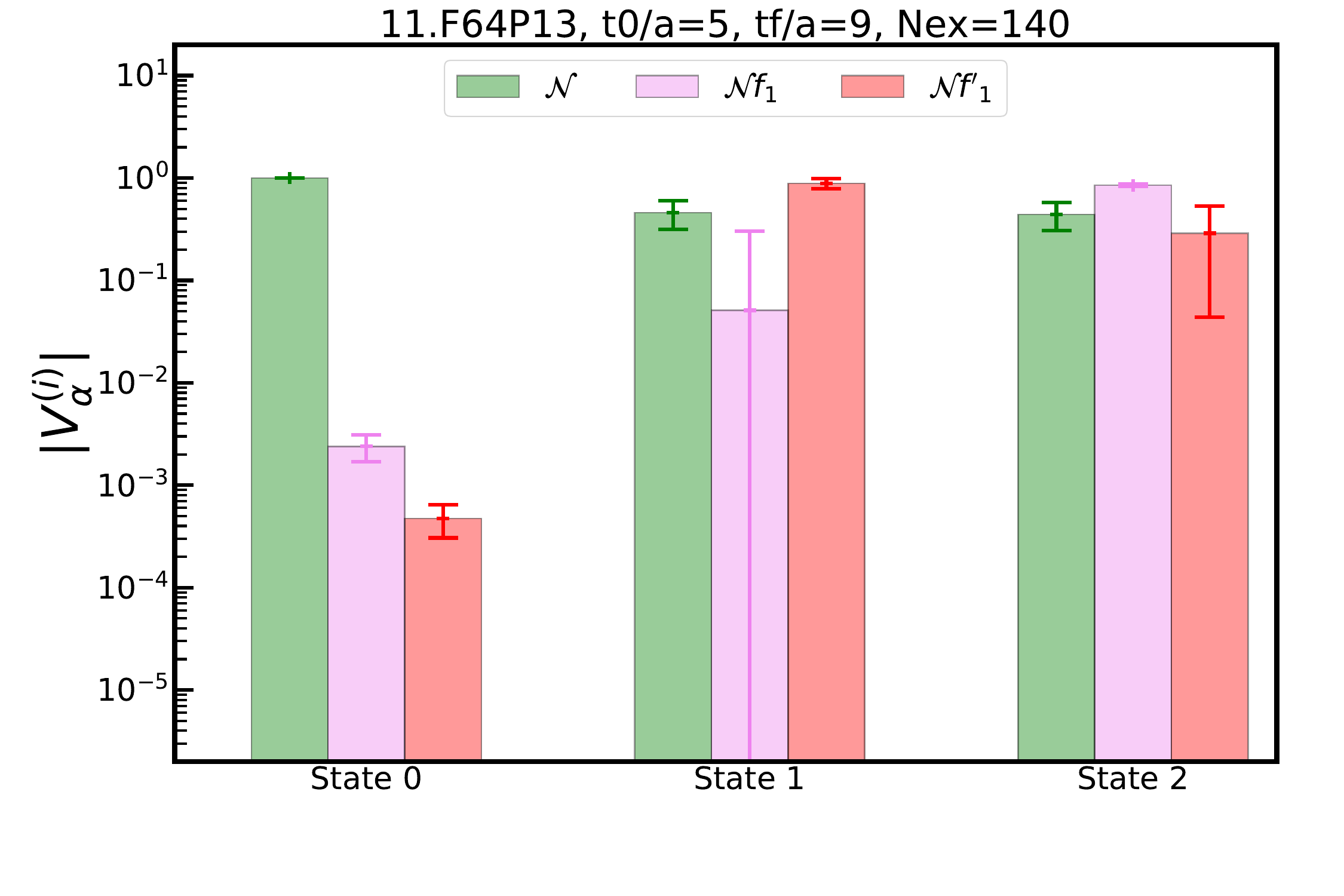}        
        \includegraphics[width=0.45 \textwidth]{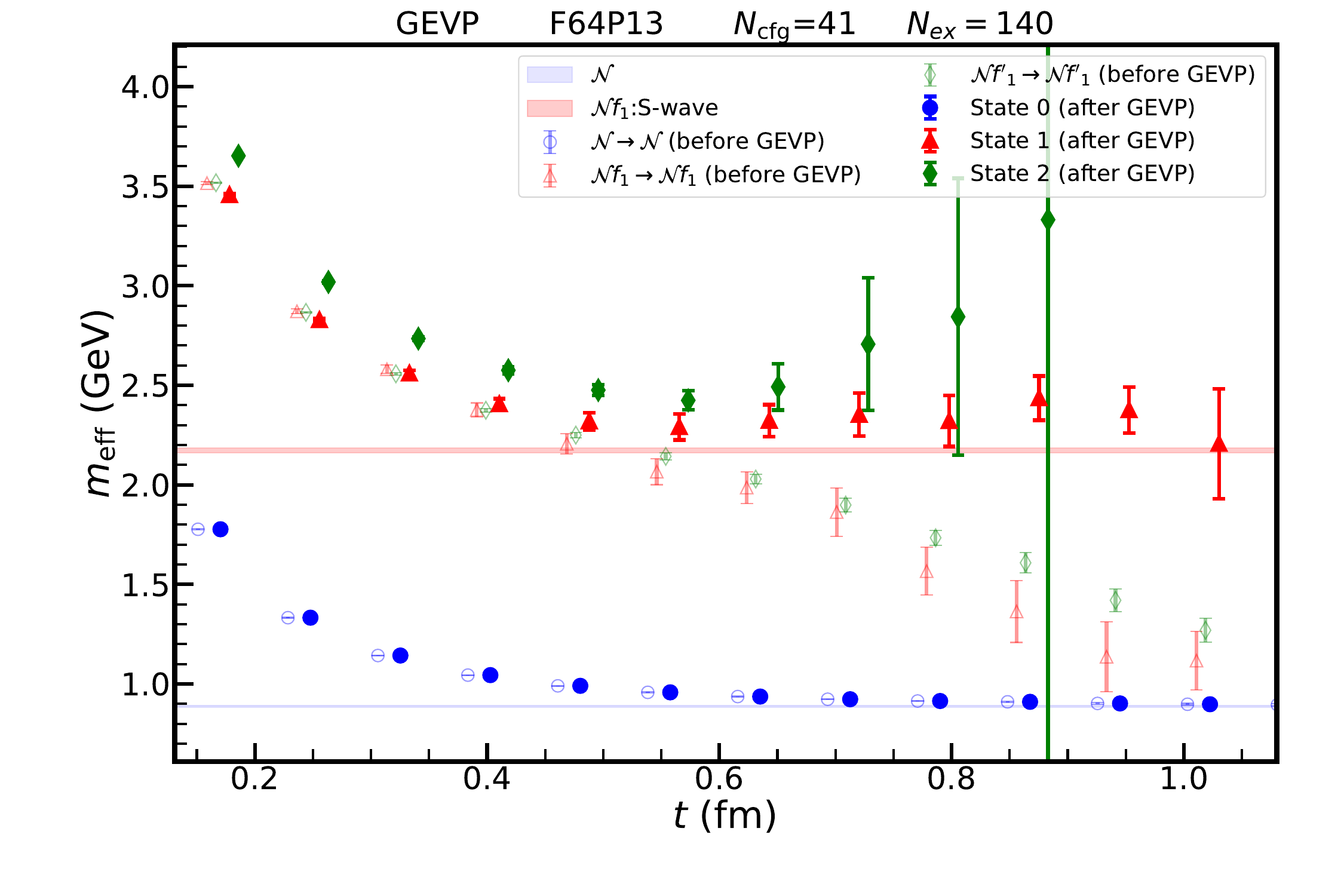}
        \caption{
        Upper Left : the ground-state eigenvector component $V_{\mathcal{N}f_1}^{(0)}$ and $V_{\mathcal{N}f'_1}^{(0)}$ as functions of $t$ for different $t_0$ on the F64P13 ensemble, showing convergence to a plateau at large Euclidean time. 
        Upper right: eigenvector composition $|V_\alpha^{(n)}|$ for each state $n=0,1,2$ at $t_0=5a$, $t=9a$. The ground state ($n=0$) is dominated by the single-nucleon operator $\mathcal{N}$ (blue), while the first ($n=1$) and second ($n=2$) excited states receive substantial contributions from the $\mathcal{N}f_1$ (orange) and $\mathcal{N}f'_1$ (green) scattering operators. 
        Lower : effective masses of the ground and first excited states before (hollow symbols) and after (filled symbols) solving the GEVP. Horizontal bands indicate the thresholds $m_N$ and $m_N+m_{f_1}$ for reference.
        }
        \label{fig.Nf1Spectrum}
    \end{figure}

The left panel of Fig.~\ref{fig.Nf1Spectrum} shows the ground-state eigenvector components $V^{(0)}_{\mathcal{N}f_1}$ and $V^{(0)}_{\mathcal{N}f'_1}$ as functions of $t$ for different reference times $t_0$. The clear convergence to a stable plateau at large $t$---independent of $t_0$---demonstrates the uniqueness of $V^{(0)}$ and confirms that the three-operator GEVP has successfully isolated the eigenstates without residual mixing. The right panel displays the eigenvector composition $|V_\alpha^{(n)}|$ at $t_0=5a$, $t=9a$: the ground state ($n=0$) is dominated by the single-nucleon operator $\mathcal{N}$ (blue, $\gtrsim 0.99$), while the first ($n=1$) and second ($n=2$) excited states are characterized by large admixtures of the $\mathcal{N}f_1$ (orange) and $\mathcal{N}f'_1$ (green) scattering operators. The lower panel compares the effective masses before (hollow symbols) and after (filled symbols) the GEVP rotation. While the ground-state effective mass receives only minor corrections, the first-excited-state effective mass shows a dramatically improved plateau, close to the $m_N+m_{f_1(f_1')}$ threshold. The distinct singlet and octet $f_1$ contributions are naturally disentangled by the three-operator basis, establishing the $\mathcal{N}f_1$ nature of the low-lying excitation.

\begin{figure}
    \centering
    \includegraphics[width=0.80 \textwidth]{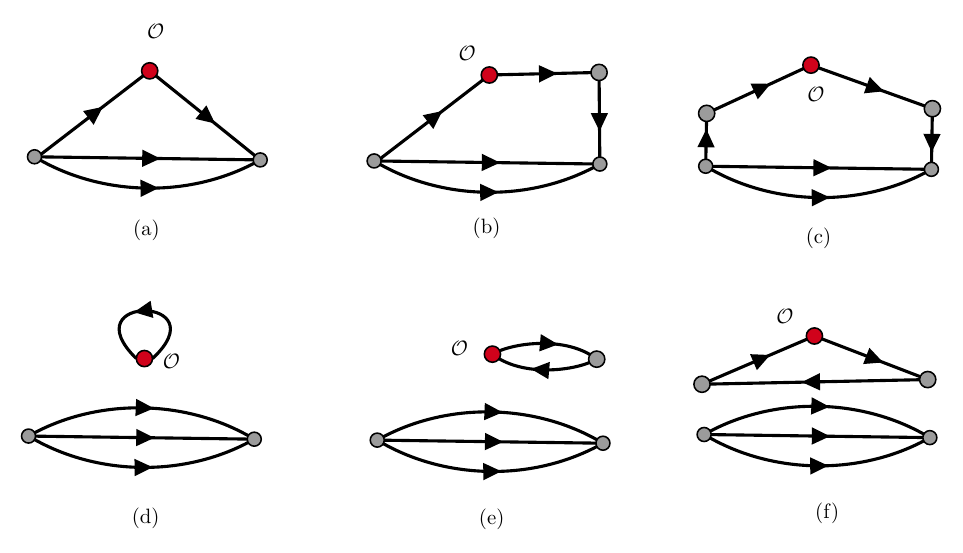}
    \caption{
        Examples of quark diagrams contributing to the 3-pf for each combination of nucleon interpolators in the three-operator basis: $\mathcal{N}$--$\mathcal{N}$ (left panels), $\mathcal{N}$--$\mathcal{N}f_1$ (middle panels), and $\mathcal{N}f_1$--$\mathcal{N}f_1$ (right panels). The upper three panels show connected diagrams, while the lower three show disconnected diagrams. Note that diagrams (e) and (f) include a vacuum contribution and therefore receive a volume enhancement relative to the other diagrams.
    }
    \label{fig.npimixFeynman}
\end{figure}

\begin{table}[htbp]
    \centering
    \caption{
        Number of quark diagrams for the 2-pf $\langle X_{\alpha} X^{\dagger}_{\beta} \rangle$ and 3-pf $\langle X_{\alpha} \mathcal{O} X^{\dagger}_{\beta} \rangle$, where the interpolator basis is $X=\{\mathcal{N}, \mathcal{N}f_1, \mathcal{N}f'_1\}$, and the current operator is either $\mathcal{O}=A_{\Sigma}$ or $A_s$.
    }
    \begin{tabular}{c|c|c|c}
    \hline
    \hline
         & 2-pt & 3-pt with $A_{\Sigma}$ & 3-pt with $A_{s}$ \\
    \hline
    $\mathcal{N} \to \mathcal{N}$         & 2    & 6   & 2   \\
    $\mathcal{N}f_1 \to \mathcal{N}$      & 12   & 84  & 14  \\
    $\mathcal{N}f_1 \to \mathcal{N}f_1$   & 84   & 612 & 112 \\
    \hline
    \hline
    \end{tabular}
    \label{tab:GEVPFeynmanNum}
\end{table}

Note that the $\mathcal{N}f_1$--$\mathcal{N}f_1$  channels contribute an equal number of diagrams; the full counting including the $f'_1$ sector is listed in Table~\ref{tab:GEVPFeynmanNum}. 
Evaluating such a large number of diagrams would be prohibitively expensive with traditional stochastic-quark-loop or sequential-source methods. However, the blending method employed in this work evaluates all of them without any additional inversion cost.

Using the ground-state eigenvectors $V^{0}$ with $t_0=5a, t_f=9a$  obtained from the GEVP of the two-point function, we construct the three-point function with the $\mathcal{N}f_1$ contamination suppressed:
\begin{align}
\tilde{C}_{3,00}(t_f,t) \equiv (V^{0}_\alpha(t,t_0))^{\dagger} \, C_{3,\alpha\beta}(t_f,t) \, V^{0}_\beta(t,t_0),\quad
C_{3,\alpha\beta}(t_f,t) \equiv \langle X_\alpha(t_f) A_{\Sigma}(t) X_\beta^\dagger(0)\rangle,
\label{Eq:GEVP_3pt}
\end{align}
where $A_{\Sigma}=\bar{u}\gamma_5S\!\!\!/ u+\bar{d}\gamma_5S\!\!\!/ d+\bar{s}\gamma_5S\!\!\!/ s$ is the flavor-singlet axial current ($S$ is the nucleon polarization vector). The resulting ratio ${\cal R}_N^{A_{\Sigma}}(t_f,t) = \tilde{C}_{3,00}(t_f,t)/\tilde{C}_{2,00}(t_f)$ is then suppresses the leading $\mathcal{N}f_1$ excited-state contamination.

The numbers of quark diagrams for the currents \(A_{\Sigma}\) and \(A_{s}\) are listed in Table~\ref{tab:GEVPFeynmanNum}, and we perform full Wick contractions including all diagrams. Evaluating such a large number of diagrams would be prohibitively expensive with traditional stochastic‑quark‑loop or sequential‑source methods. However, the blending method employed in this work evaluates all of them without any additional inversion cost.

\begin{figure}
    \centering
    \includegraphics[width=0.45 \textwidth]{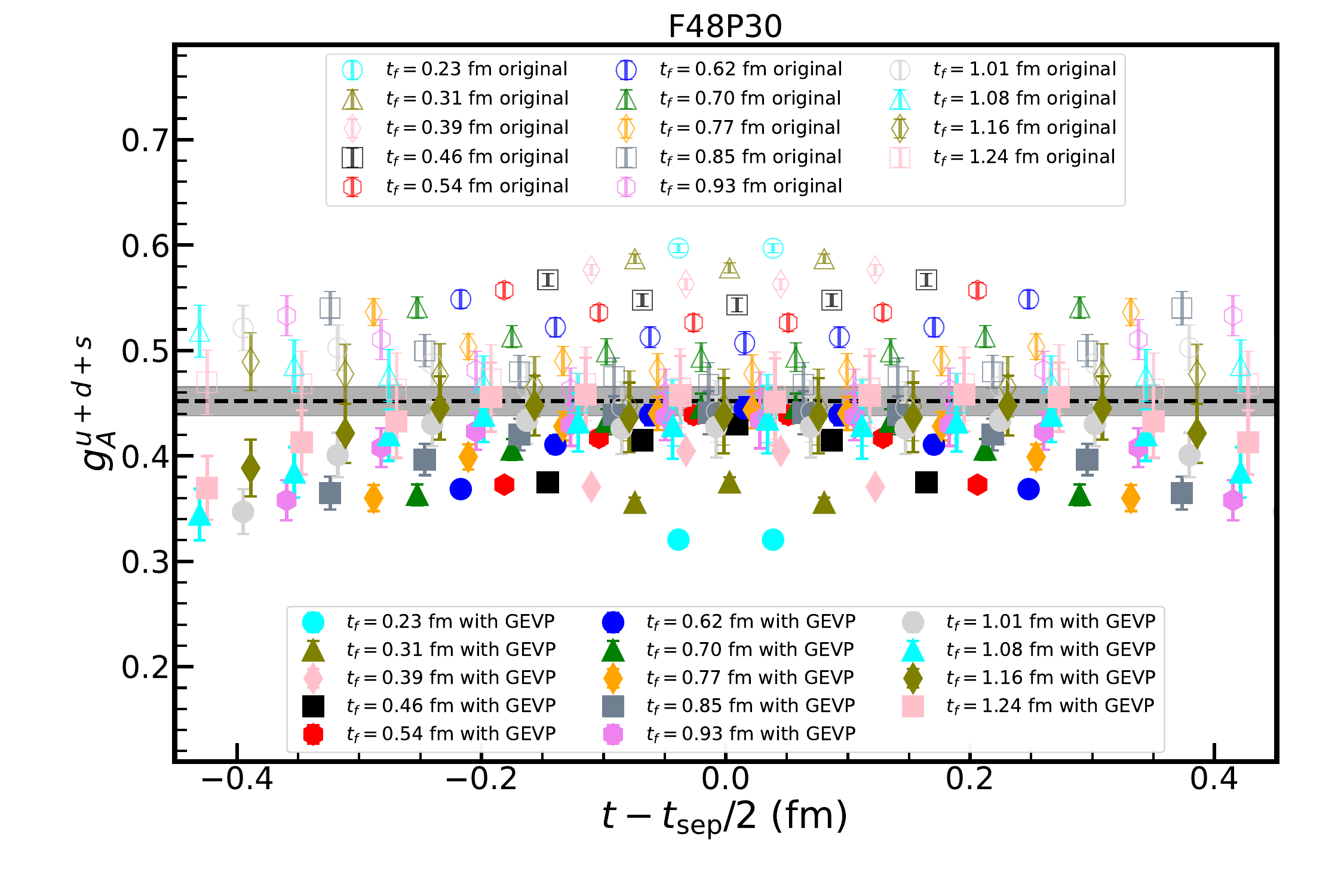}
    \includegraphics[width=0.45 \textwidth]{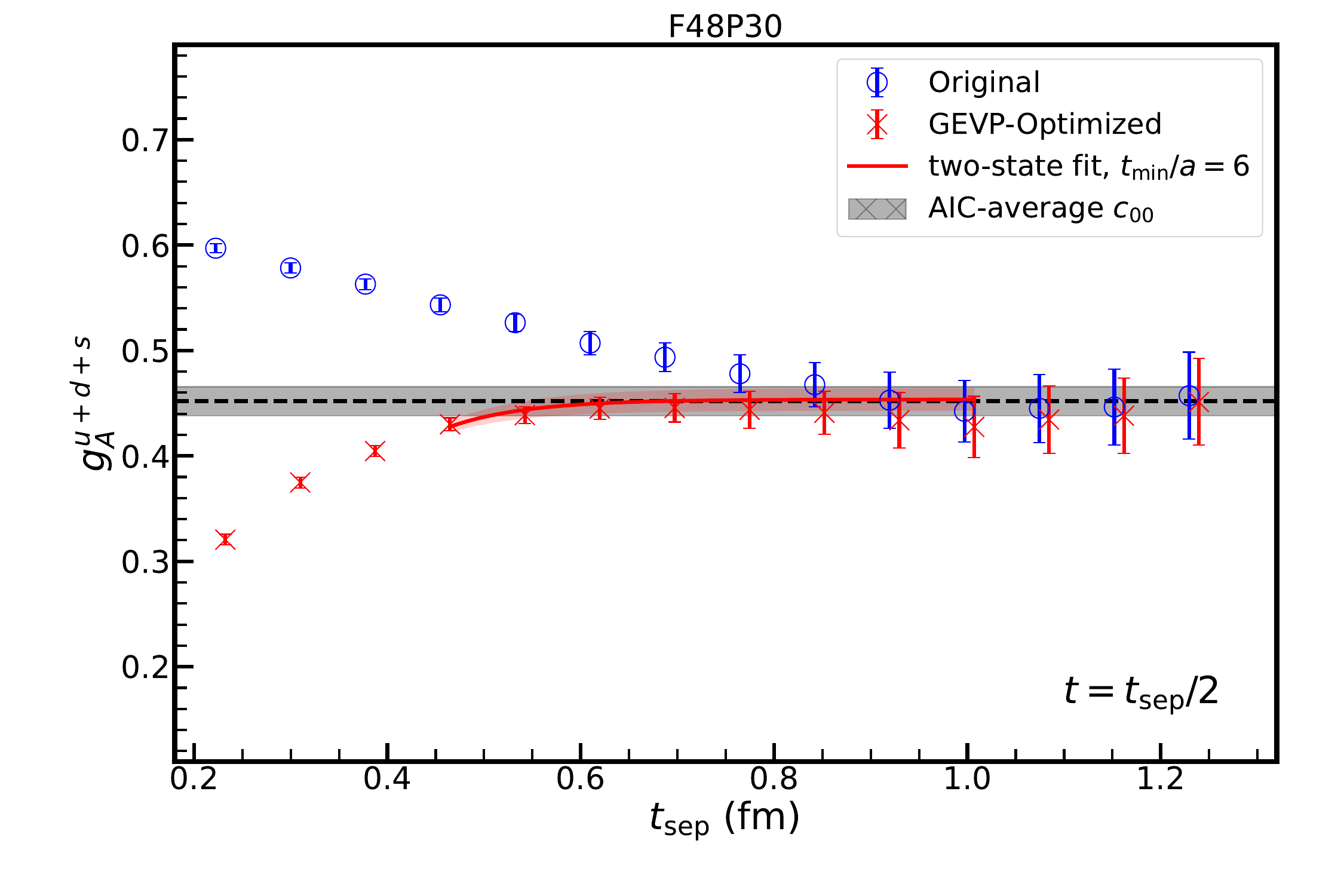}
    \includegraphics[width=0.45 \textwidth]{GAC3_GEVPope_tsep_F64P13.pdf}
    \includegraphics[width=0.45 \textwidth]{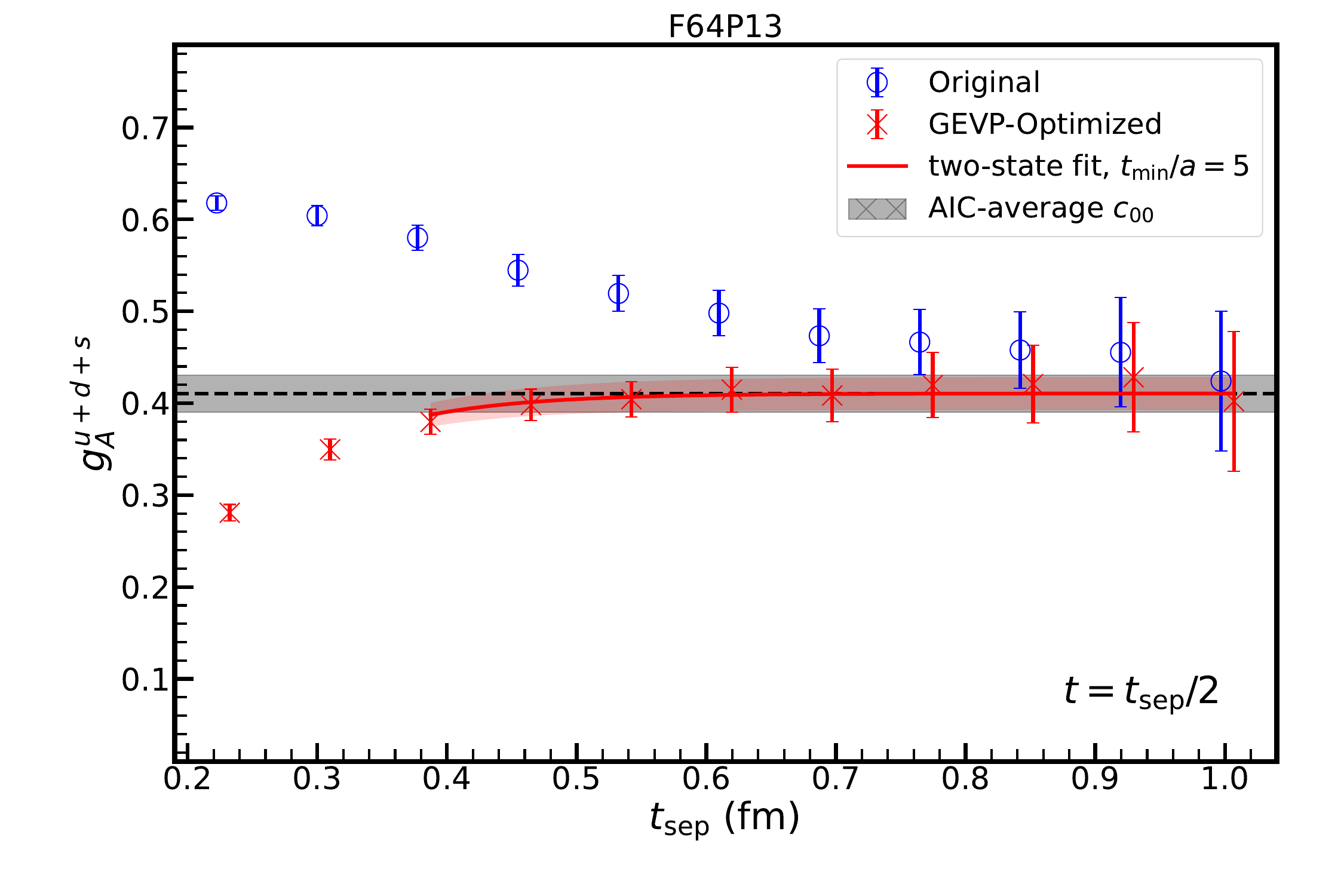}
    \caption{
    Left panels: the ratio ${\cal R}_N^{A_{\Sigma}}(t_f,t)$ for different source--sink separations $t_f$ and insertion times $t$, before (hollow symbols) and after (filled symbols) solving the GEVP with the three-operator basis $\{\mathcal{N}, \mathcal{N}f_1, \mathcal{N}f'_1\}$. Right panels: ${\cal R}_N^{A_{\Sigma}}(t_f,t_f/2)$ as a function of $t_f$.
    The gray band indicates the extracted ground-state matrix element from AIC average over different fit windows.
    Upper panels: results on the F48P30 ensemble; lower panels: results on the F64P13 ensemble.
    }
    \label{fig:ga_gevp2}
\end{figure}

As demonstrated in Fig.~\ref{fig:ga_gevp2} for both the F48P30 and F64P13 ensembles, the GEVP analysis with the traditional nucleon operator and two $\mathcal{N}f_1$ scattering operators dramatically accelerates the convergence of the ratio ${\cal R}_N^{A_{\Sigma}}(t_f,t)$: after isolating the ground state from the $\mathcal{N}f_1$ contamination, the ratio stabilizes at source--sink separations as small as $t_f \lesssim 0.5\,\text{fm}$---far earlier than without GEVP. This confirms that the $\mathcal{N}f_1$ state is indeed consistent with a dominant excited-state contamination in the axial channel, within the present operator basis, and the three-operator GEVP with $\{\mathcal{N}, \mathcal{N}f_1, \mathcal{N}f'_1\}$ provides a physically transparent and spectroscopically rigorous framework for its isolation. Crucially, this full GEVP analysis is made possible by the blending method, which evaluates all required diagrams without any additional inversion cost.

\begin{figure}[htbp]
\centering
\centering
\includegraphics[width=0.48\textwidth]{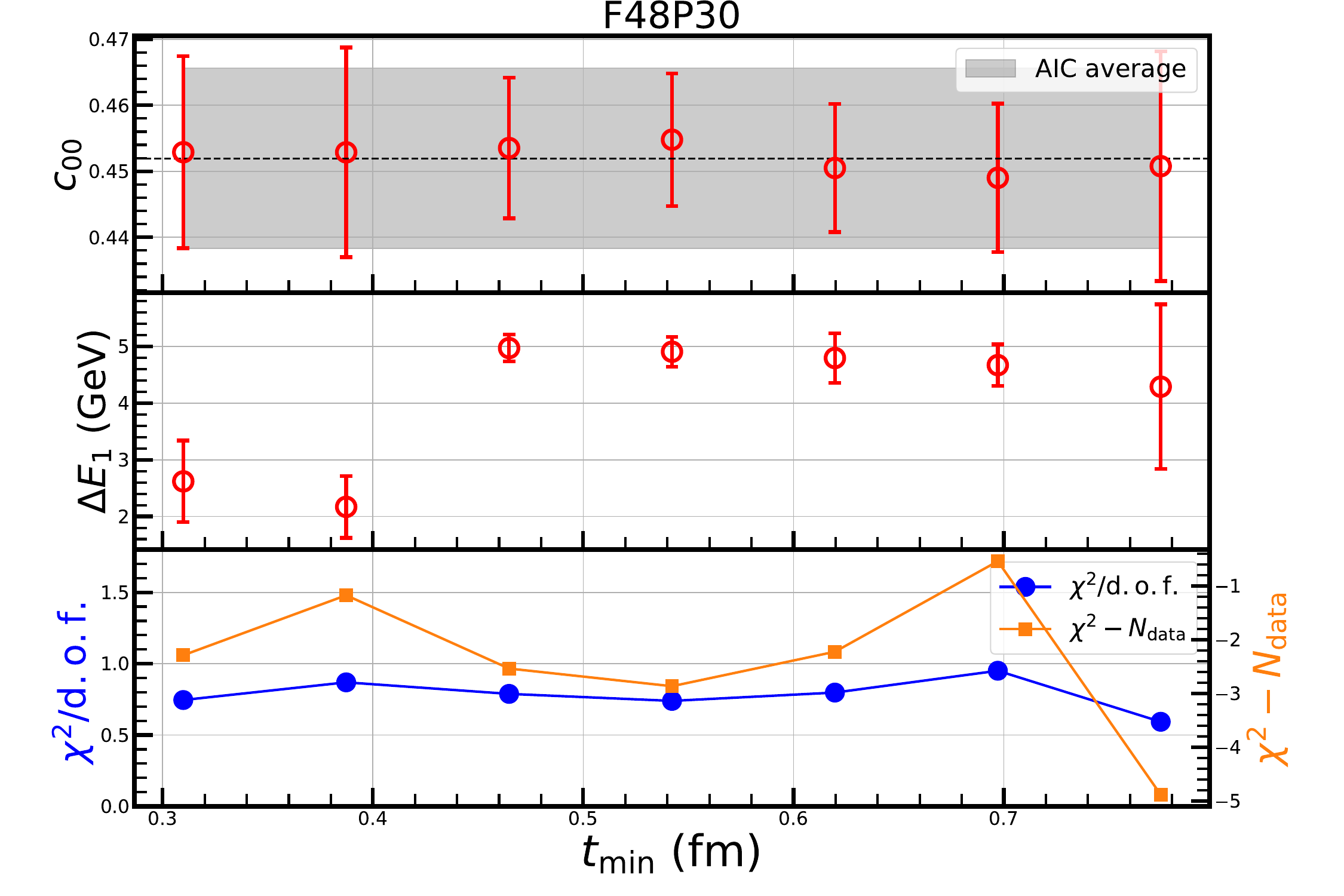}
\includegraphics[width=0.48\textwidth]{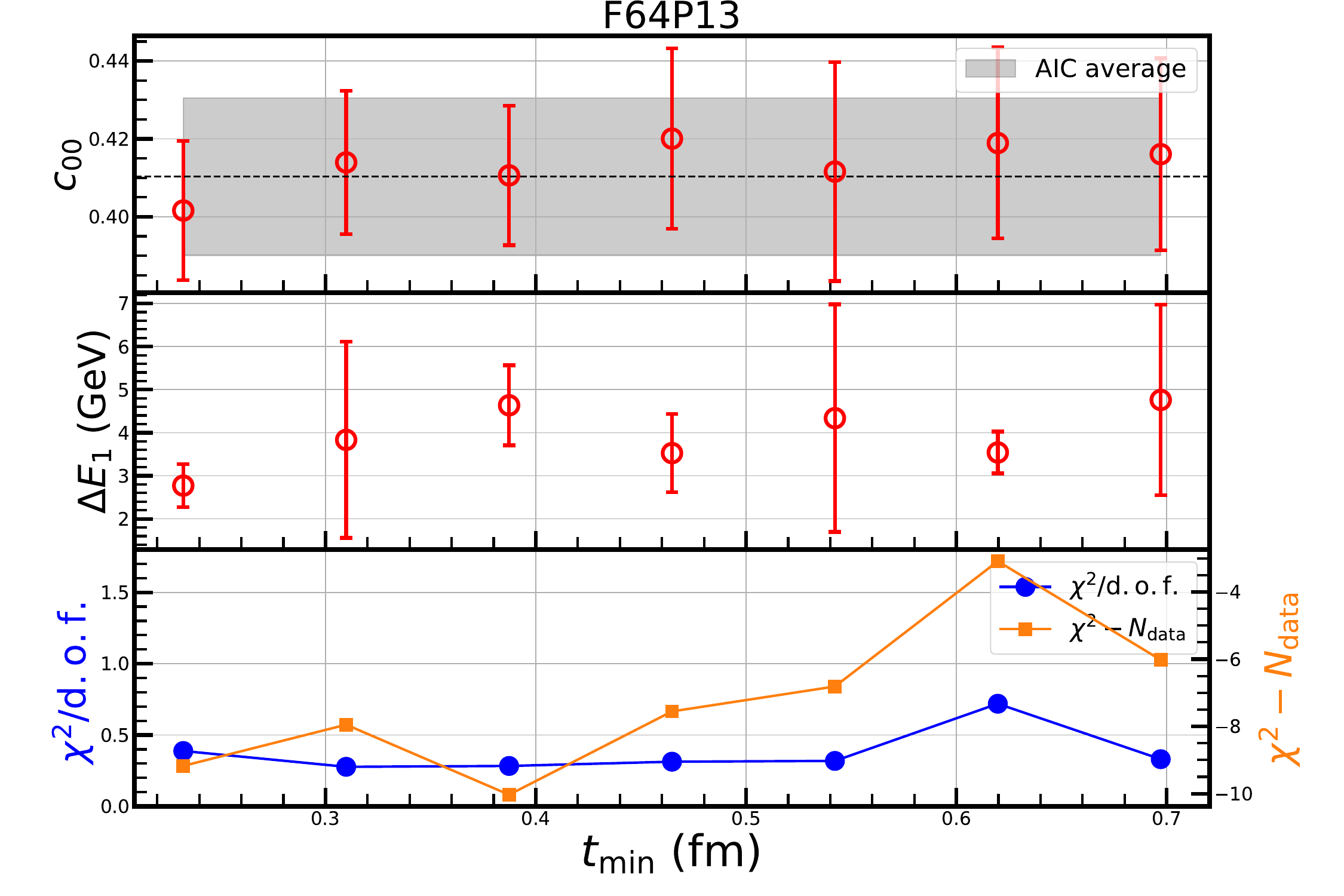}
\caption{
    AIC analysis for F48P30 (left) and F64P13 (right). 
    Top panels: $c_{00}$ as a function of $t_{\rm min}$, with the AIC-averaged result shown as the gray band. 
    Middle panels: $\Delta E_1$ in GeV. 
    Bottom panels: $\chi^2/{\rm dof}$ (blue) and $\chi^2-n_i$ (orange). 
}
\label{fig:aic}
\end{figure}

The AIC-averaged result is {$c_{00}^{\rm AIC}=0.452(13)$} for F48P30 as in Fig.~\ref{fig:aic}. The value at the physical point ensemble F64P13 is slightly smaller, as {$c_{00}^{\rm AIC}=0.410(20)$}. 
In both cases the AIC error exceeds the statistical error of any single window, as it accounts for both statistical fluctuations and the spread among fit-window choices.
GEVP analysis suppress excited-state contamination enough well that the fit result is stable against variations in the fit window, and the final uncertainty is dominated by statistics rather than systematics from fit-window choice.

 \begin{table}[htbp]
     \centering
     \caption{Final results of the flavor‑separated results for \(g_A^{q}\) (\(q = u, d, s\)).
        }
       \begin{tabular}{l|c|c|c|c}
       \hline
       \hline
            & $g_A^u$ & $g_A^d$ & $g_A^s$ & $g_A^{\Sigma}$ \\
       \hline
       F48P30, multi-$\bar{N}_{\rm e}$ fit &   0.886(14)    &  -0.350(14)     &  -0.0335(72)  & 0.503(26)                                              \\
       F48P30, GEVP with AIC               &  0.8639(53)    &   -0.3721(53)   &  -0.0405(43)   &  0.452(13)    \\
       \hline
       F64P13, GEVP with AIC &   0.8408(86)      &  -0.3929(86)    &-0.0381(57)   &   0.410(20)     \\
       \hline
       \end{tabular}%
     \label{tab:gA_final}%
   \end{table}%

Finally, we collect the flavor‑separated results for \(g_A^{q}\) (\(q = u, d, s\)) in Table~\ref{tab:gA_final}. These are obtained by combining \(g_A^{u-d}\) from the multi‑\(\bar{N}_e\) fit, with \(g_A^{\Sigma}\) and \(g_A^s\) from either the multi-$\bar{N}_{\rm e}$ fit (for F48P30) and current-involved GEVP (for both F48P30 and F64P13). 

As shown in Table~\ref{tab:gA_final}, the results obtained from the GEVP analysis are generally more precise than those from the multi‑\(N_e\) fit, owing to the substantially better suppression of excited-state contamination. The central values are also slightly lower, indicating that including the current‑involved interpolator is essential for obtaining more reliable results.

In addition, the extracted \(g_A^{u,d,s}\) on the physical‑pion‑mass ensemble F64P13 are systematically lower than those on the F48P30 ensemble (\(m_\pi = 300\) MeV), bringing the final results into better agreement with experimental constraints.
}

\red{
\subsection{The Cost Advantage of the ``Blending'' Method for baryon Matrix Elements}\label{sec:cost_compare}

\subsubsection{Uncertainty scaling on $N_{\mathrm st}$ and saturation}

Numerical tests confirm the following scaling behavior for statistical uncertainties with respect to $N_{\mathrm st}$, in agreement with theoretical expectations:

1. For blended quark bilinear operators without momentum transfer or gauge links, and also off-shell blended quark states: $\delta_1\propto N_{\mathrm st}^{-1/2}$;

2. For blended quark bilinear operators involving either momentum transfer or gauge links: $\delta_2\propto N_{\mathrm st}^{-1}$;

3. For the combinations of N objects in case 1 and M objects in case 2: $\delta \propto \delta_1^N\delta_2^M$.

     \begin{figure}[!h]
        \includegraphics[width=0.46 \textwidth]{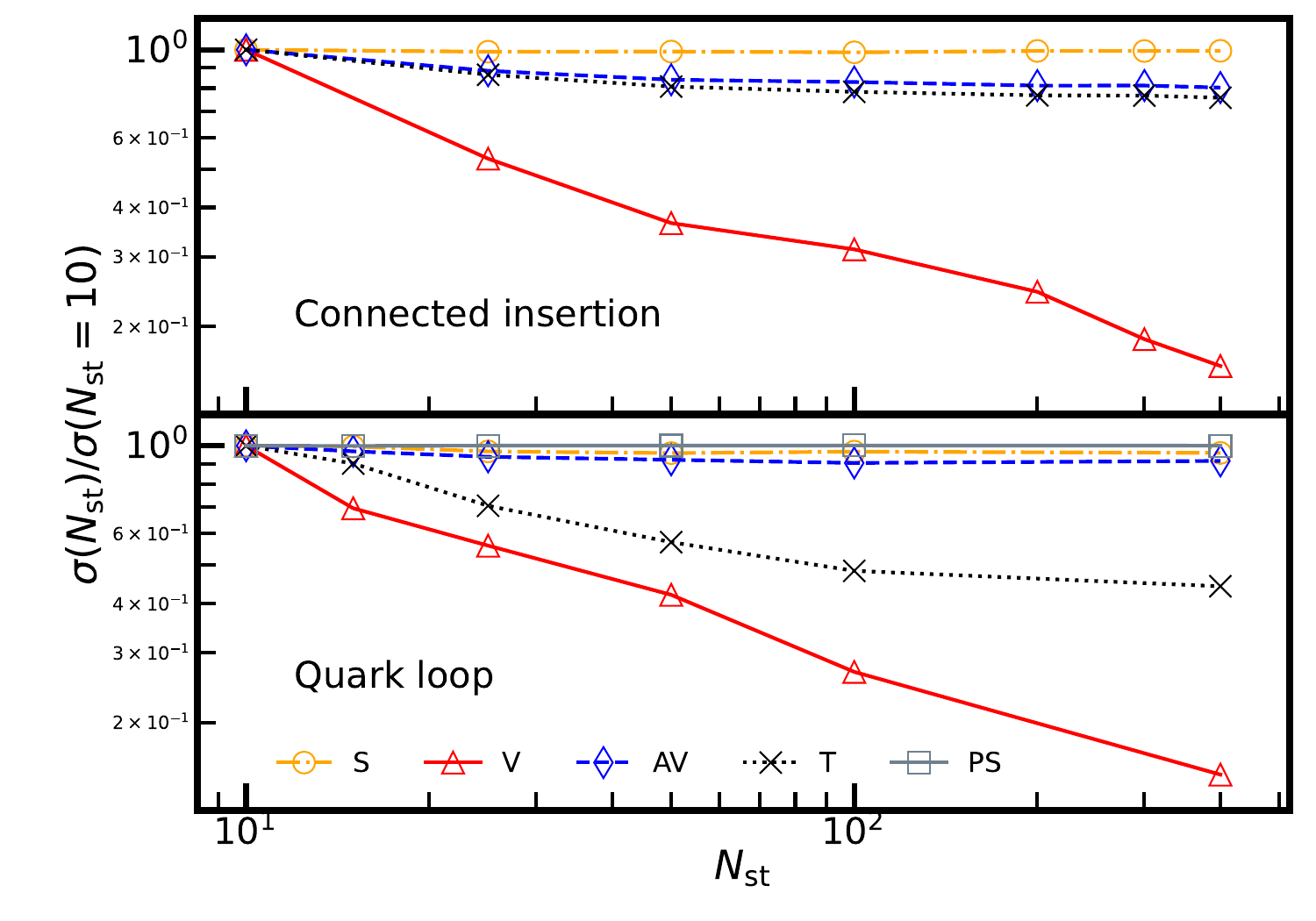}
        \caption{
            Statistical uncertainty of nucleon matrix element ${\cal R}_N^{\bar{u}\Gamma u-\bar{d}\Gamma d}$ (upper panel) and quark loop $\langle \bar{q}\Gamma q\rangle$ (lower panel) with different gamma matrices: scalar ($\Gamma={\cal I}$), pseudo-scalar ($\Gamma=\gamma_5$), vector ($\Gamma=\gamma_{\mu}$), axial-vector ($\Gamma=\gamma_5\gamma_{\mu}$) and tensor ($\Gamma=\gamma_{\mu}\gamma_{\nu}$), varies with number of noise vectors $N_{\mathrm st}$.
        }
        \label{Fig:error_local_operator}
    \end{figure}    

In Fig.~\ref{Fig:CC_C32P32}, we observe that the statistical uncertainty of \( \langle V^{cc}_4 \rangle_{\pi} \) scales approximately as \( N_{\mathrm st}^{-1} \), as the expectation value of \( {\cal R}_{\pi}^{{\mathcal V}} \) is exactly one. Since the random vectors \( \eta_j \) are sampled independently across different time slices and gauge configurations, the number of random vectors \( N_{\mathrm st} \) required to achieve sufficient precision is significantly smaller than \( [\mathcal{L}_2]\), and the vector current represents the most challenging case among quark bilinear operators with different gamma matrices. For instance, the statistical uncertainty of connected nucleon matrix elements with iso-vector currents ${\cal R}_N^{\bar{u}\Gamma u-\bar{d}\Gamma d}$ (upper panel of Fig.~\ref{Fig:error_local_operator}) and quark loops $\langle \bar{q}\Gamma q\rangle$ (lower panel) saturates to the statistical fluctuation for $N_{\rm cfg}=90$ at \( N_{\mathrm st} \leq 100 \) for all other gamma matrices. Additionally, the scalar and pseudo-scalar cases reach saturation with even fewer random vectors, typically \( N_{\mathrm st} \sim 10 \). 

The inversions cost comparison is presented in the following two subsections, using the results for \(g_A^{u-d}\) and \(g_A^{u+d+s}\) as representative cases of connected and disconnected quark diagrams, respectively. The analysis is based on physical-pion ensembles with \(a \sim 0.07\)–\(0.09\) fm and \(L = 64\). We fix \(\bar{N}_{\rm e} = 140\), a value that balances the two competing effects: excited-state contamination in \(g_A^{u-d}\) grows with \(\bar{N}_{\rm e}\), while statistical uncertainty decreases. The number of stochastic vectors is set to \(N_{\mathrm st} = 60\), which proves sufficient for most cases.

The computational advantage of the blending method becomes particularly pronounced when combinations of different interpolating operators are used, as these incur no additional inversion cost. The gain is less dramatic at heavier pion masses; nevertheless, the blending method enables far more efficient reuse of compressed all-to-all propagators.

     \subsubsection{Cost for non-singlet matrix element with the connected insertions only}

    \begin{table}[htbp]
     \centering
     \caption{
        The total inversion cost for nucleon matrix element including $g_A^{u-d},g_S^{u-d},g_T^{u-d},g_V^{u-d}$ among different collaborations at physical point.
        For a fair comparison, we list the inverse cost up to $1\%$ error of $g_A$ (Inv.to 1\% $g_A$) in the last column. The values of $g_S$ and $g_T$ come from Ref.~\cite{Wang:2025nsd} using the same compressed propagators.
     }
    \resizebox{0.99\columnwidth}{!}{
       \begin{tabular}{c|ccccc|c|ccc|c|c}
       \hline
       \hline
       Collaborations & Ensemble & L    & T    & a(fm) & $m_\pi$(MeV) & $n_{\rm cfg}$ & $g^{u-d}_A$& $g^{u-d}_S$& $g^{u-d}_T$ & Inversion & Inv.to 1\% $g^{u-d}_A$ 
       \\
        \hline
       ETM~\cite{Alexandrou:2019brg}   & cB211.072.64 & 64   & 128  & 0.08 & 139  & 750  
       & 1.290(020) &1.35(17)&0.939(027)  & 0.85~M & 2.1~M 
       \\
       RQCD~\cite{Bali:2023sdi}   & D452 & 64   & 128  & 0.076 & 156  & 1000 
       & 1.190(250) &-0.6(3.0)&0.870(110)         
       & 0.01M & 5.3~M 
       \\
       PNDME~\cite{Jang:2023zts} & a09m130 & 64   & 96   & 0.09 & 138  & 1290 
       & 1.320(034) &1.05(23)&1.010(006)   & 1.8~M & 11.9~M 
       \\
       \hline
       \textbf{This Work} & F64P13 & 64 & {{128}} & {{0.078}} & {{134}} & 41 
       & 1.234(008) &1.00(03)   &0.998(004) 
       & {0.35~M} & {{0.15~M}} 
       \\
       \hline
       \hline
       \end{tabular}%
    }
     \label{tab:Cost_CI}%
   \end{table}%

To quantify the computational efficiency of the blending method, we compare its cost—primarily measured by the number of quark propagator inversions required to reach a target statistical precision—against traditional approaches (sequential source or stochastic methods) used in recent high-precision studies.

The total number of inversions (with 12 columns per inversion) for different works are calculated below and summarised in Table~\ref{tab:Cost_CI}.

\begin{itemize}
    \item \textbf{ETM, Sequential Source Method}~\cite{Alexandrou:2019brg}:  
    \(\sum_{t_{\rm sep}} N_{t_{\rm sep}}^{\rm meas} \times N_f \times N_{\rm pol} = (750 + 1500 + 3000 + 4500 + 12000 + 36000 + 48000) \times 2 \times 4 = 0.85\ \mathrm{M}\)

    \item \textbf{RQCD, Stochastic Sequential Source Method} (divided by 12 for stochastic source)~\cite{Bali:2023sdi}:  
    \(n_{\rm cfg} \times (N_{\rm tsep} + N_{\rm noise} / 12) = 1000 \times (4 + 100 / 12) = 0.012\ \mathrm{M}\)

    \item \textbf{PNDME, Sequential Source Method}~\cite{Jang:2023zts}:  
    \(N_{\rm meas}^{\rm LP} \times (1 + n_{\rm pol} \times n_f \times N_{\rm tsep} / N_{\rm sub}) = 165120 \times (1 + 4 \times 2 \times 5 / 4) = 1.8\ \mathrm{M}\)
    \item \textbf{This work, Blending Method}:  
    \(n_{\rm cfg} \times N_T \times (N_e + N_{\mathrm st}) / N_c = 41 \times 96 \times (140 + 60) / 3 = 0.35 \mathrm{M}\)
\end{itemize}
where $N_c=3$ is the SU(3) color degree of freedom. 

As shown in Table~\ref{tab:Cost_CI}, the blending method developed in this work reduces the computational cost—measured in terms of the total number of inversions or equivalent computing resources—by at least an order of magnitude compared to results from collaborations such as ETMC~\cite{Alexandrou:2019brg}, RQCD~\cite{Bali:2023sdi}, and PNDME~\cite{Jang:2023zts} (assuming the cost of the low precision propagator to be 1/3 of the high precision one), when targeting the same statistical precision. 

    \subsubsection{Cost for singlet matrix element which involves disconnected quark loops}

   \begin{table}[htbp]
     \centering
     \caption{
        The total inversion cost of propagator for the nucleon 2-pf and quark loop for $g_A^{u/s,{\rm DI}}$ with 
        similar lattice spacing and pion mass.
    }
       \begin{tabular}{c|c|ccccc|c c|c|c}
       \hline
       \hline
            & Ensemble & L    & T    &  a(fm) & $m_\pi$ & $n_{\rm cfg}$ & $g_A^{s}$   &$g_A^{u+d+s}$ & Inversion & Inv.to 4\% $g^{u+d+s}_A$ \\
       \hline
            PNDME~\cite{Lin:2018obj}     & a09m310 & 64   & 96   & 0.09 & 138  & 847 &  -0.047(27)  & -- & 8.8~M   &--           \\
       \hline
            ETM17~\cite{Alexandrou:2017oeh}      & cA2.09.48        & 48   & 96    & 0.09 & 135  &2136 & -0.021(5) &0.402(35)   & 7.2~M & 34.1~M\\
            ETM19~\cite{Alexandrou:2019brg}     & cB211.072.64 & 64   &128   & 0.08 & 139  & 750 & -0.046(7) &  0.382(31)  &  1.37~M  & 5.63~M      \\          
        \hline
           This Work & F64P13  & 64   &128   & 0.078 & 134  & 41   & -0.038(6)   & 0.410(20) & 0.70~M   &0.70~M          \\
       \hline
       \hline
       \end{tabular}%
     \label{tab:Cost_DI}%
   \end{table}%

Disconnected quark loops are inherently noisy and computationally demanding. Traditional methods require separate inversions for the hadron two-point function (light quarks) and for the light- and strange-quark loops. A more meaningful comparison, therefore, focuses on the additional inversions cost required to obtain the singlet \(g_A^{u+d+s}\) once the connected insertion has been computed. The total inversion counts for different works are summarized in Table~\ref{tab:Cost_DI} and detailed below:

\begin{itemize}
    \item \textbf{PNDME, Stochastic Source Method}~\cite{Lin:2018obj}:  
    \(N^s_{\rm cfg}\times N^s_{\rm inv} = 8.8\ \mathrm{M}\) for \(g_A^s\) alone.

    \item \textbf{ETM17, Stochastic Source Method}~\cite{Alexandrou:2017oeh}:  
    \(N_{\rm cfg}*(N_{\rm 2pt}+N_{\rm loop}^l + N_{\rm loop}^s) = 7.2\ \mathrm{M}\)
    
    \item \textbf{ETM19, Hierarchical Probing Method}~\cite{Alexandrou:2019brg}:  
    \(N_{\rm 2pt} + N_{\rm loop}^l + N_{\rm loop}^s = 600{,}000 + 750 \times 512 \times 2 = 1.37\ \mathrm{M}\)
    
    \item \textbf{This work, Blending Method}:  
    \(n_{\rm cfg} \times N_T \times (N_e + N_{\mathrm st}) / N_c = 41 \times 128 \times (140 + 60) / 3 \times 2 = 0.70\ \mathrm{M}\) for the compressed light and strange quark propagators.
\end{itemize}

The PNDME study~\cite{Lin:2018obj} computed \(g_A^s\) only with the traditional stochastic source method. Their determination of \(g_A^{u+d+s}\) relied on constraints from other ensembles and is therefore not directly comparable here. The blending method achieves a significantly more precise determination with substantially fewer inversions. 

For comparison, the \(N_f = 2\) ETM17 work also employed a stochastic source method and obtained a smaller uncertainty for \(g_A^s\) with a comparable number of inversions. However, this apparent advantage likely benefits from their use of a smaller physical volume.

Using hierarchical probing~\cite{Stathopoulos:2013aci}—an improved variant of space-time dilution~\cite{XQCD:2013odc}—the $N_f=2+1+1$ ETM19 work~\cite{Alexandrou:2019brg} employed roughly 0.4 million strange-quark inversions to reach a comparable precision for \(g_A^s\), in addition to 0.6 million light-quark inversions for the nucleon two-point function and another 0.4 million for the light-quark loop to obtain \(g_A^{u+d+s}\). 

In contrast, the blending method uses a comparable number of strange-quark inversions (0.39 million) not only for the strange-quark loop itself, but also to construct the five-quark interpolator that includes a strange-quark pair. This enables a GEVP analysis that simultaneously incorporates over 300 quark diagrams, effectively suppressing excited-state contamination and yielding a clean ground-state signal at smaller source-sink separations. As a result, we achieve similar 4\% precision for \(g_A^{u+d+s}\) with near an order of magnitude fewer total inversions than Ref. ~\cite{Alexandrou:2019brg}, demonstrating that the precision gained through the blending method remains highly useful even in the presence of disconnected diagram noise.

Moreover, once the compressed light- and strange-quark propagators are available, the blending method enables the computation of \(g_{A,S,T,\dots}^{u,d,s}\) for other baryons (e.g., \(\Lambda\), \(\Xi\), \(\Sigma\), \(\Omega\)) without additional inversions. Such calculations would otherwise require new inversions using traditional sequential methods.}

\end{widetext}

\end{document}